\documentclass[runningheads]{llncs}
\usepackage{multirow}

\usepackage{wrapfig}

\usepackage{graphicx}
\usepackage{xcolor}
\definecolor{darkgreen}{RGB}{34, 139, 34} 
\definecolor{ipcolor}{RGB}{242,170,60}
\definecolor{bpcolor}{RGB}{55,126,247}
\usepackage{xspace}
\usepackage{amssymb,amsmath,epsfig}

\usepackage{amsthm}
\usepackage{amsthm}
\usepackage{hyperref}  % citation link
\hypersetup{
    urlcolor=blue,  % color of external links
}
\usepackage{verbatim}
\usepackage[ruled]{algorithm}
\usepackage[noend]{algpseudocode}
\algrenewcommand\algorithmicindent{0.9em}%

\usepackage{courier}
\usepackage{float}
\usepackage{caption}
\usepackage{subcaption}
\usepackage{tikz}
\usetikzlibrary{shapes,snakes}
\usepackage[capitalise]{cleveref}
\usepackage{tablefootnote}
\usepackage[numbers]{natbib}

\newcommand{\nucA}{\ensuremath{\text{\sc A}}}
\newcommand{\nucC}{\ensuremath{\text{\sc C}}}
\newcommand{\nucG}{\ensuremath{\text{\sc G}}}
\newcommand{\nucU}{\ensuremath{\text{\sc U}}}
\newcommand{\pairs}{\ensuremath{\mathit{pairs}}\xspace}
\newcommand{\unpaired}{\ensuremath{\mathit{unpaired}}\xspace}
\newcommand{\MFE}{\ensuremath{\text{\rm MFE}}\xspace}
\newcommand{\UMFE}{\ensuremath{\text{\rm uMFE}}\xspace}

\newcommand{\LP}{\ensuremath{\mathit{loops}}\xspace}
\newcommand{\CR}{\ensuremath{\mathit{critical}}\xspace}

\newcommand{\D}{\ensuremath{\mathrm{\Delta}}}
\newcommand{\DG}{\ensuremath{\mathrm{\Delta} G^\circ}}
\newcommand{\DDG}{\ensuremath{\mathrm{\Delta\Delta} G^\circ}}

\newcommand{\blangle}{\ensuremath{\boldsymbol{\langle}}}
\newcommand{\brangle}{\ensuremath{\boldsymbol{\rangle}}}
\newcommand{\proj}{\ensuremath{\vdash}}

\DeclareMathOperator{\defeq}{\stackrel{\Delta}{=}}

\renewcommand{\vec}[1]{\ensuremath{\boldsymbol{{#1}}}\xspace}

\newcommand{\vecx}{\ensuremath{\vec{x}}\xspace}
\newcommand{\vecy}{\ensuremath{\vec{y}}\xspace}
\newcommand{\vecz}{\ensuremath{\vec{z}}\xspace}

\newcommand{\ystar}{\ensuremath{\vec{y^\star}}\xspace}

\newcommand{\setxi}{\ensuremath{\hat{X}}\xspace}

\newcommand{\loops}{\ensuremath{\mathit{loops}}\xspace}

\newcommand{\mum}{minimal undesignable motif\xspace}

\newcommand{\um}{undesignable motif\xspace}
\newcommand{\m}{motif\xspace}
\newcommand{\M}{\ensuremath{\vec{m}}\xspace}
\newcommand{\vecm}{\ensuremath{\vec{m}}\xspace}
\newcommand{\ipairs}{\ensuremath{\mathit{ipairs}}\xspace}
\newcommand{\bpairs}{\ensuremath{\mathit{bpairs}}\xspace}
\newcommand{\mstar}{\ensuremath{\vec{m^\star}}\xspace}

\newcommand{\bigO}[1]{\mathcal{O}(#1)}
\renewcommand{\emptyset}{\varnothing}
\newcommand{\card}{\ensuremath{\mathit{card}}\xspace}

\newcommand{\libmum}{\ensuremath{\mathcal{M}_{\text{miniundesignable}}}\xspace}
\newcommand{\libdsm}{\ensuremath{\mathcal{M}_{\text{designable}}}\xspace}
\newcommand{\mrival}{\ensuremath{\mathcal{M}_{\text{rival}}}\xspace}

\usepackage{orcidlink}

\usepackage[T1]{fontenc}
\usepackage{threeparttable}

\begin{document}

%\title{Undesignable RNA Structure Identification via Rival Structure Generation and Structure Decomposition}
\title{Scalable and Interpretable Identification of Minimal Undesignable RNA Structure Motifs with Rotational Invariance}

\author{Tianshuo Zhou\inst{1}\orcidlink{0009-0008-4804-0825} \and
Apoorv Malik\inst{1}\orcidlink{0009-0004-3351-7097} \and
Wei Yu Tang\inst{1,6}\orcidlink{0009-0008-1141-9479} \and
\\ David H. Mathews\inst{3, 4, 5}\orcidlink{0000-0002-2907-6557}  \and
Liang Huang\inst{1,2}\orcidlink{0000-0001-6444-7045} }
\authorrunning{T.~Zhou et al.}
\institute{$^1$ School of EECS and
$^2$ Dept.~of Biochemistry \& Biophysics, Oregon State University, Corvallis OR 97330, USA \\
$^3$ Dept.~of Biochemistry \& Biophysics, 
%University of Rochester Medical Center, Rochester, NY 14642, USA\\
%\and
$^4$ Center for RNA Biology, 
%University of Rochester Medical Center, Rochester, NY 14642, USA\\
and $^5$ Dept.~of Biostatistics \& Computational Biology, University of Rochester Medical Center, Rochester, NY 14642, USA\\
$^6$
Department of Quantitative and Computational Biology, University of Southern California, CA 90089, USA
}
\maketitle              % typeset the header of the contribution
\begin{abstract}
RNA design aims to find a sequence that folds with the highest probability into a designated target structure. However, certain structures are {\em undesignable}, meaning no sequence can 
fold into the target structure under the default (Turner) RNA folding model. 
Understanding the specific local structures (i.e., ``motifs'') that contribute to undesignability is crucial for refining RNA folding models and determining the limits of RNA designability. Despite its importance, this problem has received very little attention,  % until recently.
and previous efforts are neither scalable nor interpretable.

\smallskip

%In this paper, 
We develop a new theoretical framework for motif (un-)designability, and  design scalable and interpretable algorithms to identify minimal undesignable motifs within a given RNA secondary structure. Our approach establishes motif undesignability by searching for rival motifs, rather than exhaustively enumerating all (partial) sequences that could potentially fold into the motif. 
Furthermore, we exploit rotational invariance 
in RNA structures %for the first time 
to detect, group and reuse equivalent motifs
and to construct a database of unique minimal undesignable motifs.
%inherent in RNA folding models 
%for the first time. 
To achieve that, we propose a loop-pair graph representation for %RNA structures and 
motifs and a recursive graph isomorphism algorithm for motif equivalence.

\smallskip
%\textbf{Results:} 
Our algorithms successfully identify 24 unique minimal undesignable motifs among 18 undesignable puzzles from the Eterna100 benchmark. 
Surprisingly, we also find over 350 unique minimal undesignable motifs and 663 undesignable native structures in the ArchiveII dataset, drawn from a diverse set of RNA families. 

\smallskip

\textbf{Availability:} Our source code is available at \\ {\tt\url{https://github.com/shanry/RNA-Undesign}}. \\
Our server is available at \\  {\tt\url{https://linearfold.eecs.oregonstate.edu/motifs}}.
\end{abstract}

%\newpage
% !TEX root = main.tex
%\begin{wrapfigure}{r}{0.6\textwidth}
%%\begin{figure}[h]
%\vspace{-0.6cm}
%    \centering
%    \begin{subfigure}[b]{0.27\textwidth}
%        \centering
%        \includegraphics[width=\textwidth]{figs/eterna_52_ymotif1_mode0.pdf}
%        %\caption{ Motif with the external loop}
%        \label{fig:e52m1}
%    \end{subfigure}
%    \hspace{.1cm}
%    \begin{subfigure}[b]{0.27\textwidth}
%        \centering
%        \includegraphics[width=\textwidth]{figs/eterna_52_ymotif2_mode0.pdf}
%        %\caption{Motif without the external loop}
%        \label{fig:e52m2}
%    \end{subfigure}
%    \vspace{-0.5cm}
%    \caption{Illustration of two minimal undesignable motifs from Eternal00 puzzle \#52 (motif loops in green, boundary pairs in orange, internal pairs in blue).}
%    \label{fig:e52motifs}
%    \vspace{-0.7cm}
%%\end{figure}
%\end{wrapfigure}

\section{Introduction}

%Ribonucleic acid (RNA) plays crucial roles in life by facilitating protein translation, catalyzing reactions, and regulating gene expression~\cite{doudna+cech:2002}. 
%The successful design of RNA molecules with specific functions requires an understanding of RNA structure because structure dictates function. 

RNA secondary structure plays a crucial role in various biological processes, including gene regulation, protein synthesis (translation), and RNA-protein interactions \cite{bose+:2024,chelkowska2021role}.
RNA design~\cite{zhou+:2023samfeo,bellaousov+:2018accelerated,portela:2018unexpectedly,garcia+:2013rnaifold,Zadeh+:2010} focuses on identifying one or more RNA sequences capable of folding into a target secondary structure.

The significance and complexity of RNA design have garnered widespread attention. 
Numerous methods, including SAMFEO \cite{zhou+:2023samfeo}, NEMO \cite{portela:2018unexpectedly}, RNAiFold \cite{garcia+:2013rnaifold}, NUPACK \cite{Zadeh+:2010},  and others~\cite{bellaousov+:2018accelerated}, have been developed to generate sequences based on a given target structure. 
Despite substantial improvements in empirical design quality, certain puzzles in the widely-used benchmark Eterna100 \cite{anderson+:2016} are considered undesignable, having never been successfully solved \cite{koodli+:2021redesigning}. However, limited research~\cite{aguirre+:2007computational} has been dedicated to exploring the undesignability of RNA structures.
Recently, We introduced RIGENDE \cite{zhou+:2023undesignable} to identify undesignable RNA structures in a more general manner by pinpointing rival structures, resulting in the identification of 16 puzzles in Eterna100 deemed undesignable by the unique minimum free energy (MFE) criterion under the standard Turner RNA folding model \cite{Mathews+:2004,turner+:2010nndb} implemented in ViennaRNA \cite{lorenz+:2011} . 

\begin{figure}[h]
    \centering
    \begin{subfigure}[b]{0.32\textwidth}
        \centering
        \includegraphics[width=\textwidth]{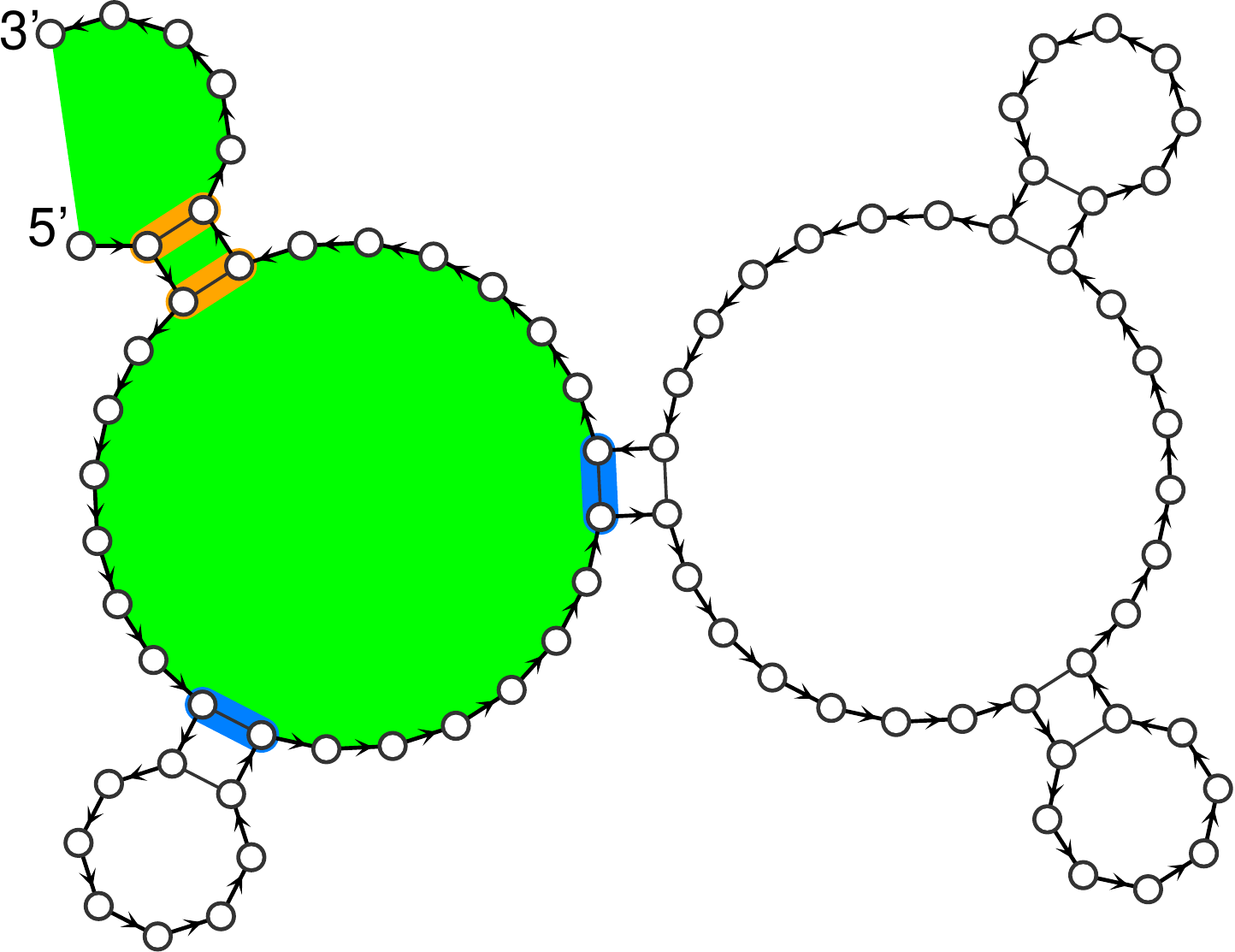}
        %\caption{ Motif with the external loop}
        \label{fig:e52m1}
    \end{subfigure}
    \hspace{.1cm}
    \begin{subfigure}[b]{0.32\textwidth}
        \centering
        \includegraphics[width=\textwidth]{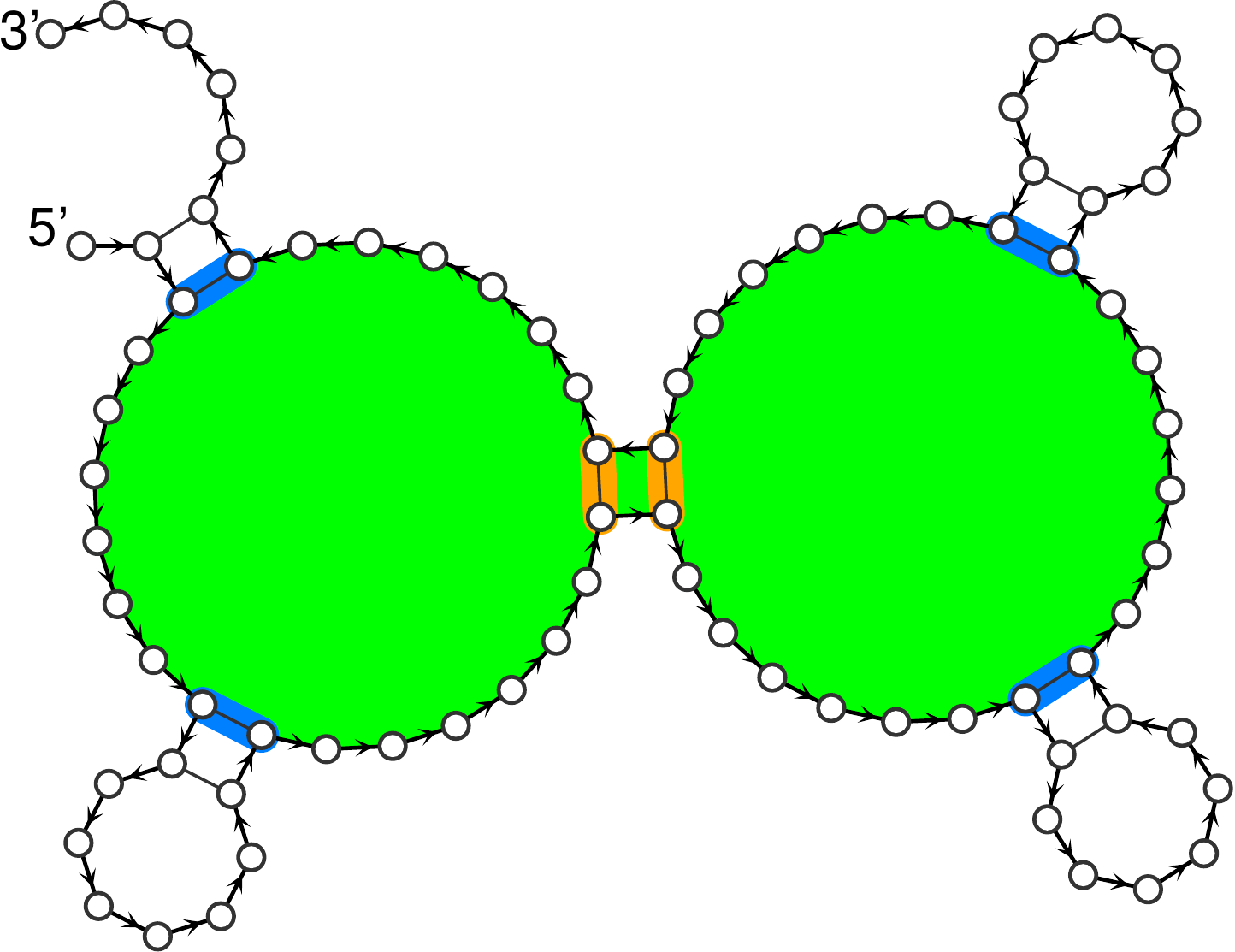}
        %\caption{Motif without the external loop}
        \label{fig:e52m2}
    \end{subfigure}
    \vspace{-0.5cm}
    \caption{Illustration of two minimal undesignable motifs from Eternal00 puzzle \#52 (motif loops in green, boundary pairs in orange, internal pairs in blue).}
    \label{fig:e52motifs}
\end{figure}

While effective, RIGENDE has some limitations  in terms of explainability and scalability. While rival structures, which consistently exhibit better folding energy compared to the target structure, can serve as compelling evidence for an undesignable structure, their interpretability is often limited to the entire structure. In practice, the undesignability of an RNA structure typically arises from some specific local region, or \emph{structure motif}. Identifying these critical motif(s) in a structure could offer deeper insight into why certain structures resist design, enhancing both interpretability and potential reusability in RNA Design. However, important as it is,
this problem has received very little attention,
and the existing efforts for motif designability~\cite{yao+:2019,yao:2021thesis}  are neither scalable nor interpretable (see Sec.~\ref{sec:related} for details).

To address these limitations, we conduct a systematic study of undesignable motifs, introducing general theories and efficient algorithms for identifying \emph{minimal undesignable motifs} (examples shown in Fig.~\ref{fig:e52motifs}) from given RNA secondary structures. Our contributions are as follows:

\begin{enumerate}
\item We develop a new theoretical framework for RNA motif (un-)designability, including new definitions and theorems.
 \item We propose fast algorithms to identify motifs that are undesignable by searching for rival motifs, offering strong explainability. %In experiments, the average time cost per motif is less than 20 seconds.
  \item We exploit rotational invariance in RNA structures to detect, group, and reuse equivalent motifs. To achieve that, we introduce a loop-pair graph representation for motifs and develop a recursive graph isomorphism algorithm to identify unique (undesignable) motifs.
 \item We develop an efficient bottom-up scan algorithm called FastMotif to identify minimal undesignable motifs from RNA structures, with an average time cost of less than 10 seconds per structure in experiments.
 
 \item We identify 24 unique minimal undesignable motifs from 18 puzzles in the Eterna100 benchmark. Moreover, we identify 331 unique minimal undesignable motifs in natural RNA structures from ArchiveII, revealing limitations in the Turner nearest neighbor energy model. In total there are 355 unique minimal undesignable motifs with explainable rivals motifs, and the majority (300+) of them were never proven undesignable before this work.
 \end{enumerate}
% !TEX root = main.tex
\section{RNA Structure and its Undesignability}
\subsection{Secondary Structure, Loop and Free Energy}

An RNA sequence $\vecx$ of length $n$ is specified as a string of  base nucleotides $\vecx_1\vecx_2\dots \vecx_n,$  where $\vecx_i \in \{\nucA, \nucC, \nucG, \nucU\}$ for $i=1, 2,...,n$. 
A secondary structure~$\mathcal{P}$  for $\vecx$ is a set of paired indices where each pair $(i, j) \in \mathcal{P}$ indicates two distinct bases $\vecx_i \vecx_j \in \{\nucC\nucG,\nucG\nucC,\nucA\nucU,\nucU\nucA, \nucG\nucU,\nucU\nucG\}$ and each index from $1$ to $n$ can only be paired once. A secondary structure is pseudoknot-free if there are no two pairs $(i, j)\in \mathcal{P}\text{ and }(k, l)\in~\mathcal{P}$ such that  $i<k<j<l$. Alternatively, $\mathcal{P}$ can be represented as a string~$\vecy=\vecy_1\vecy_2\dots \vecy_n$,  where a pair of indices $(i, j) \in~\mathcal{P}$ corresponds to $\vecy_i=``("$, $\vecy_j=``)"$ and any unpaired index $k$ corresponds to $\vecy_k=``."$. The unpaired indices in $\vecy$ are denoted as $\unpaired(\vecy)$ and the set of paired indices in $\vecy$ is denoted as $\pairs(\vecy)$, which is equal to~$\mathcal{P}$. The lengths of $\vecx$ and $\vecy$ can also be denoted as $|\vecx|$ and $|\vecy|$, respectively. While some RNA structures in nature contain  pseudoknots, we do not consider them in this work as the computational model we use does not allow these. 

\begin{figure}[htbp]
    \begin{minipage}{.40\textwidth}
        \centering
        \includegraphics[width=0.9\textwidth]{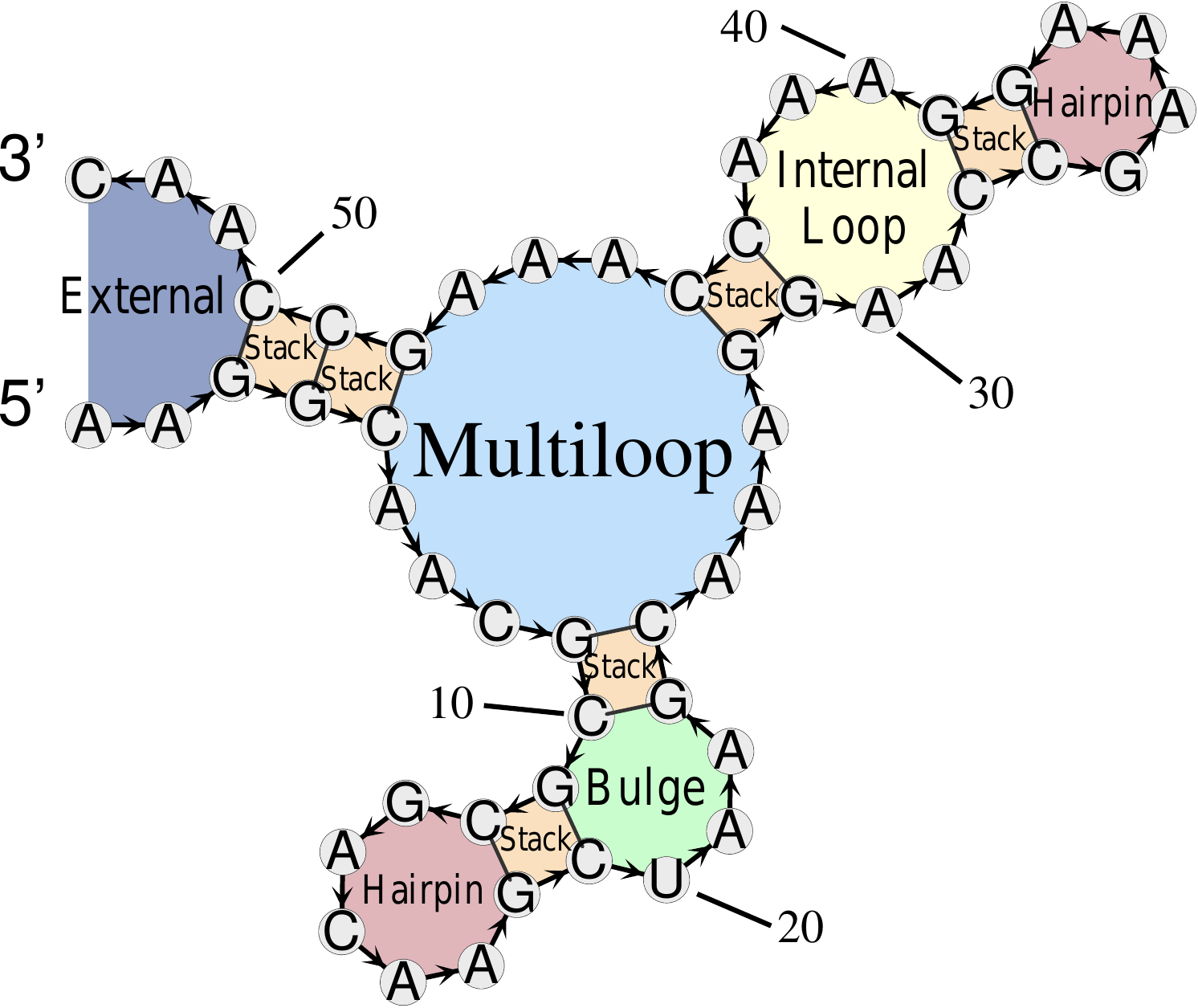}
        \captionof{figure}{Example of secondary structure and loops.} \label{fig:loop_supp}
    \end{minipage}
    \hspace{0.01\textwidth}
    \begin{minipage}{.59\textwidth}
        \centering
        \captionof{table}{Critical positions of loops in Fig.~\ref{fig:loop_supp}.}\label{tab:cr_ex}
        \resizebox{0.99\textwidth}{!}
        {
        \begin{tabular}{|c|c|c|}
            \hline
            \multirow{2}{*}{Loop Type}  & \multicolumn{2}{|c|}{Critical Positions} \\ \cline{2-3}
            & Closing Pairs & Mismatches (Unpaired) \\ \hline
            External & %({\em 5', 3'}), 
            (3, 50) & 2, 51 \\
            Stack & (3, 50), (4, 49) & ~ \\
            Stack & (4, 49), (5, 48) & ~ \\
            Multi & (5, 48), (9, 24), (28, 44) & 4, 49, 8, 25, 27, 45 \\
            Stack & (9, 24), (10, 23) & ~ \\
            Bulge & (10, 23), (11, 19) & ~ \\
            Stack & (11, 19), (12, 18) & ~ \\
            Hairpin & (12, 18) & 13, 17 \\
            Stack & (28, 44), (29, 43) & ~ \\
            Internal & (29, 43), (32, 39) & 30, 42, 31, 40 \\
            Stack & (32, 39), (33, 38) & ~ \\
            Hairpin & (33, 38) & 34, 37 \\ \hline
        \end{tabular}
        }
    \end{minipage} 
\end{figure}
\begin{table}[h]
    \centering
    \captionsetup{justification=centering} % Center-align the caption
    %\captionof{table}{Critical positions for each type of loops under Turner model implemented in ViennaRNA (special hairpins~\cite{Mathews+:2004} of triloops, tetraloops and hexaloops are not considered here)}\label{tab:cr}
    \captionof{table}{Critical positions for each type of loops under Turner model implemented in ViennaRNA (special hairpins~\cite{Mathews+:2004} of triloops, tetraloops and hexaloops are not considered here).}\label{tab:cr}
    \resizebox{\textwidth}{!}
    {
    \begin{tabular}{|c|c|c|}
       \hline
       \multirow{2}{*}{Loop Type}  & \multicolumn{2}{|c|}{Critical Positions}\\ \cline{2-3}
       & Closing Pairs & Mismatches\\ \hline
       External & $(i_1, j_1), (i_2, j_2), \ldots, (i_k, j_k)$ & $(i_1-1, j_1+1), (i_2-1,j_2+1), \ldots, (i_k-1, j_k+1)$\\
       Hairpin & $(i, j)$ & $i+1, j-1$ \\
       Stack & $(i, j), (k, l)$ & ~ \\
       Bulge & $(i, j), (k, l)$ & ~ \\
       Internal & $(i, j), (k, l)$ & $i+1, j-1, k-1, l+1$ \\
       Multi & $(i, j), (i_1, j_1), (i_2, j_2), \ldots, (i_k, j_k)$ & $i+1, j-1, i_1-1, j_1+1, i_2-1, j_2+1, \ldots, i_k-1, j_k+1$\\ \hline
    \end{tabular}
    }
 \end{table}
 
The \emph{ensemble} of an RNA sequence $\vecx$  is the set of all secondary structures that  $\vecx$ can possibly fold into, denoted as $\mathcal{Y}(\vecx)$. The \emph{free energy} $\DG(\vecx, \vecy)$ is used to characterize the stability of $\vecy \in \mathcal{Y}(\vecx)$. The lower the free energy~$\DG(\vecx, \vecy)$, the more stable the secondary structure $\vecy$ for $\vecx$. In  the nearest neighbor energy model~\cite{turner+:2010nndb},  a secondary structure is decomposed into a collection of loops, where each loop is usually a region enclosed by some base pair(s). Depending on the number of pairs on the boundary, the main types of loops include hairpin loop, internal loop and multiloop, which are bounded by 1, 2 and 3 or more base pairs, respectively. In particular, the external loop is the most outside loop and is bounded by two ends ($5'$ and $3'$) and other base pair(s). Thus each loop can be identified by a set of pairs. Fig.~\ref{fig:loop_supp}  showcases an example of secondary structure with loops, and the loops are listed in Table~\ref{tab:cr_ex}.

%\begin{enumerate}
%\item Hairpin: $H\blangle (12, 18) \brangle$.
%\item Bulge: $B\blangle (10, 23), (11, 19) \brangle$.
%\item Stack: $S\blangle (3, 50), (4, 49) \brangle$.
%\item Internal Loop: $I\blangle (29, 43), (32, 39)\brangle$.
%\item Multiloop: $M\blangle (5, 48), (9, 24), (28, 44)\brangle$.
%\item External Loop: $E\blangle(3, 50) \brangle$.
%\end{enumerate}

The function $\LP(\vecy)$ is used to denote the set of loops in a structure $\vecy$. The free energy of a secondary structure $\vecy$ is the sum of the free energy of each loop, $ \DG(\vecx, \vecy) = \sum_{\vecz \in \loops(\vecy)} \DG(\vecx, \vecz),$
%\begin{equation}
%  \DG(\vecx, \vecy) = \sum_{\vecz \in \loops(\vecy)} \DG(\vecx, \vecz),
%\end{equation}
where each term $ \DG(\vecx, \vecz)$ is the energy for one specific loop in $\LP(\vecy)$. Refer to  RIGENDE~\cite{zhou+:2023undesignable} for detailed energy functions for all types of loops in the Turner model \cite{turner+:2010nndb}. The energy of each loop is typically determined by the identity of nucleotides on the  positions of enclosing pairs and their adjacent mismatch positions, which are named as \emph{critical positions} in this article. Table~\ref{tab:cr_ex} lists the critical positions for all the loops in Fig.~\ref{fig:loop_supp} and Table \ref{tab:cr} shows the indices of critical positions for each type of loops. Additionally, some \emph{special hairpins}~\cite{Mathews+:2004} of unstable triloops and stable tetraloops and hexaloops in Turner model have a separate energy lookup table. When evaluating the energy of a loop, it suffices to input only the nucleotides at its critical positions, i.e.,
\vspace{-0.2cm}
\begin{equation}
 \DG(\vecx, \vecy) = \sum_{\vecz \in \loops(\vecy)} \DG(\vecx, \vecz) = \sum_{\vecz \in \loops(\vecy)} \DG(\vecx \proj \CR(\vecz), \vecz), \label{eq:e_sum}
 \vspace{-0.2cm}
\end{equation}
%where  $\CR(\vecz)$ denotes the critical positions of loop $\vecz$ and  $\vecx\proj \CR(\vecz)$ denotes the nucleotides from $\vecx$ that are ``projected" onto $\CR(\vecz)$. See {Supplementary Section~\ref{sec:op}} for the detailed functionality of projection operator. The projection (\proj) allows us to focus on only the nucleotides that are relevant to free energy evaluation. 
where $\CR(\vecz)$ denotes the critical positions of loop $\vecz$, and $\vecx\proj \CR(\vecz)$ represents the nucleotides from $\vecx$ that are ``projected'' onto $\CR(\vecz)$. For a detailed explanation of the projection operator, see Supplementary Section~\ref{sec:op}. This projection ($\proj$) enables us to focus exclusively on the nucleotides relevant to free energy evaluation.
\subsection{Undesignability by the Unique \MFE Criterion}
The structure with the \emph{minimum free energy} is  the most stable structure in the ensemble. A structure $\ystar$ is an \MFE structure of $\vecx$ if and only if
\vspace{-0.1cm}
\begin{equation}
%\MFE(\vecx) \defeq \argmin_{\vecy \in \mathcal{Y} (\vecx)} \DG(\vecx, \vecy).
 ~\forall \vecy \in \mathcal{Y}(\vecx)  \text{ and }\vecy \ne \ystar , \DG(\vecx, \ystar) \leq  \DG(\vecx, \vecy). \label{eq:umfe}
\end{equation}

\vspace{-0.1cm}
%Notice that ties for the $\argmin$ are broken arbitrarily {(technically, $\MFE(\vecx)$ should be a set)}, thus there could be multiple \MFE structures for  $\vecx$. 
RNA design is  the inverse problem of RNA folding. Given a target structure $\ystar$, RNA design aims to find suitable RNA sequence $\vecx$ such that $\ystar$ is an \MFE structure of $\vecx$. Here we follow a more strict definition of \MFE criterion adopted in some previous studies~\cite{bonnet+:2020designing,halevs+:2015combinatorial,yao+:2019,ward+:2022fitness,zhou+:2023samfeo} on the designability of RNA, i.e., $\vecx$ is a correct design if and only if $\vecy$ is the only \MFE structure of $\vecx$, which we call unique \MFE(\UMFE) criterion to differentiate it from the traditional \MFE criterion. Formally, $\UMFE(x) = \ystar $ if and only if
\vspace{-0.2cm}
\begin{equation}
 ~\forall \vecy \in \mathcal{Y}(\vecx)  \text{ and }\vecy \ne \ystar , \DG(\vecx, \ystar) <  \DG(\vecx, \vecy). \label{eq:umfe}
\end{equation}

\vspace{-0.1cm}
Following previous work~\cite{halevs+:2015combinatorial,zhou+:2023undesignable} on undesignability, we define the undesignability based the \UMFE criterion, i.e., the condition in Eq.\ref{eq:umfe} can not be satisfied for any RNA sequence $\vecx$ given a target structure $\ystar$. Alternatively, we give the formal definition of undesignability as follows.
\vspace{-0.1cm}
\begin{definition}\label{def:umfe}
 An RNA secondary structure $\ystar$ is undesignable by \UMFE criterion if and only if
 \vspace{-0.1cm}
 \begin{equation}
 \forall \vecx, \exists \vec{y'} \neq \ystar, \DG(\vecx, \vec{y'}) \leq \DG(\vecx, \ystar). \vspace{-0.37cm}
 \end{equation}
\end{definition}
% !TEX root = main.tex
\section{Motif and its Undesignability}
Recent work, RIGENDE~\cite{zhou+:2023undesignable}, has demonstrated that some structures (puzzles) in Eterna100 are undesignable by the \UMFE criterion. For instance, the puzzle ``\texttt{[RNA] Repetitive Seqs. 8/10}'' in Fig.~\ref{fig:e52motifs} is proven undesignable because a rival structure  always has a lower free energy than the target structure . 
%Specifically, for any possible RNA sequence, at least one structure from the rival set will have a lower free energy than the target structure of the puzzle ``\texttt{[RNA] Repetitive Seqs. 8/10}''. 
However, such explainability remains at the whole-structure level rather than a local level. Ideally, an undesignability identification method should not only verify that a structure is undesignable but also pinpoint specific local regions within structures that causes undesignability. We refer to these regions as \emph{undesignable motifs}.

The smaller undesignable motifs we identify, the deeper we can understand undesignability or designability of secondary structures. Thus, the goal is to identify {\em minimal undesignable motifs}. For example, our proposed algorithm identified two minimal undesignable motifs within the puzzle ``\texttt{[RNA] Repetitive Seqs. 8/10}'', highlighted in Fig.~\ref{fig:e52motifs}. Previous efforts~\cite{yao+:2019,yao:2021thesis} based on exhaustive search failed to identify them due to scalability limitations.
%Both motifs were missed by CountingDesign, as its method would take years to enumerate and fold all possible (partial) sequences. (CountingDesign is only feasible for very short motifs.)
Moreover, there are two major issues with the motif definition proposed by previous works~\cite{yao+:2019,yao:2021thesis}:
 \begin{enumerate} 
 \item Their definition excludes external loop regions, as it requires a motif to begin with a base pair. Consequently, the minimal undesignable motifs (as defined by us) shown in Fig.~\ref{fig:e52motifs} would not be recognized as motifs in their works.
 \item They define a motif as a rooted tree, where each node represents either a base pair or an unpaired base. This definition translates the bases direclty into motifs, disregarding the concept of loops and lacking a meaningful abstraction at the physical level.
 \end{enumerate}

%We argue that the most effective way to define a motif is by focusing on loop composition, aligning with how RNA structures are composed of different loop types.
%In this way,  a motif serves as a generalization of RNA structure. Therefore, we introduce our formal definition of motifs in the following subsection.
We propose that the most effective way to define a motif is by emphasizing loop composition, reflecting the fundamental organization of RNA structures into distinct loops. In this sense, a motif generalizes RNA structure. Accordingly, we introduce our formal definition of motifs in the following subsection.
\begin{wrapfigure}{l}{0.46\textwidth}
\vspace{-0.4cm}
    \centering
  \begin{subfigure}[b]{0.2\textwidth}
    \includegraphics[width=\textwidth]{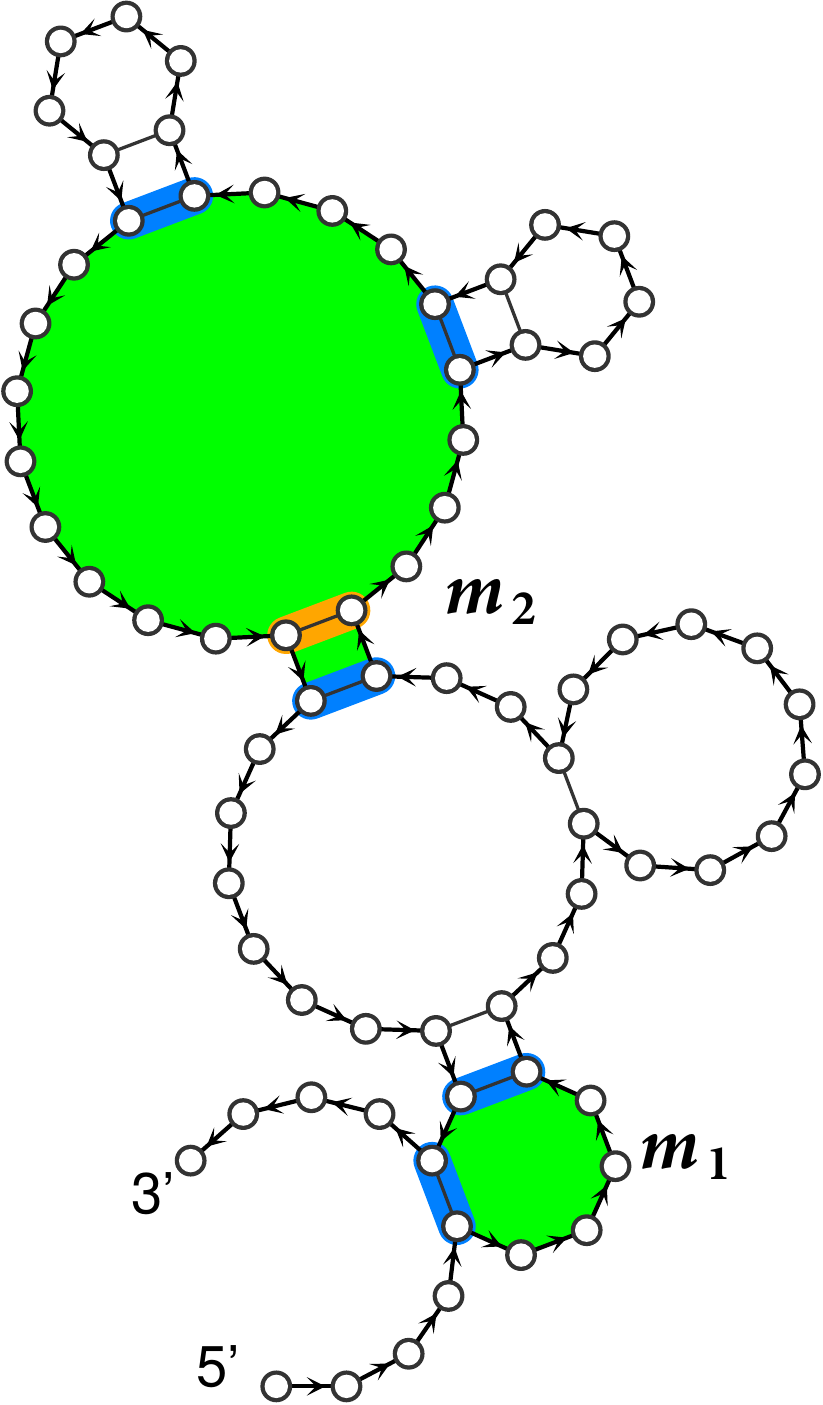}
  \end{subfigure}
  \begin{subfigure}[b]{0.2\textwidth}
    \includegraphics[width=\textwidth]{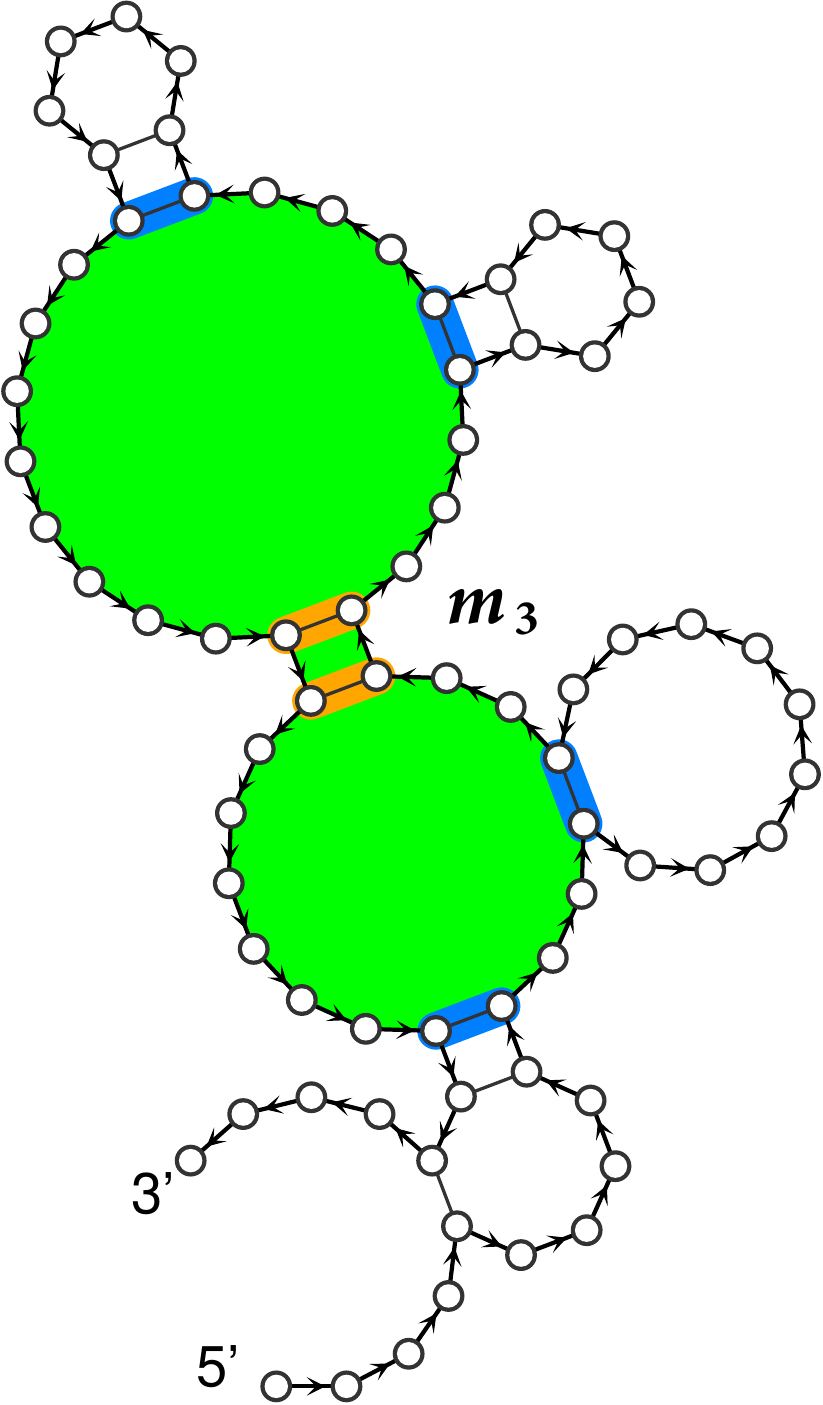}
  \end{subfigure}
  \caption{Motifs with various cardinalities (numbers of loops):
  $\card(\M_1)\!=\!1$, $\card(\M_2)\!=\!2$, $\card(\M_3)\!=\!3$.
  Loops are in green, internal pairs ($\ipairs$) in orange and {boundary pairs} ($\bpairs$) in blue.}\label{fig:card}
  \vspace{-1.2cm}
\end{wrapfigure}
\vspace{-0.4cm}
\subsection{Motif is a Generalization of Structure}\label{subsec:motif}
\begin{definition}\label{def:motif-loops}
A \m \vecm is a contiguous (sub)set of loops in an RNA secondary structure \vecy, notated $\vecm\subseteq\vecy$.
 \end{definition}
Many functions defined for secondary structures can also be applied to motifs. For example, $\LP(\M)$ represents the set of loops within a motif \M, while $\pairs(\M)$ and $\unpaired(\M)$ represent the sets of base pairs and unpaired positions, respectively. 
We define the \emph{cardinality} of \vecm  as the number of loops in \vecm , i.e., $\card(\vecm) = |\loops(\vecm)|$.
Fig.~\ref{fig:card} illustrates three motifs, $\M_1, \M_2$, and $\M_3$, in a structure adapted from the Eterna puzzle ``\texttt{Cat's Toy}''. These motifs contain 1, 2, and 3 loops, respectively. 
We also define the \emph{length} of a motif $|\vecm|$ as the number of bases it contains, which is consistent with the length of a secondary structure $|\vecy|$.

Since motifs are defined as sets of loops, we can conveniently use set relations to describe their interactions.
A motif $\M_A$ is a \emph{sub-motif} of another motif $\M_B$ if $\M_A$ is contained within $\M_B$, denoted as $\M_A \subseteq \M_B$. For the motifs in Fig.~\ref{fig:card}, we observe the relation $\M_2 \subseteq \M_3$. We further use $\M_A \subset \M_B$ to indicate that $\M_A$ is a proper sub-motif of $\M_B$, meaning $\M_A \neq \M_B$. Therefore, $\M_2 \subset \M_3$. The entire structure $\vecy$ can be regarded as the largest motif within itself, and accordingly, $\M \subseteq \vecy$ signifies that motif \M is a part of structure $\vecy$, with $\M \subset \vecy$ implying \M is strictly smaller than \vecy.

%The loops in a motif are connected via base pairs. Each pair in $\pairs(\M)$ either connects two loops within the motif, or one loop in the motif and the other loop outside the motif. We call them \emph{internal pair} and \emph{boundary pair}, and use $\ipairs(\M)$ and $\bpairs(\M)$ to denote the set of {internernal pairs} and {boundary pairs} in a motif \M respectively. Apparently,
The loops in a motif \M are connected by base pairs. Each base pair in $\pairs(\M)$ is classified as either an \emph{internal pair} linking two loops in \M or a \emph{boundary pair} connecting one loop inside \M to one outside. These two types of pairs in \M are denoted as disjoint sets $\ipairs(\M)$ and $\bpairs(\M)$, respectively:
\begin{align}
\ipairs(\M) \cap \bpairs(\M) = \emptyset,~\ipairs(\M) \cup \bpairs(\M) = \pairs(\M).
\end{align}
%Utilizing the widely adopted nearest neighbor model for RNA folding introduces the possibility that certain motifs may remain elusive in structures folded from RNA sequences. For instance, the motif $\M_3$ is deemed undesignable, as the removal of its two internal pairs consistently results in lower free energy. In this context, we present the definition of \emph{\um}.
Utilizing the commonly accepted nearest neighbor model for RNA folding, it becomes evident that certain motifs may be absent from structures folded from RNA sequences. For instance, motif $\M_3$ in Fig.~\ref{fig:card} is considered undesignable, as the removal of its two internal pairs consistently reduces the free energy. This brings us to the definition of an  \emph{\um}.

\subsection{Motif Ensemble from Constrained Folding}
The designability of motifs is based on \emph{constrained folding}. Given a sequence $\vecx$, a structure in its ensemble $\vecy \in \mathcal{Y}(\vecx)$, we can conduct constrained folding by constraining the loops outside $\M$, i.e., $\vec{c} = \vecy\setminus\M$. We generalize the concept of (structure) \emph{ensemble} to \emph{motif ensemble} as the set of motifs  that \vecx can possibly fold into given the constraint $\vecy\setminus\vecm$, denoted as $\mathcal{M}(\vecx, \vecy\setminus\vecm)$. Motifs in $\mathcal{M}(\vecx, \vecy\setminus\vecm)$ have the same boundary pairs, i.e., \vspace{-0.1cm}
\vspace{-0.1cm}
\begin{equation}
\forall \vecm', \vecm'' \in \mathcal{M}(\vecx, \vecy\setminus\vecm), \bpairs(\vecm') = \bpairs(\vecm'') = \bpairs(\vecm). \vspace{-0.2cm}
\end{equation}
\vspace{-0.1cm}
The \emph{free energy} of a motif $\vecm$ is the sum of the free energy of the loops in \vecm,
\vspace{-0.1cm}
\begin{equation}\label{eq:emotif}
\DG(\vecx, \vecm) = \sum_{\vecz\in \loops(\vecm)}  \DG(\vecx, \vecz).
\vspace{-0.2cm}
\end{equation} 
%\vspace{-0.2cm}
The definitions of \MFE and \UMFE can also be generalized to motifs via constrained folding.
\begin{definition}
A motif $\mstar \subseteq \vecy$ is an \MFE motif of folding $\vecx$ under constraint $\vecy\setminus\mstar$ , i.e., $\MFE(\vecx, \vecy\setminus\mstar)$, if and only if 
\begin{equation}
%\vspace{-0.1cm}
\forall \vecm \in \mathcal{M}(\vecx, \vecy\setminus\mstar)  \text{ and }\vecm \ne \mstar , \DG(\vecx, \mstar) \leq  \DG(\vecx, \vecm). \label{def:mfe-motif}
%\vspace{-0.1cm}
\end{equation}
\end{definition}
\begin{definition} \label{def:umfe-motif}
A motif $\mstar \subseteq \vecy$ is an \UMFE motif of folding $\vecx$ under constraint $\vecy\setminus\mstar$ , i.e., $\UMFE(\vecx, \vecy\setminus\mstar)$, if and only if 
\begin{equation}
\forall \vecm \in \mathcal{M}(\vecx, \vecy\setminus\mstar)  \text{ and }\vecm \ne \mstar , \DG(\vecx, \mstar) <  \DG(\vecx, \vecm). \label{eq:umfe-motif} 
\end{equation}
\end{definition}
\subsection{Undesignability of Motif}
The designability and undesignability of motifs by \UMFE criterion can be defined based on Def.~\ref{def:umfe-motif}.
\begin{definition}\label{def:um}
A motif $\mstar \subseteq \vecy$ is an \um by \UMFE criterion if and only if \vspace{-0.28cm}
\begin{equation}
 \forall \vecx, \exists \vecm' \neq \mstar, \DG(\vecx, \vec{m'}) \leq \DG(\vecx, \mstar). %\vspace{-0.28cm}
 \end{equation}
\end{definition}
Similarly, we can establish the definition of designable motifs.
\begin{definition}
A motif  $\mstar \subseteq \vecy$ is a designable motif by \UMFE criterion if and only if \vspace{-0.28cm}
\begin{equation}
 \exists \vecx,  \mstar = \UMFE(\vecx, \vecy\setminus\mstar). \label{eq:motif_umfe} %\vspace{-0.28cm}
 \end{equation}
\end{definition}
Moreover, if $\M$ is undesignable, any motif or structure containing \M will be undesignable.

\begin{theorem}\label{theorem:m2m}
 If a motif $\mstar$ is undesignable, then any motif $\vecm$ such that $\mstar \subseteq \vecm$ is undesignable.
 \end{theorem}
 \begin{proof}
 By Def.~\ref{def:um}, $ \forall \vecx, \exists \vecm' \neq \mstar, \DG(\vecx, \vec{m'}) \leq \DG(\vecx, \mstar)$. We can construct a motif $\vecm'' \neq \vecm$ by substituting  the loops of $\mstar$ within $\vecm$ with $\vecm''$ such that $\loops(\vecm'')=\loops(\vecm)\setminus\loops(\mstar)\cup\loops(\vecm')$. As a result, $ \forall \vecx, \exists \vecm'' \neq \vecm, \DG(\vecx, \vecm'') - \DG(\vecx, \vecm)=  \DG(\vecx, \vecm') -  \DG(\vecx, \mstar) \leq 0$.
 \end{proof}
 \begin{corollary}\label{coro:m2y}
 If a motif $\mstar$ is undesignable, then any  structure $\vecy$ such that $\mstar \subseteq \vecy$ is also undesignable. 
 \end{corollary}
\begin{proof}
%As stated in Sec.~\ref{subsec:motif}, 
 The structure $\vecy$ is the largest motif in $\vecy$. Thus the correctness of Corollary~\ref{coro:m2y} follows Theorem~\ref{theorem:m2m}. 
 \end{proof}
 \vspace{-0.28cm}
 \begin{corollary}\label{coro:y2m}
 If a motif $\vecm$ is designable, then any sub-motif $\vecm_{\text{sub}} \subseteq \vecm$ is also designable. As a special case, if a structure \vecy is designable, then any motif \vecm within \vecy, i.e., $\vecm \subseteq \vecy$ is also designable.
 \end{corollary}
 \begin{proof}This follows as the contrapositive of Theorem~\ref{theorem:m2m}, thus completing the proof. \end{proof}
According to Theorem~\ref{theorem:m2m}, we can access the undesignability of a big motif by inspecting its sub-motifs. 
Therefore, it is crucial and valuable to determine the minimality of an undesignable motif. We introduce the concept of \emph{minimal undesignable motif}.
\begin{definition}
A motif \M is a minimal undesignable motif if and only if the two conditions both hold: (1) \M is an undesignable motif, and (2) $\forall \M' \subset \M$, $\M'$ is designable.
\end{definition}

By this definition, the motif $\M_3$ in Fig.~\ref{fig:card} is a \mum because all its proper sub-motifs are designable. Since the minimality is based on the concept of loops, one fundamental question is that what's the least number of loops a \mum can contain. Therefore, it is worthwhile to prove that any motif composed of a single loop is designable.
\begin{theorem}\label{theorem:m1loop}
 If a motif $\mstar$ is composed of one loop, i.e., $\card(\mstar) = 1$, then $\mstar$ is designable.
 \end{theorem}
 %Proof is detailed in Supplementary Section~\ref{sec:oneloop}.
 \begin{proof}
If $|\loops(\mstar)| = 1$, then there is no internal pairs, $\ipairs(\mstar) = \emptyset$. Let each paired position in $\mstar$ have a base pair \texttt{($\nucC, \nucG$)} or \texttt{($\nucG, \nucC$)} and unpaired position in $\mstar$ have a nucleotide \texttt{$\nucA$}, then no internal pair can be formed. $\mstar$ is the only motif in the motif ensemble of constrained folding, i.e.,  $ \mathcal{M}(\vecx, \vecy\setminus\mstar)= \{\mstar\}$.
\end{proof}
It turns out that it is possible for a two-loop motif to be minimal undesignable (as illustrated in Table~\ref{table:puzzles}).
The primary issue is how to identify undesignable motifs. A trivial solution is to enumerate all the (partial) sequences for the target motif \mstar and check the when \mstar is a \UMFE motif after constrained folding.  However, it is impractical for long motifs because of exponentially high time cost. See Supplementary Section~\ref{sec:bf} for a detailed discussion. The existing work CountingDesign \cite{yao+:2019,yao:2021thesis} is better than exhaustively enumerating sequences for each motif one by one, yet in essence it is still an exhaustive enumeration method. As a result, it cannot scale to long motifs and the found undesignable motifs lack interpretability. To provide scalability but also interpretability, we borrow the philosophy of rival structure from RIGENDE~\cite{zhou+:2023undesignable} and  propose to utilize rival motif to establish the undesignability of motifs, which is discussed in the following Section \ref{sec:rival}.
 
% !TEX root = main.tex
\vspace{-0.4cm}
\section{Rival Motifs Identification}\label{sec:rival}
\vspace{-0.25cm}
 \subsection{Identify Single Rival Motif}
 \vspace{-0.2cm}
 It is possible that there is another motif $\vecm' $ which always has lower free energy than the target motif $\mstar$, if we can find such a \emph{rival motif}, $\mstar$ is undesignable. For instance, removing the internal pairs of the motif $\vecm_3$ highlighted in Fig.~\ref{fig:card} will yield a rival motif $\vecm'$ that is always energetically favored than $\vecm_3$, proving $\vecm_3$ in Fig.~\ref{fig:card} undesignable by the following theorem. Another example of single rival motif is shown in Fig.~\ref{fig:rm1}.
 \begin{theorem}\label{theorem:onerm}
 If $\exists \vecm' \neq \mstar, \forall \vecx,  \DG(\vecx, \vec{m'}) \leq \DG(\vecx, \mstar)$, then \mstar is undesignable. \vspace{-0.2cm}
\end{theorem}
 The correctness of Theorem~$\ref{theorem:onerm}$ follows Def.~\ref{def:um}.
 According to RIGENDE~\cite{zhou+:2023undesignable}, the energy difference of two structures $\DDG$ is totally decided by their \emph{differential positions}, which can also be applied to two motifs. The condition in Theorem~$\ref{theorem:onerm}$ can be written as \vspace{-0.20cm}
\begin{equation}
%\begin{aligned}
 \exists \vecm' \neq \mstar, \forall \vec\vecx'=\vecx \proj \D(\vecm', \mstar), \DDG(\vec\vecx', \vecm', \mstar) \leq 0~, \label{eq:diff_set2} \vspace{-0.28cm}
% \end{aligned}
\end{equation}
where \vspace{-0.2cm}
\begin{equation}
 \D(\vecm', \mstar) \defeq \bigcup_{z \in \loops(\mstar)\ominus \loops(\vecm')} \CR(z) \label{eq:dp} \vspace{-0.28cm}
\end{equation}
denotes the \emph{differential positions}\footnote{The operator $\ominus$ denotes the symmetric difference (or XOR) between the two sets $A \ominus B = (A \setminus  B) \cup (A \setminus  B)$.}, and \vspace{-0.10cm}
\begin{equation}
\begin{aligned}
&\DDG(\vec\vecx', \vecm', \mstar) \\
\defeq &\hspace{-0.5cm}  \sum_{z' \in \LP(\vecm')\setminus  \LP(\mstar)} \hspace{-0.9cm} \DG(\vecx'\proj \CR(z'), z') 
-  \hspace{-0.5cm} \sum_{z^\star \in \LP(\mstar)\setminus \LP(\vecm')} \hspace{-0.9cm} \DG(\vecx' \proj \CR(z^\star), z^\star) \label{eq:diff_long} 
\end{aligned}
\end{equation}
denotes the \emph{free energy difference} between $\vecm$ and \mstar. Therefore, we can verify a single rival motif by inspecting possible nucleotides  on only these differential positions. Refer to the RIGENDE~\cite{zhou+:2023undesignable} for details. 
\vspace{-0.4cm}
\subsection{Identify Multiple Rival Motifs}
\vspace{-0.2cm}
\begin{figure}
   \centering
    % First group: two figures on the left
    \vspace{-0.5cm}
    \begin{minipage}[b]{0.25\textwidth}
        \centering
        \begin{subfigure}[b]{1.0\textwidth}
            \includegraphics[width=\textwidth, angle=0]{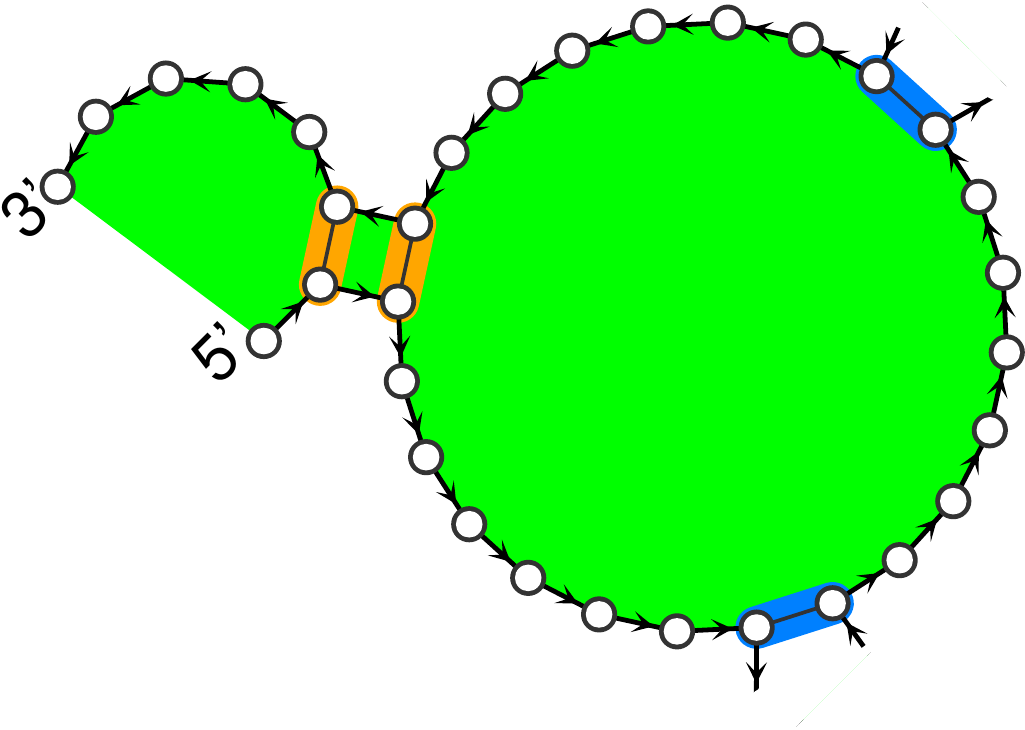}
            \caption{Target motif}
            \label{fig:tm1}
        \end{subfigure}   
        \vspace{-0.2cm}     
        \begin{subfigure}[b]{\textwidth}
	   \includegraphics[width=0.9\textwidth, angle=-90]{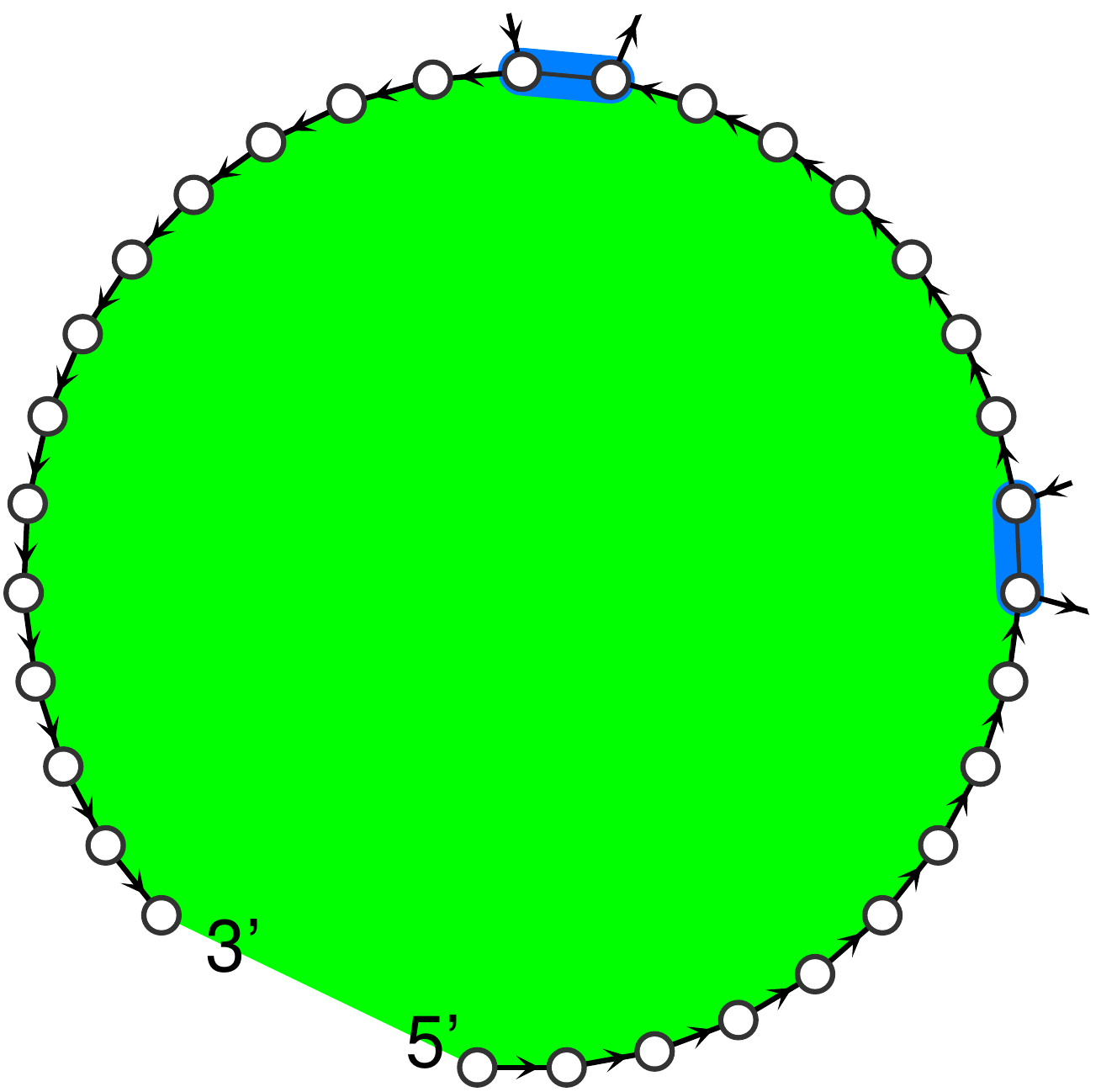}
            \caption{Single rival motif}\label{fig:rm1}
        \end{subfigure}
    \end{minipage}
    \hspace{0.0\textwidth} % Space between minipage and the vertical line
    \vline % Draw a vertical line
    %\hspace{0.02\textwidth} % Space between minipage and the vertical line
    % Second group: six figures on the right
    \begin{minipage}[b]{0.7\textwidth}
        \centering
        \begin{subfigure}[b]{0.3\textwidth}
            \includegraphics[width=\textwidth, angle=-90]{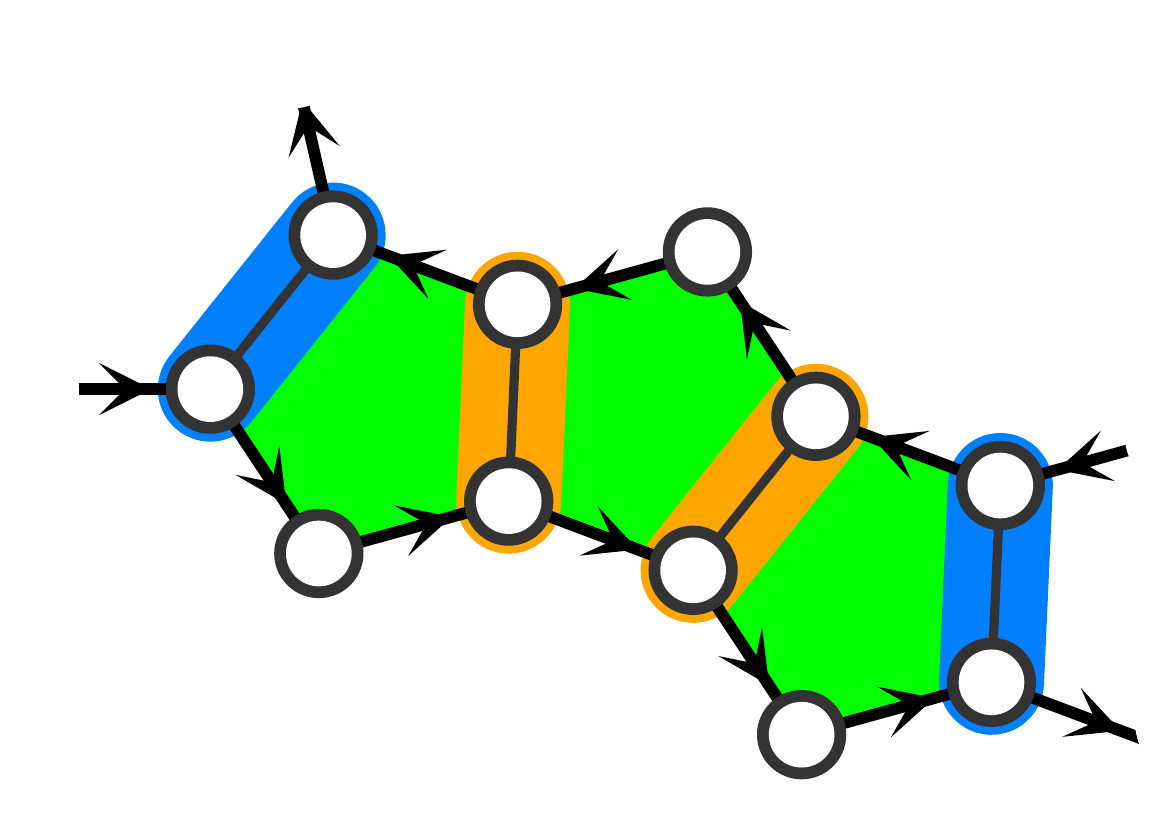}
            \caption{Target motif}
            \label{fig:tm2}
        \end{subfigure}
        \hspace{-0.1cm}
        \begin{subfigure}[b]{0.3\textwidth}
            \includegraphics[width=\textwidth, angle=-90]{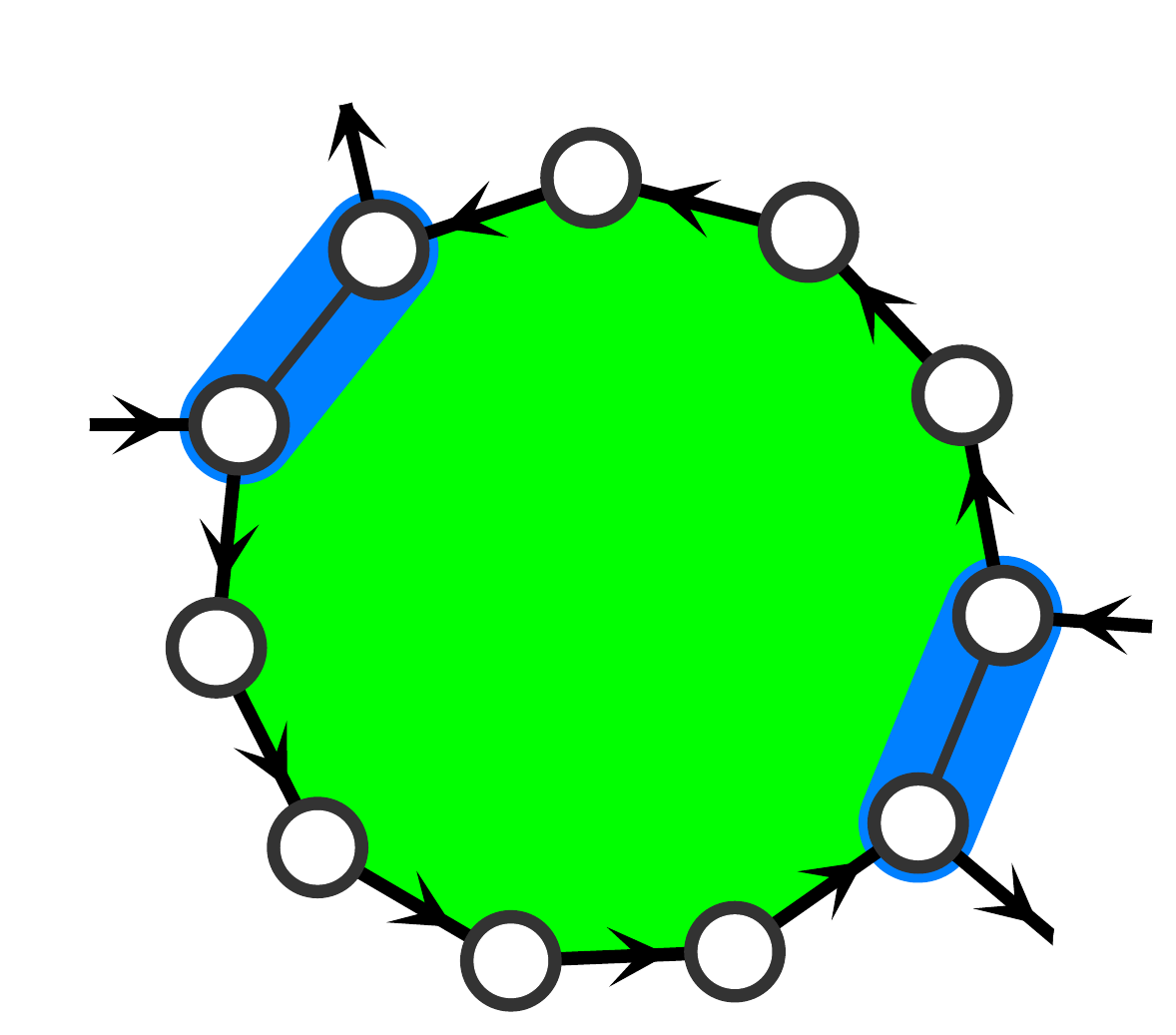}
            \caption{Rival motif 1}
        \end{subfigure}
        \hspace{-0.1cm}
        \begin{subfigure}[b]{0.3\textwidth}
            \includegraphics[width=\textwidth, angle=-90]{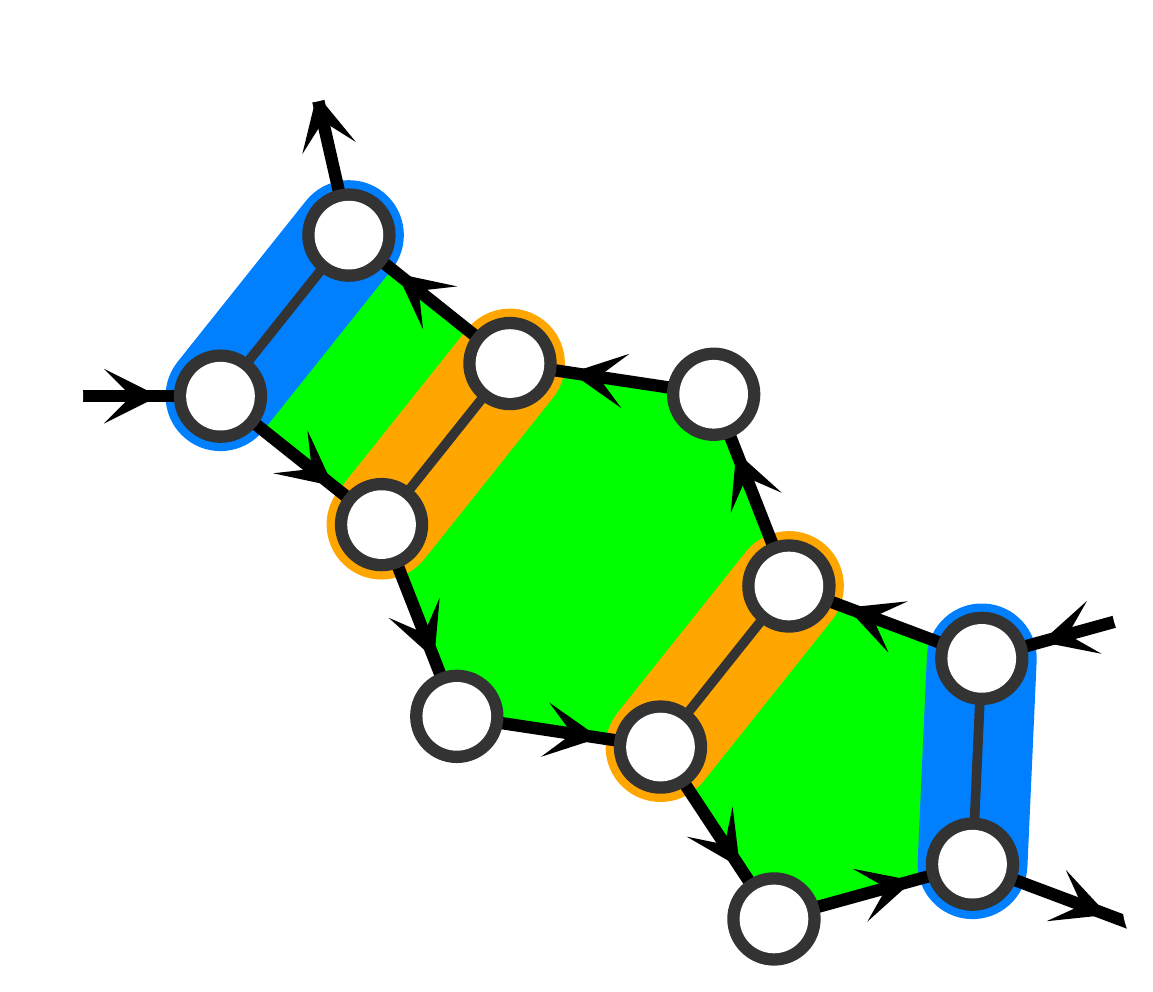}
            \caption{Rival motif 2}
        \end{subfigure}
        
        \begin{subfigure}[b]{0.3\textwidth}
            \includegraphics[width=\textwidth, angle=-90]{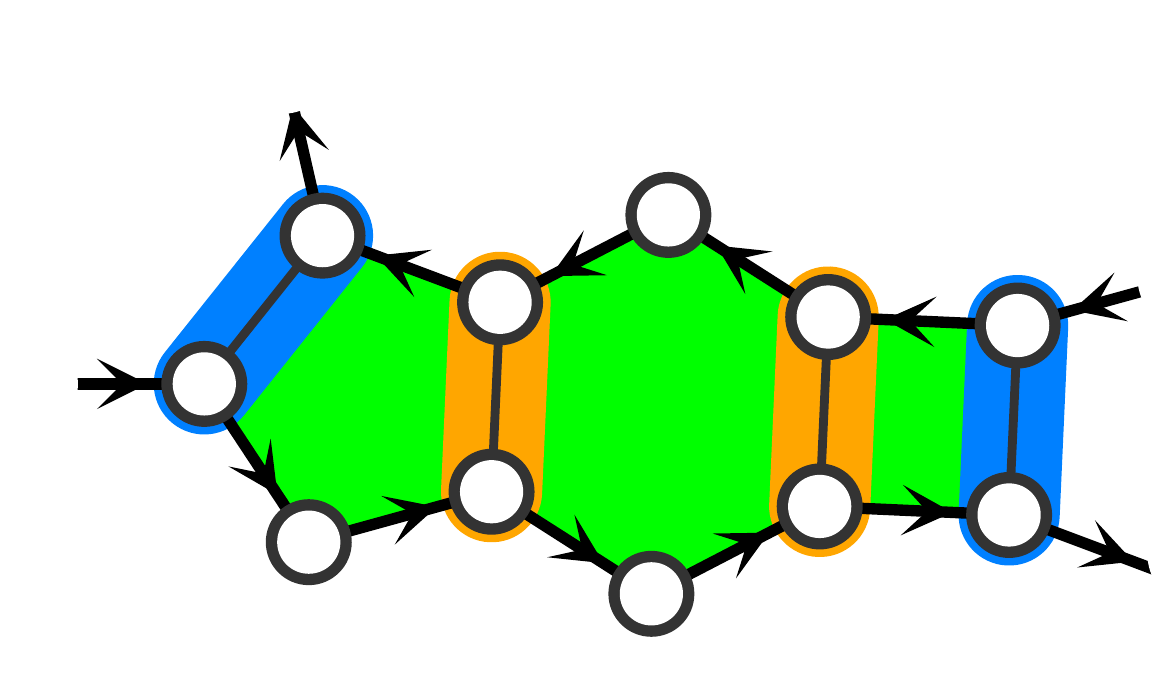}
            \caption{Rival motif 3}
        \end{subfigure}
        \hspace{-0.1cm}
        \begin{subfigure}[b]{0.3\textwidth}
            \includegraphics[width=\textwidth, angle=-90]{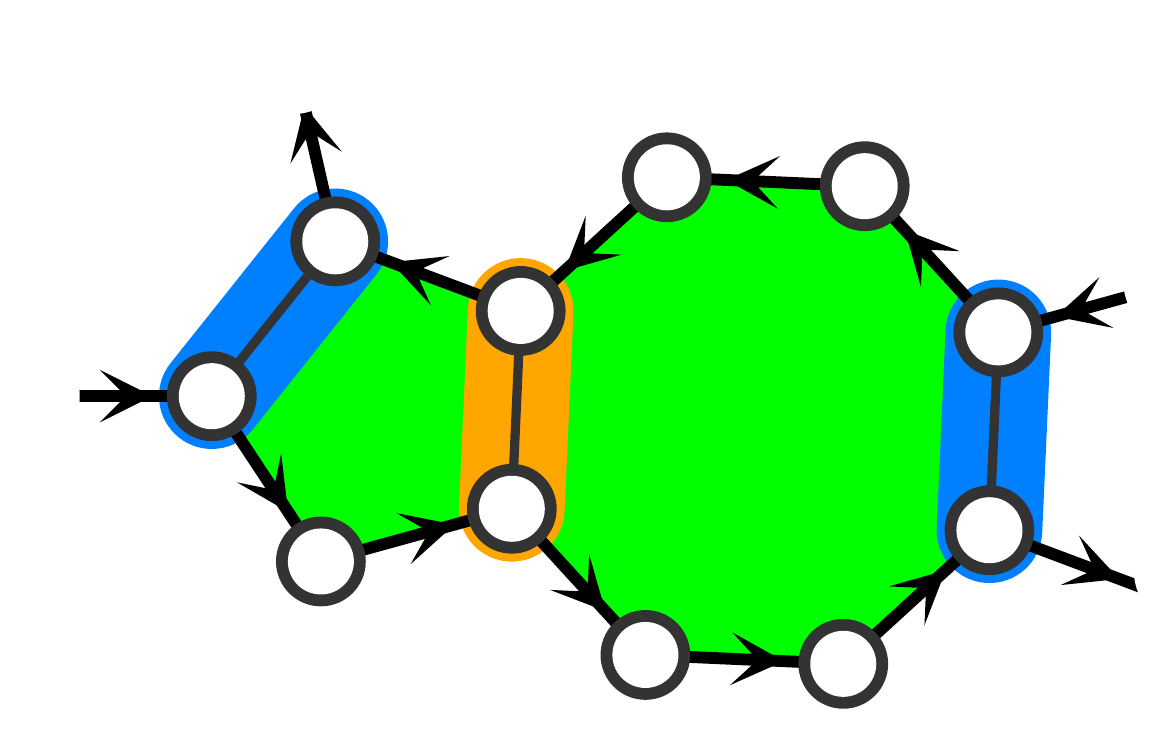}
            \caption{Rival motif 4}
        \end{subfigure}
        \hspace{-0.1cm}
        \begin{subfigure}[b]{0.3\textwidth}
            \includegraphics[width=\textwidth, angle=-90]{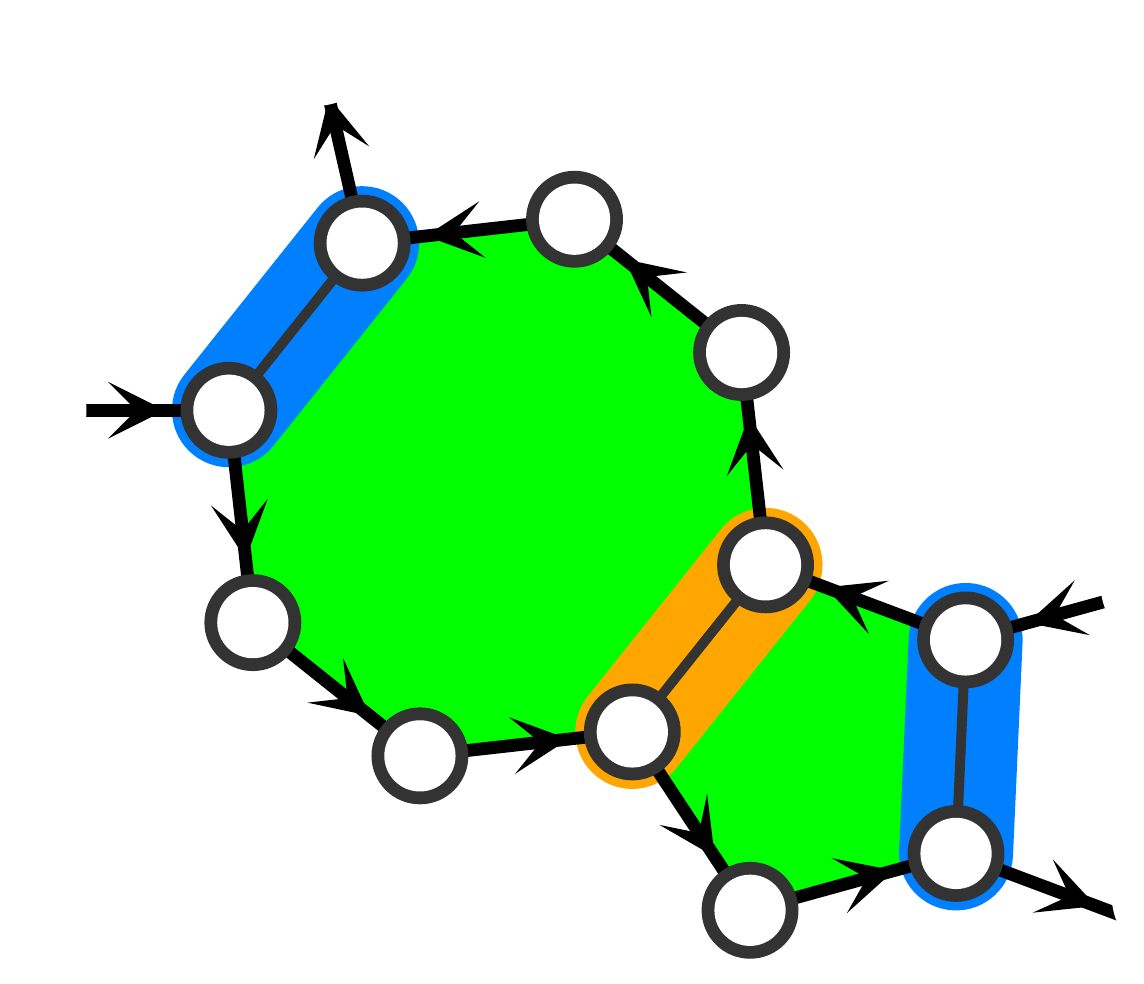}
            \caption{Rival motif 5}
        \end{subfigure}
    \end{minipage}
    \caption{Example of target motif and rival motif(s). The target motif~\ref{fig:tm1} is from the structure in Fig.~\ref{fig:e52motifs}, the target motif~\ref{fig:tm2} is from Eterna100 puzzle ``\texttt{Mat - Elements
\& Sections}" as plotted in Table Fig.~\ref{table:puzzles}.}\label{fig:rivals}
\end{figure}

 %As proven in the work RIGEND, in the context of entire structure design, when there exists a single rival structure $\vec\vecm$ for a target structure $\ystar$, they must satisfy $\pairs(\vec\vecm) \subset \pairs(\ystar)$. Such relation also holds for a single rival motif. 
For a rival motif $\vecm'$ satisfying Theorem~\ref{theorem:onerm}, we have $\pairs(\vecm') \subset \pairs(\mstar)$. %(proven in Supplementary Section~\ref{sec:onerival}). 
\begin{proof}
Suppose there exists a pair $(i, j)$ such that $(i, j) \in \pairs(\vecm')$ but $(i, j) \notin \pairs(\mstar)$. For any sequence $\vecx$ where $\vecx_i\vecx_j$ is not among allowed to pair, i.e. $\vecx_i\vecx_j \notin  \{\nucC\nucG,\nucG\nucC,\nucA\nucU,\nucU\nucA, \nucG\nucU,\nucU\nucG\}$, $\vecx$ cannot fold into $\vecm'$ because $\DG(\vecx, \vecm')$ would be $\infty$. Therefore, if $\vecx$ prefers $\vecm'$ to $\mstar$, then $\vecm'$ cannot have any pair $(i, j)$ not in $\pairs(\mstar)$. Since $\vecm' \neq \mstar$, it follows that $\pairs(\vecm') \subset \pairs(\mstar)$.
\end{proof}
However, if any motif $\vecm'$ satisfying $\pairs(\vecm') \subset \pairs(\mstar)$ is not a qualified rival motif, more rival motifs will be required as evidence for undesignability. 
 \begin{theorem}\label{theorem:multi-rival}
%A structure $\mstar \subseteq \vecy$ is undesignable,  
If $ \exists M=\{ \vecm_1, \vecm_2, .., \vecm_k \} \text{ and } \mstar \notin M, \text{such that }  \forall \vecx, \exists \vecm \in M, \\ \DG(\vecx, \vecm) \leq \DG(\vecx, \mstar)$, then \vecm is undesignable. 
\end{theorem}
The right side of Fig.~\ref{fig:rivals} shows an example of target motif with multiple rival motifs. Identifying a set of multiple rival motifs is more complicated than finding a single rival motif. 
%This requires the introduction of an advanced data structure named \emph{Design Space}, along with associated operations such as \emph{Constraint Intersection}. 
 We present a high-level Algorithm~\ref{alg:rivalmotif} to elucidate the fundamental procedures involved in identifying rival motifs, providing readers with an overview of the essential steps.
 In Algorithm~\ref{alg:rivalmotif} there are three parameters $M, N,$ and $K$. They limit the number of differential positions, rival structures and sampled sequences, preventing the algorithm running forever. 
 The overall complexity is $\bigO{NM+NK|\vecm^\star|^3}$.
 We omit the intricate details for conciseness, and encourage readers to consult the literature of RIGENDE~\cite{zhou+:2023undesignable} for a comprehensive description.

\begin{algorithm}[tb]
\caption{Rival Motifs Search Algorithm (high-level version)}\label{alg:rivalmotif}
\begin{algorithmic}[1]
  \Statex $\mathcal{X}(\mstar < \vecm') = \{ \vecx \mid \DDG(\vecx, \mstar, \vecm') < 0\}$\Comment{design space: excluding sequences impossible for successful design}
   \Function{RivalMotifSearch}{$\mstar, \vecy$}\Comment{motif \mstar in a structure \vecy}
    \State $\mrival \gets \emptyset$\Comment{define a set of rival motifs}
    \While{$\bigcap_{\vecm' \in \mrival} \mathcal{X}(\mstar < \vecm') \neq \emptyset$} \Comment{design space is not empty}
    	\If{$|\mrival| > N$}  \Return \texttt{\color{blue}unkown} \Comment{{\color{red}early stop}} \EndIf 
	\For{$i = 1$ \textbf{to} $K$}
        		\State Draw $\vecx \in \bigcap_{\vecm' \in \mrival} \mathcal{X}(\mstar < \vecm')$
        		\If{$\vecm'' = \UMFE(\vecx, \vecy \setminus \mstar)$}  \State \Return \texttt{\color{blue}designable} \Comment{found successful \UMFE design} \EndIf 
        		\State $\vecm'' \gets \text{MFE}(\vecx)$ \Comment{one of MFE structures in case of multiple}
		\If{$ 6^{|\pairs(\D(\vecm', \mstar))|}\times4^{|\unpaired(\D(\vecm', \mstar))| } < M$} 
        			\State $\mrival \gets \mrival \cup \{\vecm''\}$ \Comment{limit the size of differential positions} % 4^{|\D(\vecm', \mstar)|
		\EndIf 
        \EndFor
    \EndWhile
    \State \Return $\texttt{\color{blue}undesignable}$    
    \EndFunction
\end{algorithmic}
\end{algorithm}

% !TEX root = main.tex
\section{Rotational Invariance}
\vspace{-0.3cm}
\subsection{Invariance of Motif Energy}
\vspace{-0.2cm}
%In nearest neighbor model, the free energy of a loop does not dependent on the absolute position or specific orientation in structures.
%As a result, the free energy of motifs adhere to \emph{translational invariance}, \emph{rotational invariance} (but not \emph{symmetrical invariance}).
%Fig.~\ref{fig:rv} show two group of equivalent motifs by rotational invariance found in Eterna100 puzzles. 
%$\vecm_a, \vecm_b, \vecm_c$ are from the puzzles of IDs \#60, \#81, \#88 respectively.
%$\vecm_d, \vecm_e$ are from the puzzles of ID \#86. (Refer to Table~\ref{table:puzzles} for the detailed structures with highlighted motifs) 
%Among them, $\vecm_a$ and $\vecm_b$ can be rotated to each other, $\vecm_d$ and $\vecm_e$ can also be rotated to each other. However, $\vecm_b$ and $\vecm_c$ can not rotated to each other despite they both contains three bulge loops.
In the nearest-neighbor model, the free energy change of a loop is independent of its absolute position or specific orientation within the structure. 
As a result, the free energy of motifs adheres to both \emph{translational invariance} and \emph{rotational invariance}, though not to \emph{symmetrical invariance}. 
Fig.~\ref{fig:rv} shows two groups of equivalent motifs found in the Eterna100 puzzles, demonstrating rotational invariance. 
The motifs $\vecm_a$, $\vecm_b$, and $\vecm_c$ come from puzzles with IDs \#60, \#81, and \#88, respectively, while $\vecm_d$ and $\vecm_e$ are from the puzzle with ID \#86 (refer to Table~\ref{table:puzzles} for detailed structures with highlighted motifs). 
Among them, $\vecm_a$ and $\vecm_b$ can be rotated into each other, as can $\vecm_d$ and $\vecm_e$. However, despite both containing three bulge loops, $\vecm_b$ and $\vecm_c$ cannot be rotated into one another.

To identify and represent such rotational equivalence, we propose a noval representation for structures and motifs, referred to as \emph{loop-pair graph}. 

\vspace{-0.3cm}
\subsection{Loop-Pair Graph}
\begin{figure}[t]
 \centering
 \begin{minipage}[b]{0.32\textwidth}
    \begin{subfigure}[b]{0.20\linewidth}  % Adjust subfigure width to fit inside the minipage
        \centering
        \includegraphics[width=\textwidth]{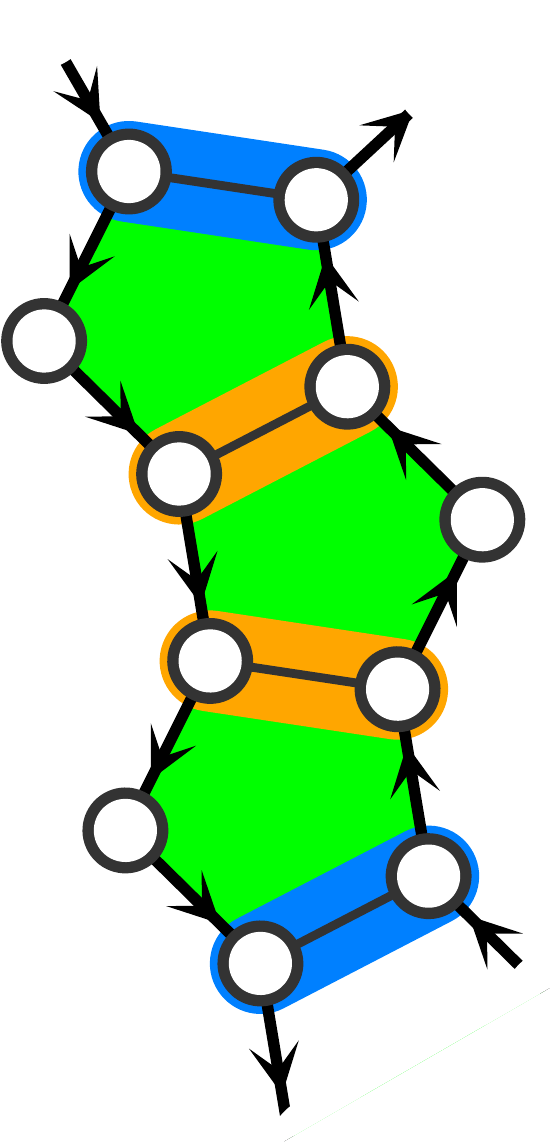}
        \vspace{-0.5cm}
        \caption{$\vecm_a$}\label{fig:ma}
    \end{subfigure}
    \hspace{-0.4cm}    \
    \begin{minipage}[b]{0.05\linewidth}  % Narrow minipage for the arrow
        \centering
        \raisebox{1.3cm}{  % Adjust the vertical position of the arrow
            \begin{tikzpicture}
                %\draw[->, thick] (0,0) -- (0.4,0);  % Draw rightward arrow
                \draw[->, thick] (-0.1,-0.1) arc [start angle=80, end angle=320, radius=0.2cm];
%                \draw[thick] (-0.4,-0.4) -- (0.0,0.0);  % Diagonal from bottom-left to top-right
%    		\draw[thick] (0.0,-0.4) -- (-0.4,0.0);  % Diagonal from top-left to bottom-right
            \end{tikzpicture}
        }
    \end{minipage}
     \hspace{0.cm}
    \begin{subfigure}[b]{0.20\linewidth}  % Adjust subfigure width to fit inside the minipage
        \centering
        \includegraphics[width=\textwidth]{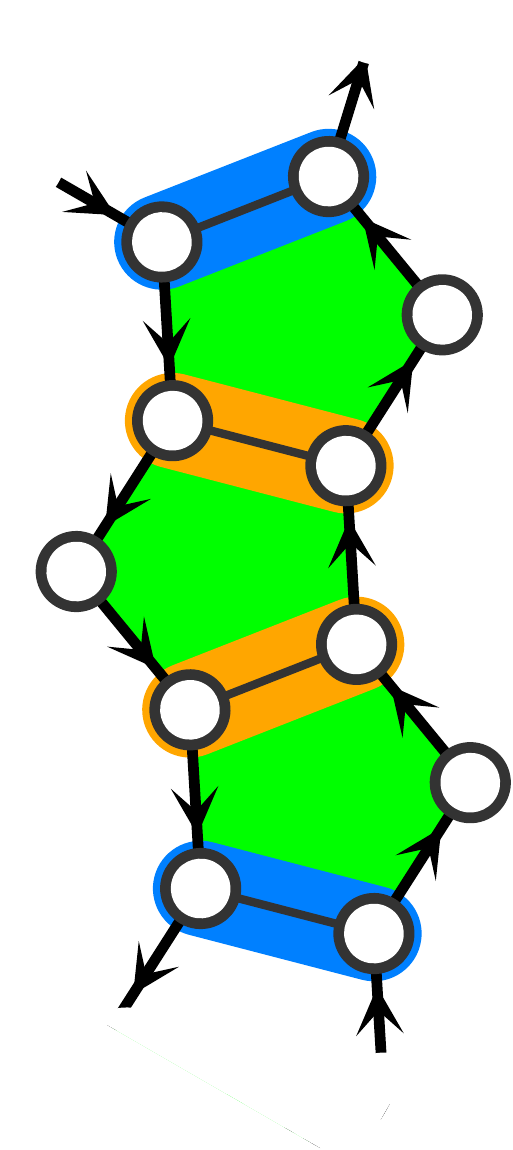}
        \vspace{-0.5cm}
        \caption{$\vecm_b$}\label{fig:mb}
    \end{subfigure}
    \vline 
    \hspace{0.2cm}  % Ensure subfigures are placed on the same line
    \begin{subfigure}[b]{0.20\linewidth}  % Adjust subfigure width to fit inside the minipage
        \centering
        \includegraphics[width=1.1\textwidth]{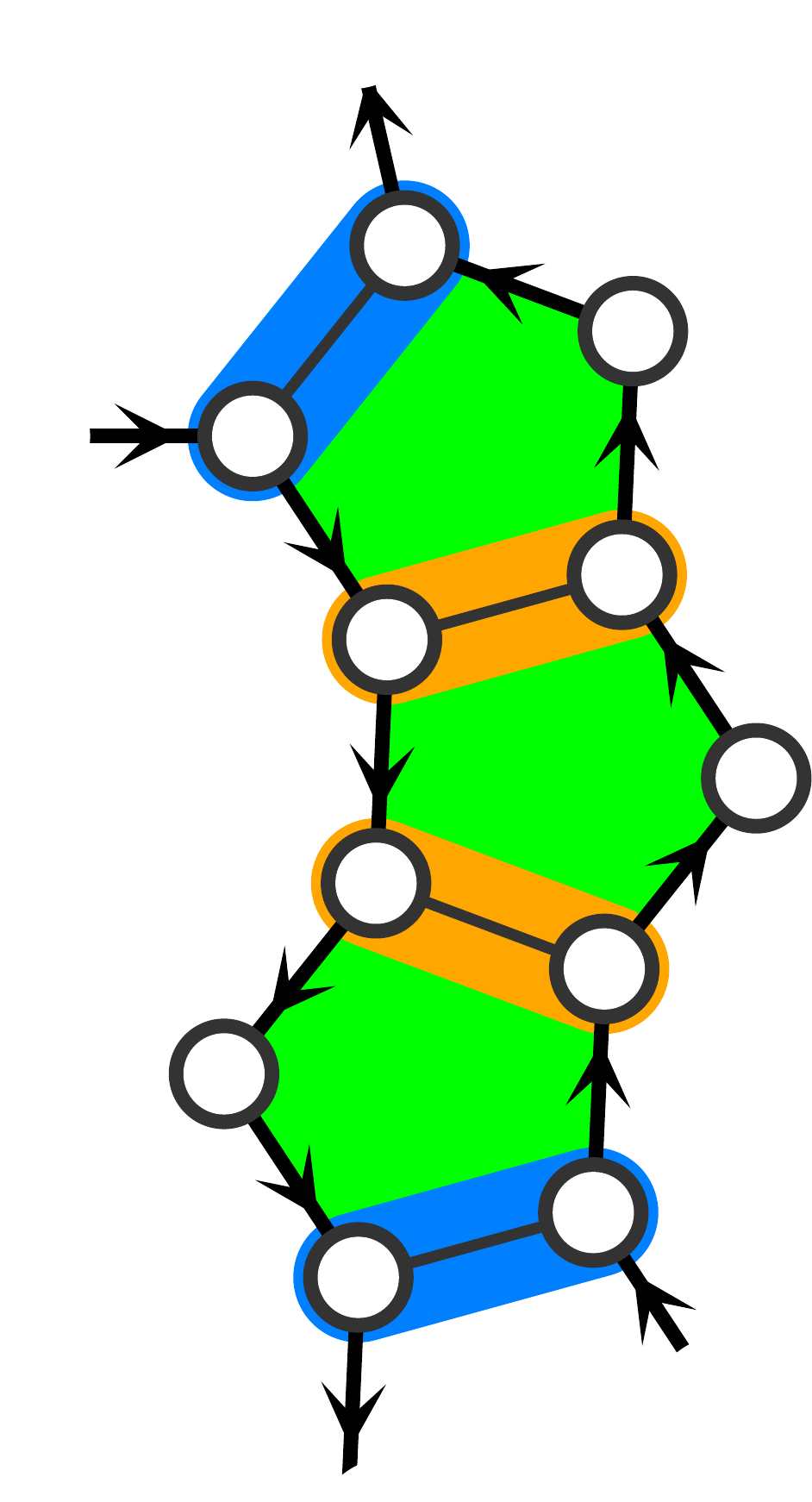}
        \vspace{-0.5cm}
        \caption{$\vecm_c$}\label{fig:zz}
    \end{subfigure}
      \begin{subfigure}[b]{0.4\linewidth}  % Adjust subfigure width to fit inside the minipage
        \centering
        \includegraphics[width=\textwidth]{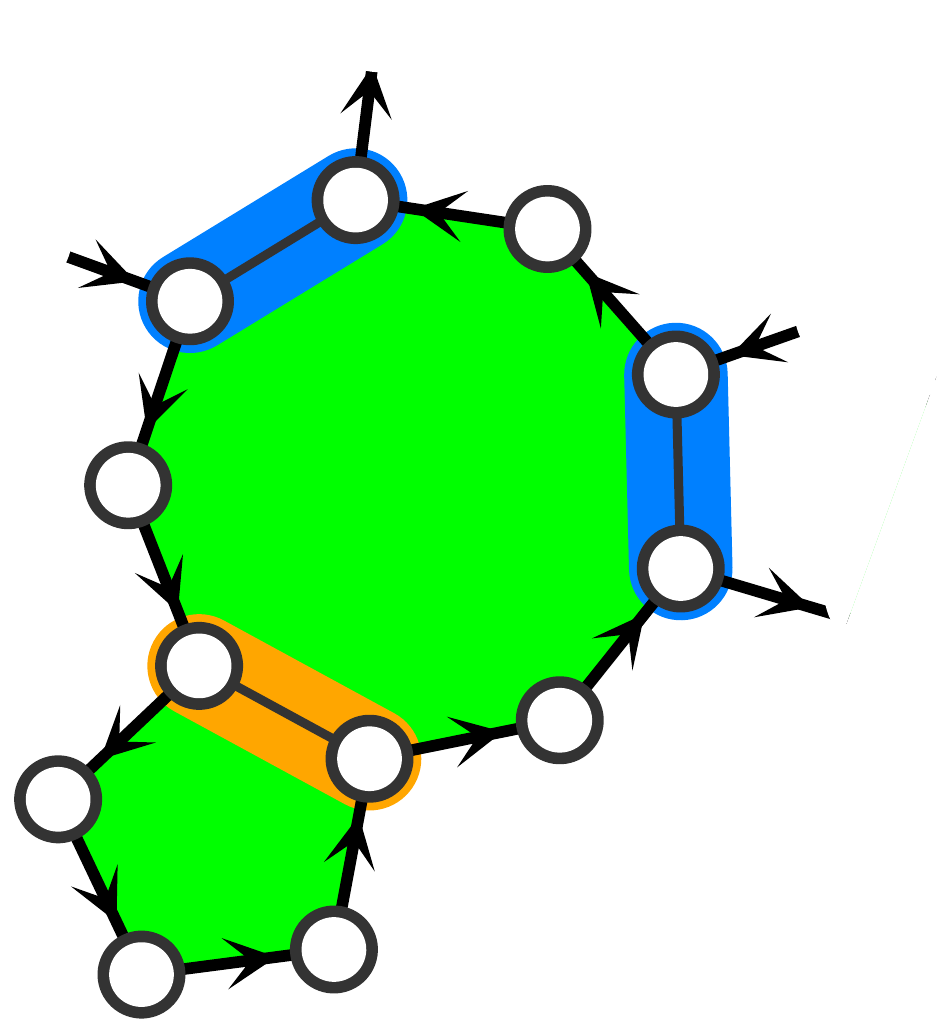}
        \vspace{-0.4cm}
        \caption{$\vecm_d$}\label{fig:zz}
    \end{subfigure}
    \begin{minipage}[b]{0.05\linewidth}  % Narrow minipage for the arrow
        \centering
        \raisebox{1.3cm}{  % Adjust the vertical position of the arrow
            \begin{tikzpicture}
                %\draw[->, thick] (0,0) -- (0.4,0);  % Draw rightward arrow
                \draw[->, thick] (0,0) arc [start angle=80, end angle=320, radius=0.2cm];
            \end{tikzpicture}
        }
    \end{minipage}
    \hfill  % Ensure subfigures are placed on the same line
    \begin{subfigure}[b]{0.4\linewidth}  % Adjust subfigure width to fit inside the minipage
        \centering
        \includegraphics[width=\textwidth]{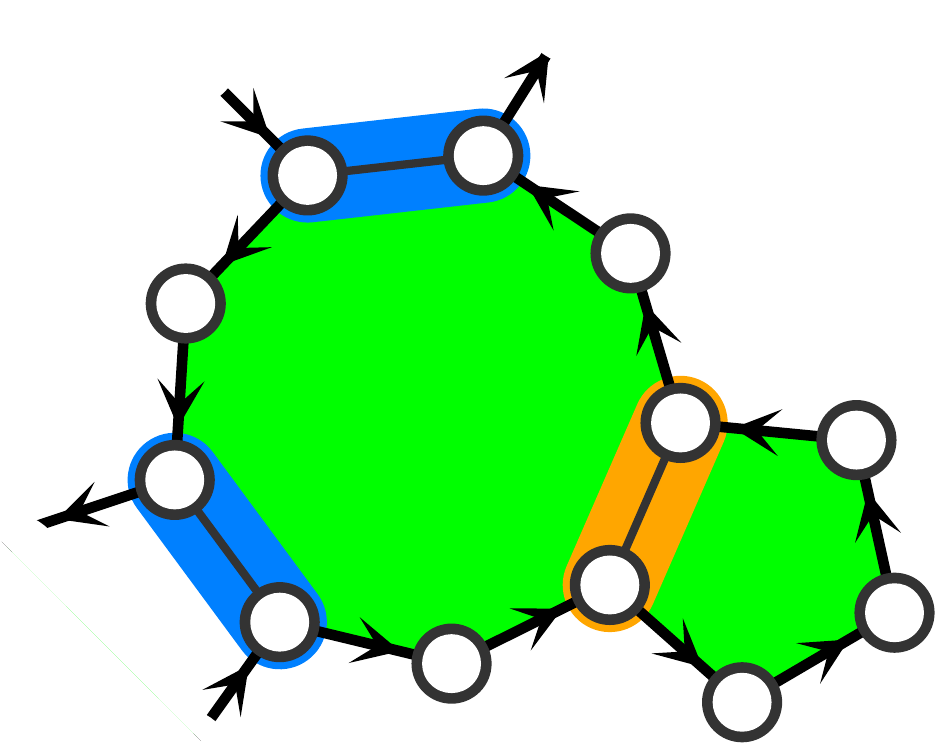}
        \vspace{-0.4cm}
        \caption{$\vecm_e$}\label{fig:zz}
    \end{subfigure}
    \\
\hspace{0.15cm}
\begin{subfigure}[b]{0.3\linewidth}  % Adjust subfigure width to fit inside the minipage
        \centering
        \begin{tikzpicture}[main/.style = {inner sep=0.05cm, draw, circle}, node distance={8mm}, diam/.style = {draw, diamond, inner sep=0.05cm, text width=0.15cm, minimum width=0.1cm, minimum height=0.1cm}, weight/.style={draw}] 
\node[diam] (1) [fill=bpcolor, scale=0.7] {\textit{p}};
\node[main] (2) [fill=green, below of=1] {M};
\node[diam] (3) [fill=ipcolor, below left of=2, scale=0.7] {\textit{p}};
\node[diam] (4) [fill=bpcolor, below right of=2, scale=0.7] {\textit{p}};
\node[main] (5) [fill=green, below left of=3] {H};
\draw (1) -- node[left, scale=0.7]{1} (2);
\draw (2) -- node[above left, scale=0.7]{1} (3);
\draw (2) -- node[above right, scale=0.7]{1} (4);
\draw (3) -- node[above left, scale=0.7]{3} (5);
\end{tikzpicture} 
        \vspace{-0.50cm}
        \caption{G($\vecm_d$)}\label{fig:zz}
    \end{subfigure}
    \hspace{0.5cm}
    \begin{minipage}[b]{0.05\linewidth}  % Narrow minipage for the arrow
        \centering
        \raisebox{2.1cm}{  % Adjust the vertical position of the arrow
            \begin{tikzpicture}
                %\draw[->, thick] (0,0) -- (0.4,0);  % Draw rightward arrow
                \draw[->, thick] (0,0) arc [start angle=80, end angle=320, radius=0.2cm];
            \end{tikzpicture}
        }
    \end{minipage}
    \hspace{0.15cm}
    \begin{subfigure}[b]{0.3\linewidth}  % Adjust subfigure width to fit inside the minipage
        \centering
        \begin{tikzpicture}[main/.style = {inner sep=0.05cm, draw, circle}, node distance={8mm}, diam/.style = {draw, diamond, inner sep=0.05cm, text width=0.15cm, minimum width=0.1cm, minimum height=0.1cm}, weight/.style={draw}] 
\node[diam] (1) [fill=bpcolor, scale=0.7] {\textit{p}};
\node[main] (2) [fill=green, below of=1] {M};
\node[diam] (3) [fill=bpcolor, below left of=2, scale=0.7] {\textit{p}};
\node[diam] (4) [fill=ipcolor, below right of=2, scale=0.7] {\textit{p}};
\node[main] (5) [fill=green, below right of=4] {H};
\draw (1) -- node[left, scale=0.7]{1} (2);
\draw (2) -- node[above left, scale=0.7]{1} (3);
\draw (2) -- node[above right, scale=0.7]{1} (4);
\draw (4) -- node[above right, scale=0.7]{3} (5);
\end{tikzpicture} 
        \vspace{-0.50cm}
        \caption{G($\vecm_e$)}\label{fig:zzzz}
    \end{subfigure}
    \hspace{-2.0cm}\caption{Rotational invariance examples from Eterna100. $\vecm_a\simeq \vecm_b, \vecm_b \not\simeq \vecm_c,\vecm_d \simeq \vecm_e$.}\label{fig:rv} % :\#81,$\vecm_c$:\#88,$\vecm_d\&\vecm_e$:\#86
\end{minipage}
%%---------------------------------------------------------------------------------------------------------------------------------------------------------------------------------------------------------------------------------------
\hspace{.20cm}
 \begin{minipage}[b]{0.5\textwidth}
     \begin{minipage}[b]{0.45\textwidth}
         \begin{subfigure}[b]{\linewidth}
             \centering
             \includegraphics[width=.72\textwidth]{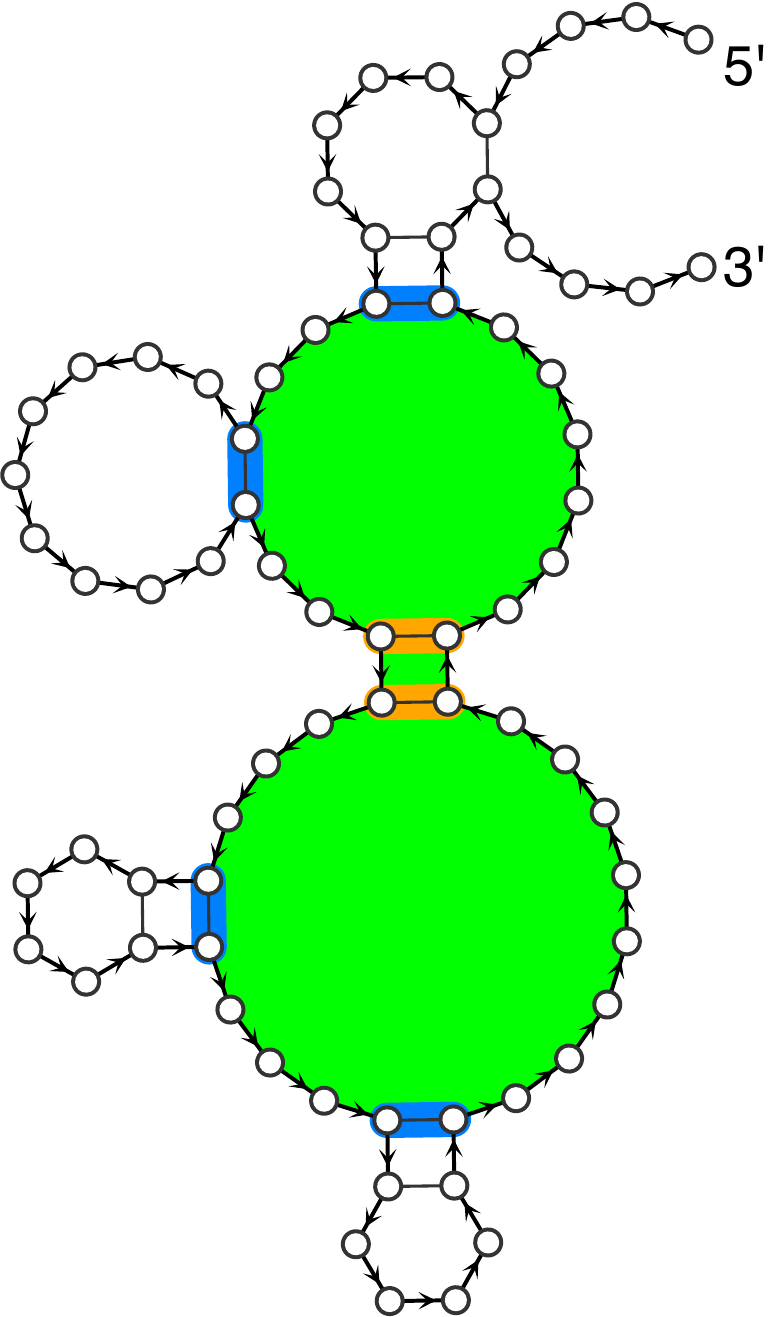} % j87_C3_m0_labeled.pdf figs/i87_C1_1_mode0_160.pdf
            %\vspace{-0.1cm}
             \caption{Polymer graph}\label{fig:m3_2}
         \end{subfigure}
          \\[.1cm]
         \begin{subfigure}[b]{1.0\linewidth}
             \centering
             \begin{tikzpicture}[main/.style = {inner sep=0.05cm, draw, circle}, node distance={6mm}] 
              \node[main] (1) [] {B};
            %  \node[main] (2) [right of=1] {S};
            %  \node[main] (3) [right of=2] {S}; 
              \node[main] (4) [right of=1] {E};
              \node[main] (5) [below of=1] {S};
              \node[main] (6) [fill=green, below of=5] {M};
              \node[main] (7) [left of=6] {H};
              \node[main] (8) [fill=green, below of=6] {S};
              \node[main] (9) [fill=green, below of=8] {M};
              \node[main] (10) [left of=9] {S};
              \node[main] (11) [left of=10] {H};
              \node[main] (12) [below of=9] {S};
              \node[main] (13) [below of=12] {H};
            %  \draw (1) -- (2);
            %  \draw (2) -- (3);
              \draw (1) -- (4);
              \draw (1) -- (5);
              \draw [preaction={draw,color={rgb,255:red,55;green,126;blue,247},-, double=blue,
              double distance=4\pgflinewidth, opacity=0.8}] (5) -- (6);
              \draw [preaction={draw,color={rgb,255:red,55;green,126;blue,247},-, double=blue,
              double distance=4\pgflinewidth, opacity=0.8}] (6) -- (7);
              \draw [preaction={draw,color={rgb,255:red,242;green,170;blue,60},-, double=orange,
              double distance=4\pgflinewidth, opacity=1.0}] (6) -- (8);
              \draw [preaction={draw,color={rgb,255:red,242;green,170;blue,60},-, double=orange,
              double distance=4\pgflinewidth, opacity=1.0}] (8) -- (9);
              \draw [preaction={draw,color={rgb,255:red,55;green,126;blue,247},-, double=blue,
              double distance=4\pgflinewidth, opacity=0.8}] (9) -- (10);
              \draw (10) -- (11);
              \draw [preaction={draw,color={rgb,255:red,55;green,126;blue,247},-, double=blue,
              double distance=4\pgflinewidth, opacity=0.8}] (9) -- (12);
              \draw (12) -- (13);
            \end{tikzpicture}
             \caption{Loop graph}
             \label{fig:lg}
         \end{subfigure}
     \end{minipage}
    \raisebox{6.5cm}{\Large$\leftrightarrow$}
    \hspace{-1.0cm} %%%%%%% c
\begin{subfigure}[b]{0.4\textwidth}
        \centering
     \begin{tikzpicture}[main/.style = {inner sep=0.05cm, draw, circle}, node distance={5.7mm}, diam/.style = {draw, diamond, inner sep=0.05cm, text width=0.15cm, minimum width=0.1cm, minimum height=0.1cm}, weight/.style={draw}] 
      \node[main] (1) [] {B};
      \node[diam] (14) [right of=1, scale=0.7] {\textit{p}};
    %  \node[main] (3) [right of=2] {S}; 
      \node[main] (4) [right of=14] {E};
    %  \node[diam] (40) [above of=4, scale=0.7] {\textit{p}};
       \node[diam] (0) [above of=4, scale=0.7] {\textit{r}};
      \node[diam] (15) [below of=1, scale=0.7] {\textit{p}};
      \node[main] (5) [below of=15] {S};
      \node[diam] (16) [fill={rgb,255:red,55;green,126;blue,247}, below of=5, scale=0.7] {\textit{p}};
      \node[main] (6) [fill=green, below of=16] {M};
      \node[diam] (17) [fill={rgb,255:red,55;green,126;blue,247}, left of=6, scale=0.7] {\textit{p}};
      \node[main] (7) [left of=17] {H};
      \node[diam] (18) [fill={rgb,255:red,242;green,170;blue,60}, below of=6, scale=0.7] {\textit{p}};
      \node[main] (8) [fill=green, below of=18] {S};
      \node[diam] (19) [fill={rgb,255:red,242;green,170;blue,60}, below of=8, scale=0.7] {\textit{p}};
      \node[main] (9) [fill=green, below of=19] {M};
      \node[diam] (20) [fill={rgb,255:red,55;green,126;blue,247}, left of=9, scale=0.7] {\textit{p}};
      \node[main] (10) [left of=20] {S};
      \node[diam] (21) [left of=10, scale=0.7] {\textit{p}};
      \node[main] (11) [left of=21] {H};
      \node[diam] (22) [fill={rgb,255:red,55;green,126;blue,247}, below of=9, scale=0.7] {\textit{p}};
      \node[main] (12) [below of=22] {S};
      \node[diam] (23) [below of=12, scale=0.7] {\textit{p}};
      \node[main] (13) [below of=23] {H};
      \draw (1) -- node[midway, above, scale=0.7] {4} (14);
      \draw (14) -- node[midway, above, scale=0.7] {4} (4);
      \draw (4) -- node[midway, left, scale=0.7] {4} (0);
    %  \draw (40) -- node[midway, left, scale=0.7] {4} (0);
      \draw (1) -- node[midway, left, scale=0.7] {0} (15);
      \draw (15) -- node[midway, left, scale=0.7] {0} (5);
      \draw (5) -- node[midway, left, scale=0.7] {0} (16);
      \draw (16) -- node[midway, left, scale=0.7] {2} (6);
      \draw (6) -- node[midway, above, scale=0.7] {2}  (17);
      \draw (17) -- node[midway, above, scale=0.7] {9} (7);
      \draw (6) -- node[midway, left, scale=0.7] {6} (18);
      \draw (18) -- node[midway, left, scale=0.7] {0} (8);
      \draw (8) -- node[midway, left, scale=0.7] {0} (19);
      \draw (19) -- node[midway, left, scale=0.7] {3} (9);
      \draw (9) -- node[midway, above, scale=0.7] {3} (20);
      \draw (20) -- node[midway, above, scale=0.7] {0} (10);
      \draw (10) -- node[midway, above, scale=0.7] {0} (21);
      \draw (21) -- node[midway, above, scale=0.7] {4} (11);
      \draw (9) -- node[midway, left, scale=0.7] {8} (22);
      \draw (22) -- node[midway, left, scale=0.7] {0} (12);
      \draw (12) -- node[midway, left, scale=0.7] {0} (23);
      \draw (23) -- node[midway, left, scale=0.7] {4} (13);
    \end{tikzpicture}
      \caption{Loop-pair graph}
     \label{fig:lpg}
  \end{subfigure}
  \caption{Graph representations of RNA structure.  
  A motif is a connected subgraph in a loop-pair graph (c).
    $\leftrightarrow$ denotes bijection.}\label{fig:reps}
  \end{minipage}  
  \end{figure}

%For brevity, we introduce the definition of loop-pair graph for RNA structure \vecy, and each motif $\vecm \subseteq \vecy$ corresponds to a subgraph.
\begin{definition}\label{def:lpg}
A {\bf loop-pair graph} for a pseudoknot-free RNA secondary structure \vecy
is a weighted undirected graph $G(\vecy)=\langle V(\vecy), E(\vecy)\rangle$ where each node $v\in V$ is
either a loop in \vecy or a pair in \vecy or the pseudo-pair node $r$ (representing $5'$ and $3'$ ends rather than a base pair), (i.e., $V(\vecy)=\loops(\vecy)\cup \pairs(\vecy) \cup \{r\}$)
and each edge $e = (u, v, w) \in E(\vecy)$ connects one loop node and one pair node, with the edge weight $w$
denoting the number of unpaired bases of the segment of the loop between the connected pair and next pair according to the direction from $5'$ to $3'$. 
%In order to recover the RNA structure, 
For each loop node $v$, there is an {\bf ordered list} $N(v)$ of neighbor nodes. \vspace{-0.2cm}
\end{definition}
The loop-pair graphs of the structure and motifs in Fig.~\ref{fig:card} is shown in Fig.~\ref{fig:lpg}. We also show Fig.~\ref{fig:m3_2} and Fig.~\ref{fig:lg} for comparison. 
The advantages of loop-pair graphs include:
%\begin{enumerate}
%\item Bijective. A loop-pair graph contains all the information of loops, base pairs and unpaired bases to recover the original structure. In comparison, the \emph{loop-graph} in Fig.~\ref{fig:lg}  cannot recover the original structure because of the lack of the information for unpaired bases.
%\item Abstract. Loop-pair graphs highlight the connection and topology of loops and base pairs in motifs while hide the nuanced backbone of all bases.
%\item Compact. The number of unpaired bases are labeled as edge weights, which is saved more space compared to \emph{polymer graph} in Fig.~\ref{fig:m3_2}.
%\end{enumerate}
\begin{enumerate} 
\item \textbf{Bijective.} A loop-pair graph contains all necessary information about loops, base pairs, and unpaired bases to fully reconstruct the original structure. In contrast, \emph{loop-graph} in Fig.~\ref{fig:lg} or
RNA-as-graph representations~\cite{kim2004+:candidates,gan2003+:exploring,gan+:1987rag,zorn+:2004structural}
cannot recover the original structure due to missing information about unpaired bases or helices, respectively.  
ignore helix stackings and can not recover the original structure.
\item \textbf{Abstract.} Loop-pair graphs emphasize the connections and topology of loops and base pairs within motifs, while abstracting away the finer details of the backbone structure. While other representations~\cite{benedetti+:1996graph,leontis+:2006building,le+:1989tree} also provide abstraction of structures, they serve for different applications.
\item \textbf{Compact.} The number of unpaired bases is encoded as edge weights, which makes the representation more space-efficient than the \emph{polymer graph} in Fig.\ref{fig:m3_2}. 
\end{enumerate}

%Secondary structures are intrinsically recursive, making loop-pair graphs connected and singly connected, and thus a tree.
%Any motif $\vecm \subseteq \vecy$ corresponds to an \emph{induced subgraph} of $G(\vecy)$, encompassing the loop nodes for each loop in $\loops(\vecm)$, the pair nodes for each pair in $\pairs(\vecm)$, and all the edges connecting those nodes.
%In order to determine the rotational isomorphism of two motifs, we can recursively rotate the graph representation of a motif using a boundary pair node as pivots, which is described in Algorithm~\ref{alg:rotate}. 
RNA secondary structures are inherently recursive, making loop-pair graphs singly connected, essentially forming a tree. Any motif $\vecm \subseteq \vecy$ corresponds to an \emph{induced subgraph} of $G(\vecy)$, containing the loop nodes for each loop in $\loops(\vecm)$, the pair nodes for each pair in $\pairs(\vecm)$, and all edges connecting these nodes.
To assess the \emph{uniqueness} of motifs under rotational isomorphism, we recursively rotate the graph representation of a motif using a boundary pair node as a pivot, as described in Algorithm~\ref{alg:rotate} (linear complexity). 

\begin{algorithm}[bt]
\caption{Rotate Loop-Pair Graph via Node (see Fig.~\ref{fig:rv} for examples)}
\label{alg:rotate}
%\begin{algorithmic}[1]
%\Function{TreeAt}{$v$, $p$} \Comment{boundary pair node and parent node}
%    \State $t \gets \text{Tree}(v)$\Comment{start a new tree (graph) from $v$}
%    \If{$p \neq nil$}
%    \State $t.\text{children} \gets [\ ]$ \Comment{initialize a list of children}
%    \State $i \gets \text{ChildIndex}(v,\ p)$ \Comment{get the child index of $v$}
%    \State $t.\text{children} = p.children[i+1:] + \Call{TreeAt}{p,\ p.parent} + p.children[:i]$ \Comment{rotate}
%%    \ForAll{$u$ \textbf{in} $N(v)[i+1:] + N(v)[:i]$} \Comment{rotate using $i$ as pivot; slicing in python syntax}
%%        \State $t.\text{children}.\text{append}\big(\Call{TreeAt}{u,\ v}\big)$
%%    \EndFor
%    \EndIf
%\State \Return $t$
%\EndFunction
%\end{algorithmic}
\begin{algorithmic}[1]
\Function{NewTree}{$v$, $\text{child}\_\text{id}$} \Comment{\footnotesize start with a leaf node $v$ and $\text{child}\_\text{id}$ 0}
    \State $t \gets \Call{TreeNode}{ }; p  \gets v.\text{parent}$\Comment{create a new (sub)tree (graph) from $v$}
    \If{$p \neq nil$}
    %\State $t.\text{children} \gets [\ ]$ \Comment{initialize a list of children}
    %\State $i \gets \text{ChildIndex}(v,\ p)$ \Comment{get the child index of $v$ in $p.\text{children}$}
    $newchild \gets [\Call{NewTree}{p,  v.\text{child}\_\text{id}}]$ \textbf{else} \State {$newchild \gets [] $}  \Comment{if $v$'s parent exists, recursive call}
    \EndIf
    \State $t.\text{children} \gets v.\text{children}[\text{child}\_\text{id}+1:] + newchild +v.\text{children}[:\text{child}\_\text{id}]$ \hspace{-0.10cm}\Comment{rotate}
\State \Return $t$
\EndFunction
\end{algorithmic}
\end{algorithm}
% !TEX root = main.tex
\begin{algorithm}[tb]
\caption{\mbox{\small Bottom-Up Scan for Identifying Minimal Undesignable Motif}}
\label{alg:bottom-up}
\begin{algorithmic}[1]
\Function{BottomUpMotifDesignabilityScan}{$\vecy$} \Statex \hspace{\algorithmicindent} \Comment{input a secondary structure \vecy}
%\State $\mathcal{M}_{designable}  \gets \emptyset,
\State $\libmum  \gets \emptyset$\Comment{ a set to store identified minimal undesignable motifs}
%\For{$i = 2$ to $|\loops(\vec{y})|$} \label{line:enum}
    \ForAll{non-singleton $\vecm \subseteq \vecy$ in increasing order of $\card(\vecm)$} \Statex \hspace{\algorithmicindent} \Comment{$\card(\vecm)=2,3,\ldots,|\loops(\vecy)|$}
         \If{$\exists \vecm' \in \libmum$ and $\vecm' \subset \vecm$} \textbf{continue} \label{alg1:notminimal} \Statex \hspace{\algorithmicindent}  \Comment{undesignable but not minimal} \EndIf
            \State $designablity$ $\gets$ \Call{Decide}{\vecm} \Comment{either \texttt{\textcolor{blue}{designable}} or \texttt{\textcolor{blue}{undesignable}}} 
                      \If{$designablity$ = {\texttt{\textcolor{blue}{undesignable}}}} \label{alg1:designable}
                 %\State $\mathcal{M}_{designable}  \gets\libdsm  \cup \{\vec{m}\}$
                 %\Else 
                 \State {$\libmum  \gets \libmum   \cup \{\vec{m}\}$}
            \EndIf
    \EndFor
%\EndFor
\State \Return $\libmum $
\EndFunction
\end{algorithmic}
\end{algorithm}
\begin{algorithm}[tb]
\caption{FastMotif (see Fig.~\ref{fig:neighbors} for an example of powerset)}\label{alg:scanp2}
\label{alg:powerset}
\begin{algorithmic}[1]
%\Require  $\mathcal{M}_{\text{miniundesignable}}$, $\mathcal{M}_{\text{designable}}$, $\mathcal{M}_{\text{undesignable}}$ \Comment{Global (minimal) undesignable and designable motifs}
\State \textbf{global:}  $\libmum$, $\libdsm$ \Comment{Global (minimal) undesignable and designable motifs} % $\mathcal{M}_{\text{undesignable}}$
\Function{FastMotif}{$\vecy$}\Comment{Input is a structure}
%\State $\mathcal{M} \gets \emptyset$; $\mathcal{D} \gets \emptyset$ \Comment{Designable motifs in $\vec{y}$}
\ForAll{loop node $u \in G(\vec{y})$}
    %\For{$i = 1$ to $|N(u)|$} \label{line2}
        \ForAll{non-empty $s \in 2^{N(u)}$  in increasing order of $|s|$} \Statex \hspace{\algorithmicindent} \Comment{$2^{N(u)}$: powerset of $N(u)$;$|s|=1,2,\ldots,|N(u)|$} %such that $|s| = i$} \setminus \{\varnothing\}$
            \State $\vecm \gets \{u\} \cup s$ \Comment{motif \vecm has loop $u$ and its neighbors in $s$, so $\card(\vecm) \geq 2$}
            \If{$\vec{m} \in \libmum$} \label{line:inlm}
            	\textbf{continue} \Statex \hspace{\algorithmicindent} \Comment{check every rotated version of \vecm; already identified}
            \Else
             \If{$\exists \vecm' \in \libmum$ and $\vecm' \subset \vecm$} \textbf{continue} \Statex \hspace{\algorithmicindent} \Comment{undesignable but not minimal} \EndIf
            \State $designablity$ $\gets$ \Call{RivalMotifSearch}{\vecm} \Statex \hspace{\algorithmicindent} \Comment{{{return \texttt{\color{blue}designable} or \texttt{\color{blue}undesignable} or \texttt{\color{blue}unkown}}}}
            \If{$designablity$ = \texttt{\color{blue}designable}} \label{alg1:ifud}
                  {$\libdsm \gets\libdsm  \cup \{\vec{m}\}$}
                 \ElsIf {$designablity$ = \texttt{\color{blue}undesignable}}  
                 %\State $\mathcal{M}_{undesignable}  \gets \mathcal{M}_{undesignable}   \cup \{\vec{m}\}$
                  \If{$\forall \vecm_{\text{sub}} \subset \vecm$,  $\vecm_{\text{sub}} \in \mathcal{M}_{\text{designable}}$} 
                  	\State $\libmum  \gets \libmum  \cup \{\vec{m}\}$  \Comment{minimal}
	              \EndIf
            \EndIf
            \EndIf
        \EndFor
    %\EndFor
\EndFor
\EndFunction
\end{algorithmic}
\end{algorithm}

\section{\scalebox{0.95}{Bottom-Up Scan of Motif Designability within Structures}}
\begin{figure}[h]
    \centering
    \vspace{-0.2cm}
    \includegraphics[width=0.25\textwidth]{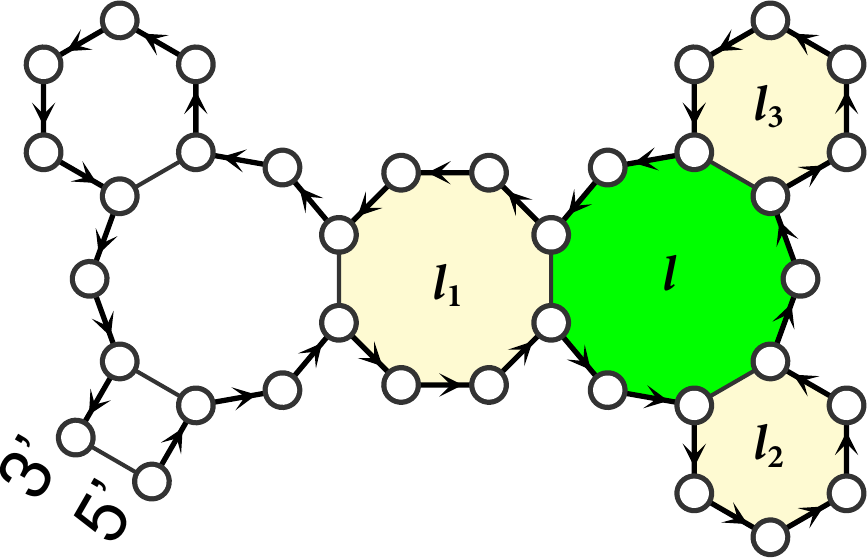}
    \captionsetup{singlelinecheck=off}
   	\caption[Example of powerset in Algorithm~\ref{alg:powerset}.]{Example of powerset in Algorithm~\ref{alg:powerset}. Among subsets $2^{N(l)} \setminus \varnothing$, i.e., $\{l_1\}, \{l_2\}, \{l_3\},  \allowbreak \{l_1, l_2\}, \{l_2, l_3\}, \allowbreak \{l_3, l_1\}, \{l_1, l_2, l_3\}$, only $\{l_2, l_3\}$ and $\{l_1, l_2, l_3\}$, when combined with $l$, are undesignable. However, only $\{l, l_2, l_3\}$ is minimal undesignable   (see Tab.~\ref{table:puzzles} for Eterna \#57).} 
	\label{fig:neighbors}   
\end{figure}

%\vspace{-0.1cm}
An ideal algorithm should be capable of identifying all minimal undesignable motifs within a given secondary structure. It is important to note that the sub-motifs of $\mathcal{M}$ always possess smaller cardinalities compared to $\mathcal{M}$. Therefore, a straightforward approach involves enumerating all motifs in the structure according to their cardinality and determining whether each motif is designable or undesignable, as outlined in Algorithm~\ref{alg:bottom-up}. This algorithm assumes the existence of an oracle function, \Call{DECIDE}{\vec{m}}, which returns whether \vec{m} is designable or undesignable. While such an ideal function \Call{DECIDE}{\vec{m}} theoretically exists (exhaustive search could achieve this), it may not be practical due to its high complexity. Nevertheless, it serves as a conceptual foundation for developing more practical algorithms. \vspace{-0.05cm}
\begin{theorem}\label{theorem:algideal}
 Given a secondary structure $\vecy$, Algorithm~\ref{alg:bottom-up} outputs a set $ \libmum $ containing all and only the minimal undesignable motifs in $\vecy$. \vspace{-0.2cm}
\end{theorem}
%The proof is conducted by induction (detailed in Supplementary Section~\ref{sec:algideal}). 
\begin{proof}
 The proof can be conducted by induction.
 \begin{enumerate}
 \item Base case. At the iteration of $i=2$, the found undesignable motifs at line~\ref{alg1:designable} are minimal because of Theorem~\ref{theorem:m1loop}. At the end of the iteration,  $\mathcal{M}$ consists of all minimal undesignable motifs of 2 loops.
 \item Induction hypothesis. Suppose when $i=k'\geq2$, the found undesignable motifs at line~\ref{alg1:designable} are minimal undesignable motifs, and all minimal undesignable motifs of cardinality equal or less than $i$ are included in $\mathcal{M}_{miniundesignable}$ at the end of the iteration.
 \item Induction step. When  $i=k'+1$, each $\vecm$ satisfying $|\loops(\vecm)|=k'+1$ will be checked. If $\vecm$ contains other undesignable motifs, line~\ref{alg1:notminimal} will stop it from being further considered. If $\vecm$ is designable,  line~\ref{alg1:designable} will prevent it from being added to $\mathcal{M}_{miniundesignable}$. As a result, $\vecm$ will be added to $\mathcal{M}_{miniundesignable}$ if and only if  $\vecm$ is minimal undesignable. 
\end{enumerate}
 \end{proof}
The total number of motifs in a structure can grow exponentially with cardinalities, making Algorithm~\ref{alg:bottom-up} impractical for large structures. However, empirical observations suggest minimal undesignable motifs typically involve a small number of loops. To address this, we propose a variant of Algorithm~\ref{alg:bottom-up}, referred to as FastMotif (Algorithm~\ref{alg:scanp2}), designed to identify as many minimal undesignable motifs as possible within a given structure $\vecy$, while maintaining computational efficiency.
In particular, we limit our evaluation to motifs composed of a loop and any (non-empty) subset of its neighboring loops (Fig.~\ref{fig:neighbors}). This approach offers several key advantages:  \vspace{-0.2cm}
\begin{enumerate}
  \item Each motif has a limited number of loops, making undesignability easier to decide.
  \item For a loop $v$ with $|N(v)|$ neighbor loops, the size of $N(v)$'s power set $2^{N(v)}$, excluding $\varnothing$, is $2^{N(v)}-1$. Since most loops have 2 or 3 neighbors, the number of motifs considered remains relatively small, while still covering most of the relevant small motifs. 
    \item Each loop and its neighbor loops can be seen as a small subgraph on the loop-pair graph $G(\vecy)$. Enumerating motifs in the power set would be equivalent to running Algorithm~\ref{alg:bottom-up} on a local subgraph.\vspace{-0.0cm}
\end{enumerate}
%Algorithm~\ref{alg:powerset} shows the pseudocode of bottom-up motif designability scan within power set of loop neighbors.
%To avoid spending time on scrutinizing motifs with known designability, we maintain a global libraries of minimal undesignable motifs and designable motifs.
%To further optimize performance, we maintain global libraries of undesignable motifs and designable motifs, avoiding repeated identifications of  designabilities. 
To enhance performance, we exclude motifs with more than 3 neighbor loops. The  complexity of FastMotif is determined by the product of the number of motifs scanned and the complexity of Algorithm~\ref{alg:rivalmotif}, making it polynominal. 
Additionally, FastMotif can be adapted to scan larger motifs.  In Sec.\ref{sec:exp}. We incorporated an undesignable \emph{substructure} in Eterna puzzles (``\texttt{Chicken feet}” and ``\texttt{Mutated chickenfeet}”) from RIGENDE~\cite{zhou+:2023undesignable} and proved it is a minimal undesignable motif. 

% !TEX root = main.tex
%\vspace{-0.25cm}
\section{Related Work}
\label{sec:related}
To the best of our knowledge, CountingDesign~\cite{yao+:2019,yao:2021thesis} is the existing method that has investigated undesignable motifs. However, it exhaustively enumerates and folds all RNA sequences for each motif group, identifying designable motifs and taking their complement to find undesignable ones. As a result, it is only applicable to short motifs, reporting motifs of up to 14 nucleotides. Another drawback is the limited interpretability of the undesignable motifs, as CountingDesign identifies them by taking the complement of designable motifs rather than directly characterizing the undesignable motifs themselves. Moreover, CountingDesign defines motifs as rooted trees starting from a base pair, making it unable to handle external loops in RNA structures. Additionally, it overlooks rotational invariance of motifs, leading to redundancy in the identified undesignable motif set. See also Sec.~\ref{sec:efficiency} for detailed efficiency comparisons.

% !TEX root = main.tex
\section{Empirical Results}\label{sec:exp}
\subsection{Settings}
We applied our algorithm FastMotif to two public RNA structure benchmarks: Eterna100~\cite{anderson+:2016} and ArchiveII~\cite{Cannone+:2002}. 
Eterna100 consists of 100 structures\footnote{We used the 21 structures that do not have successfully designed sequences by unique \MFE criterion}, 
artificially designed by human experts, and serves as a primary benchmark for RNA design. ArchiveII, comprising native RNA structures, spans 10 families~\cite{gutell:1994collection,szymanski+:19985s,damberger+:1994comparative,brown:1998ribonuclease,sprinz+:l1998compilation,zwieb+:2000tmrdb} of naturally occurring RNA and is used to evaluate RNA folding~\cite{huang+:2019linearfold}.\footnote{Pseudoknots are removed before running our algorithm (tRNA, 5S rRNA, and SRP do not contain pseudoknots).}
Following prior work in RNA design~\cite{garcia+:2013rnaifold,portela:2018unexpectedly,bellaousov+:2018accelerated,zhou+:2023samfeo} and undesignability~\cite{zhou+:2023undesignable,yao:2021thesis,yao+:2019,aguirre+:2007computational}, we used the RNA folding model parameters of ViennaRNA (v2.5.1)~\cite{lorenz+:2011}. 
Our code was written in C++, utilizing a 3.40 GHz Intel Xeon E3-1231 CPU and 32 GB memory. Parallelization was achieved by OpenMP for 8 CPU cores.
The parameters %$M, N, K$ 
during rival motif search (Alg.~\ref{alg:rivalmotif}) are set as $M = 10^{10}, N = 10^5, K = 100$. 
Our source code is available at {\tt\footnotesize\url{https://github.com/shanry/RNA-Undesign}}. % and {\tt\url{http://linearfold.org/motifs}}.\vspace{-0.25cm}
\begin{figure}[htb]
%\vspace{-0.12cm}
    \begin{minipage}{\textwidth}
        \centering
        \captionof{table}{Undesignable ({\em undes.}) structures and minimal undesignable ({\em m.~u.}) motifs  in Eterna100 puzzles \& native structures from ArchiveII.\vspace{-0.2cm}} \label{tab:stats} 
 \resizebox{\textwidth}{!}
 {
\begin{tabular}{|l|l||r|r|r|r|r||r|r||r|}
\hline
\multicolumn{2}{|l||}{\multirow{2}{*}{\em Dataset / family}}  	& 
\multirow{2}{*}{\em seqs.} &
\multirow{2}{*}{\shortstack{\em uniq.\\\em seqs.}} & 
\multicolumn{3}{c||}{\em structures} 		& \multicolumn{2}{c||}{\em m.~u.~motifs} & \multirow{2}{*}{\shortstack{\em time per\\\em structure} }	\\
\cline{5-9}
\multicolumn{2}{|l||}{}  	& &	& \em uniq.   	& \em avg.~len.  	& \em undes. & \em total 	 		& \em unique & 	\\ \hline\hline
\multicolumn{2}{|l||}{Eterna100 (Tab.~\ref{table:puzzles})} 		& - & - & 100   	& 218.3  		& 18  			& 36   	 			& 24				& 59.7 s\\
\hline
\parbox[t]{2mm}{\multirow{10}{*}{\rotatebox[origin=c]{90}{ArchiveII}}}
& tRNA   	(Fig.~\ref{fig:trna})	& 557 & 492	& 175	&77.1  		& 1      		& 1   	   		& 1 				&	0.2 s	 \\ \cline{2-10}
& 5S rRNA   		& 1,283 & 1,147	& 643	&118.7  		& 23      		& 31   	   		& 17 				&	0.5	 s \\ \cline{2-10}
& SRP   			& 928 & 702	& 661 & 183.9  		& 261     		& 458  	   		& 	118		&	9.0	s \\ \cline{2-10}
& RNaseP   		& 454 & 429 &	396 & 332.1  		& 99      		& 110  	   		& 60 			&	2.1 s	 \\ \cline{2-10}
& tmRNA   		& 462 & 404 &	348	& 366.0  		& 46      		& 58   	   		& 31 				&	15.9 s		 \\ \cline{2-10}
& Group I Intron 		& 98 & 93 & 93		& 426.4  		& 46      		& 50   	   		& 49 				&	18.4 s	 \\ \cline{2-10}
& telomerase   		& 37 & 37	& 37	& 444.6 		& 4      			& 4   	   		& 4 				&	0.7	 s \\ \cline{2-10}
& Group II Intron 		& 11 & 11& 11		& 716.5  		& 0      		& 0   	   		& 0 				&	9.1 s	 \\ \cline{2-10}
& 16S rRNA   		& 22 & 22 	& 22		& 1547.9 		& 22      		& 79   	   		& 30 				&   	502.8 s			 \\ \cline{2-10}
& 23S rRNA   		& 5 & 5 	& 5		& 2927.4 		& 5      			& 86   	   		& 36 				&   	129.8	 s	 \\
\hline
\multicolumn{2}{|l||}{\em All}    			& \em 3,857 & \em 3,342 & \em 2,491	& \em 207.6		& \em 525     		& \em 913		& \em 370			& \em 8.2  s    \\
\hline
\multicolumn{10}{|c|}{\shortstack{unique minimal undesignable motifs across all families: {\bf 355}.\\length: [5, 203] (avg 39.2); cardinality: [2, 5]}} \\ \hline
\end{tabular}
}
\end{minipage}
 \end{figure}
 
\begin{figure}[h]
	\vspace{-0.2cm}
	\begin{minipage}{\textwidth}
        \centering
	\begin{subfigure}[b]{0.32\textwidth}
            \centering
            \includegraphics[width=0.42\textwidth]{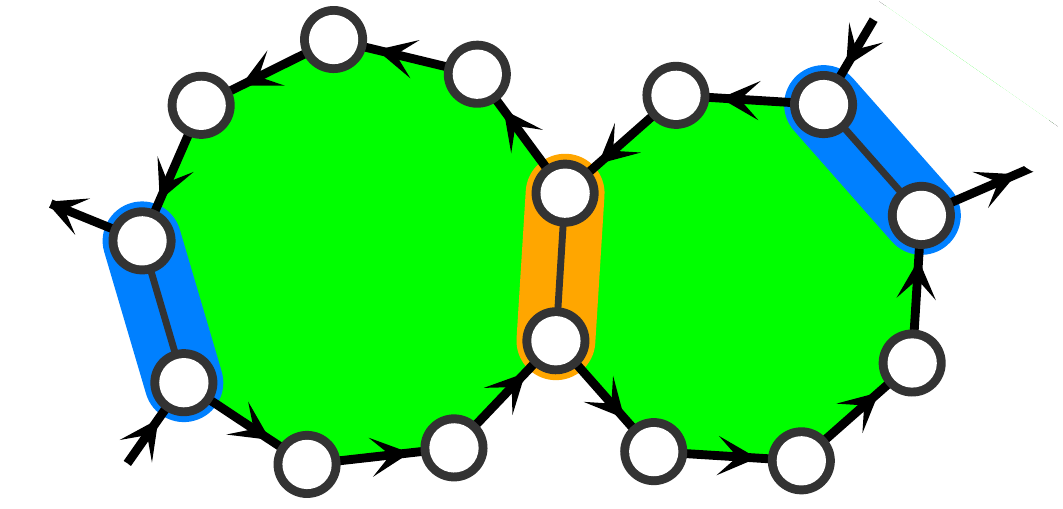}
            \begin{minipage}[b]{0.1\linewidth}  % Narrow minipage for the arrow
                \centering
                \raisebox{0.2cm}{  % Adjust the vertical position of the arrow
                    \begin{tikzpicture}
                        \draw[->, thick] (1,1) arc [start angle=80, end angle=320, radius=0.2cm];
                    \end{tikzpicture}
                }
            \end{minipage}
            \includegraphics[width=0.42\textwidth]{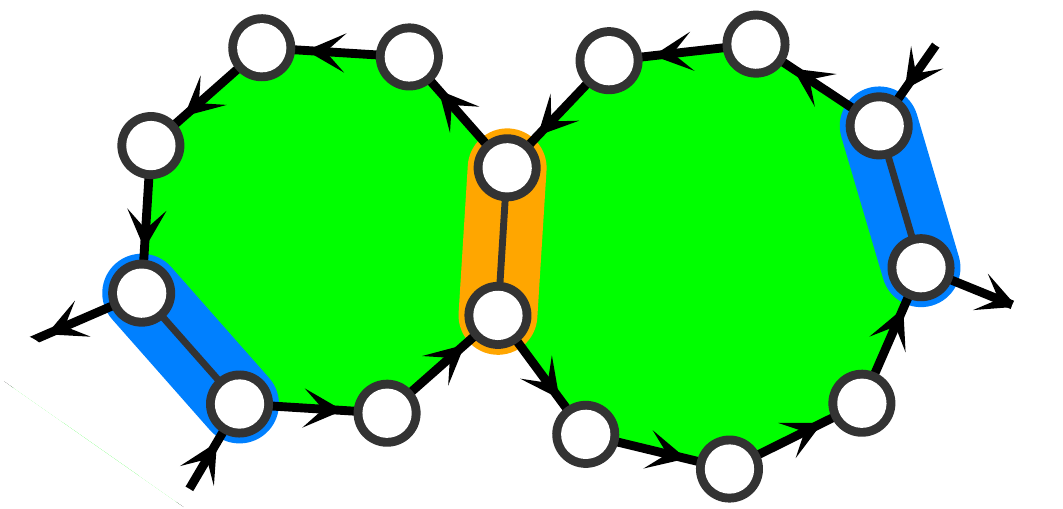}
            \caption{Rotational variants}
            \label{fig:rotational}
        \end{subfigure}
        \hfill % Add space between subfigures
        % Second subfigure: 5S rRNA
        \begin{subfigure}[b]{0.33\textwidth}
            \centering
            \includegraphics[width=0.9\textwidth]{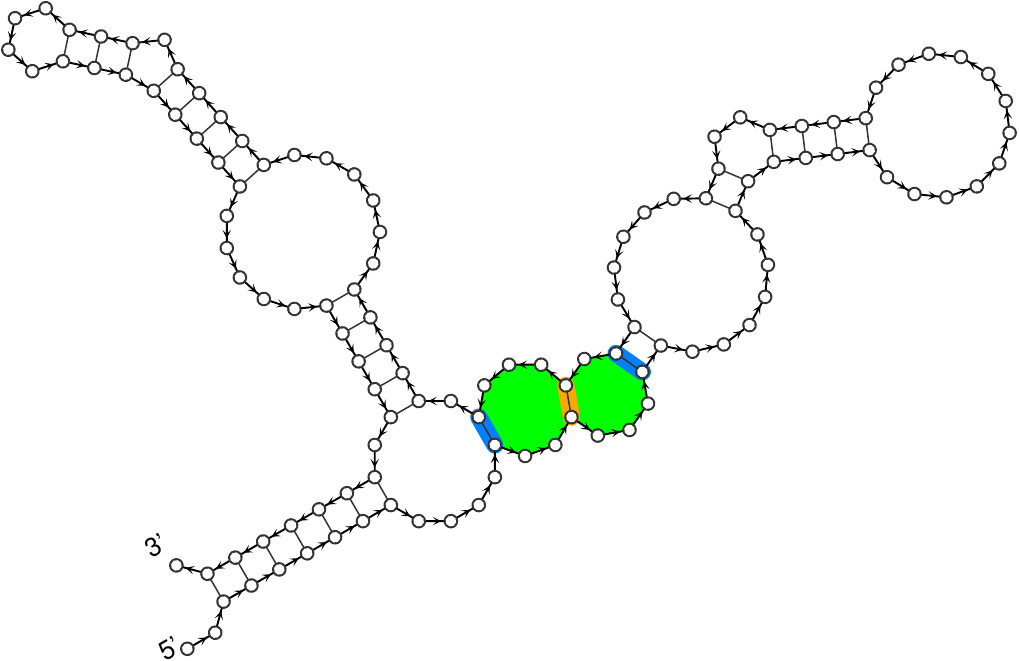}
            \caption{5S rRNA}
            \label{fig:5s}
        \end{subfigure}
        \hfill
        % Third subfigure: SRP
        \begin{subfigure}[b]{0.33\textwidth}
            \centering
            \includegraphics[width=0.95\textwidth, angle=0]{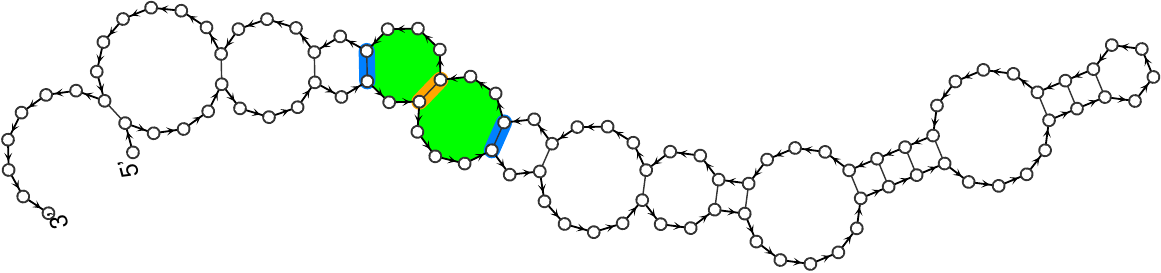}
            \caption{SRP}
            \label{fig:srp}
        \end{subfigure}
    \end{minipage}
    \caption{\mbox{\small Rotational variants of a minimal undesignable motif in 5S rRNA and SRP families.}}
    \label{fig:shared}
\end{figure}

\subsection{\mbox{Undesignable Structures and Unique Minimal Undesignable Motifs}}
Table~\ref{tab:stats} summarizes the statistics of undesignable structures and (minimal) undesignable motifs, as well as the running time,
for both Eterna
 puzzles and native structures from ArchiveII. 
%Note in Eterna100 puzzle of ID \#50, one motif is undesignable if heuristic energies for special hairpins are ignored.
Among Eterna100 puzzles, we found 24 unique minimal undesignable motifs 
(36 occurrences), resulting in 18 undesignable puzzles
(Table.~\ref{table:puzzles}).
This result is stronger than that of RIGENDE, which identified 16 undesignable puzzles along with rival (sub-)structures for each puzzle.

\begin{figure}[h]
    \vspace{-0.9cm}
    \centering
        \includegraphics[width=0.40\textwidth,angle=10]{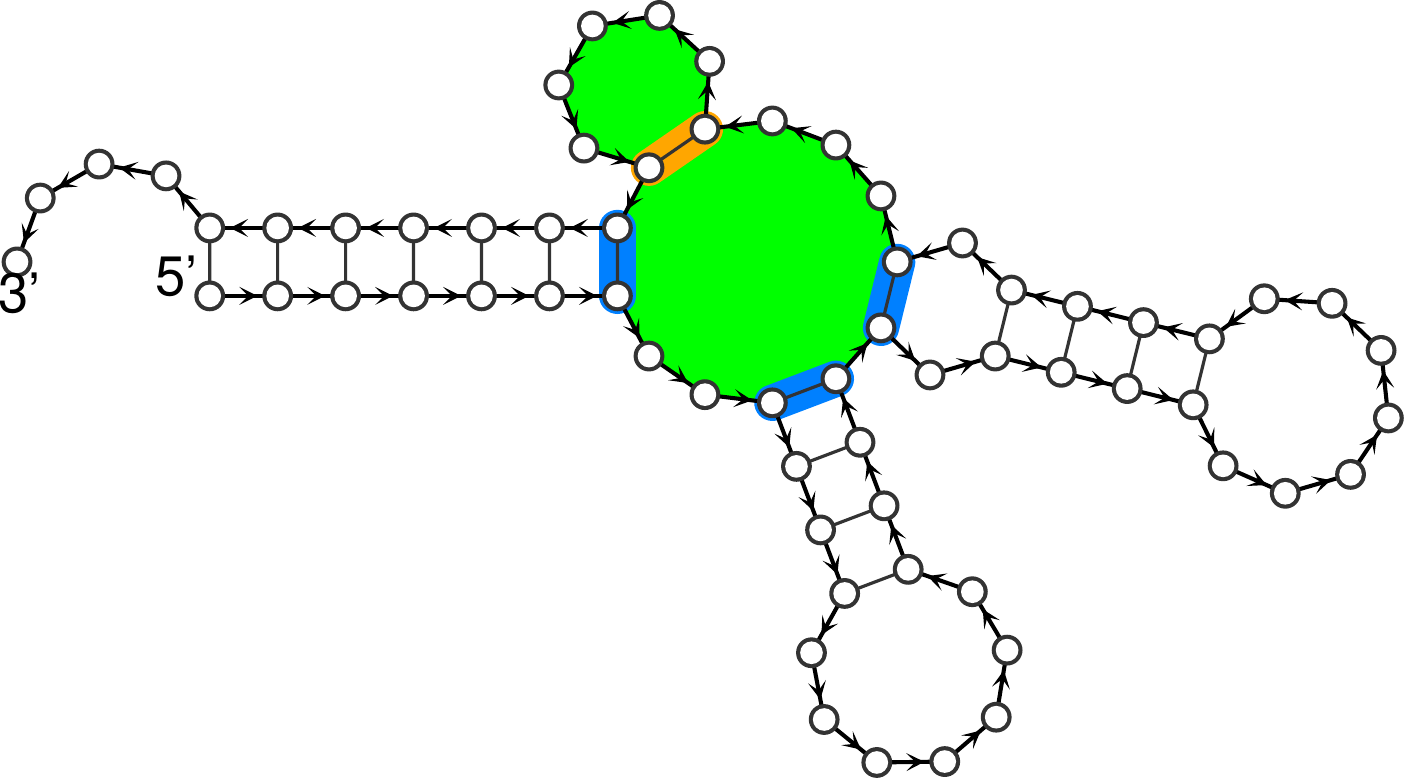}\vspace{-0.4cm}
   \caption{The minimal undesignable motif identified in a tRNA secondary structure.}\label{fig:trna}
   \vspace{-0.9cm}
\end{figure}

The structures in ArchiveII, on the other hand, are high-quality native structures,
which intuitively should be designable.
%originating from nature, led us to anticipate a scarcity of undesignable motifs within them. 
Surprisingly, there are about 900 occurrences of undesignable motifs from almost all ArchiveII families (except for Group II Intron). 
In total, we found 331 unique minimal undesignable motifs in ArchiveII,
some of them shared across families,
and more than 500 undesignable structures.
For example, Fig.~\ref{fig:shared} shows a minimal undesignable motifs shared across two families and Fig.~\ref{fig:trna} shows
the only undesignable tRNA structure and motif.
% shown in Fig.~\ref{fig:trna}, which we found very counter-intuitive. 
%Refer to our web server for more motif examples in other families such as 16S \&  23S  rRNA structures.
We suspect the large number of undesignable structures and motifs are due to
the energy model (ViennaRNA 2) being imperfect 
and pseudoknots playing a role in designability which is beyond our work.
Interestingly, no motifs are shared between Eterna100 and ArchiveII, and
in total we found 355 unique minimal undesignable motifs.
%which can be browsed on our web server {\tt\footnotesize\color{blue}\url{http://linearfold.org/motifs}}.
%Users can even upload a new structure and the server 
%will find minimal undesignable motifs in it on the fly.
\subsection{Efficiency}
\label{sec:efficiency}
\begin{figure}[hbtp]
    \centering
    \includegraphics[width=0.58\textwidth]{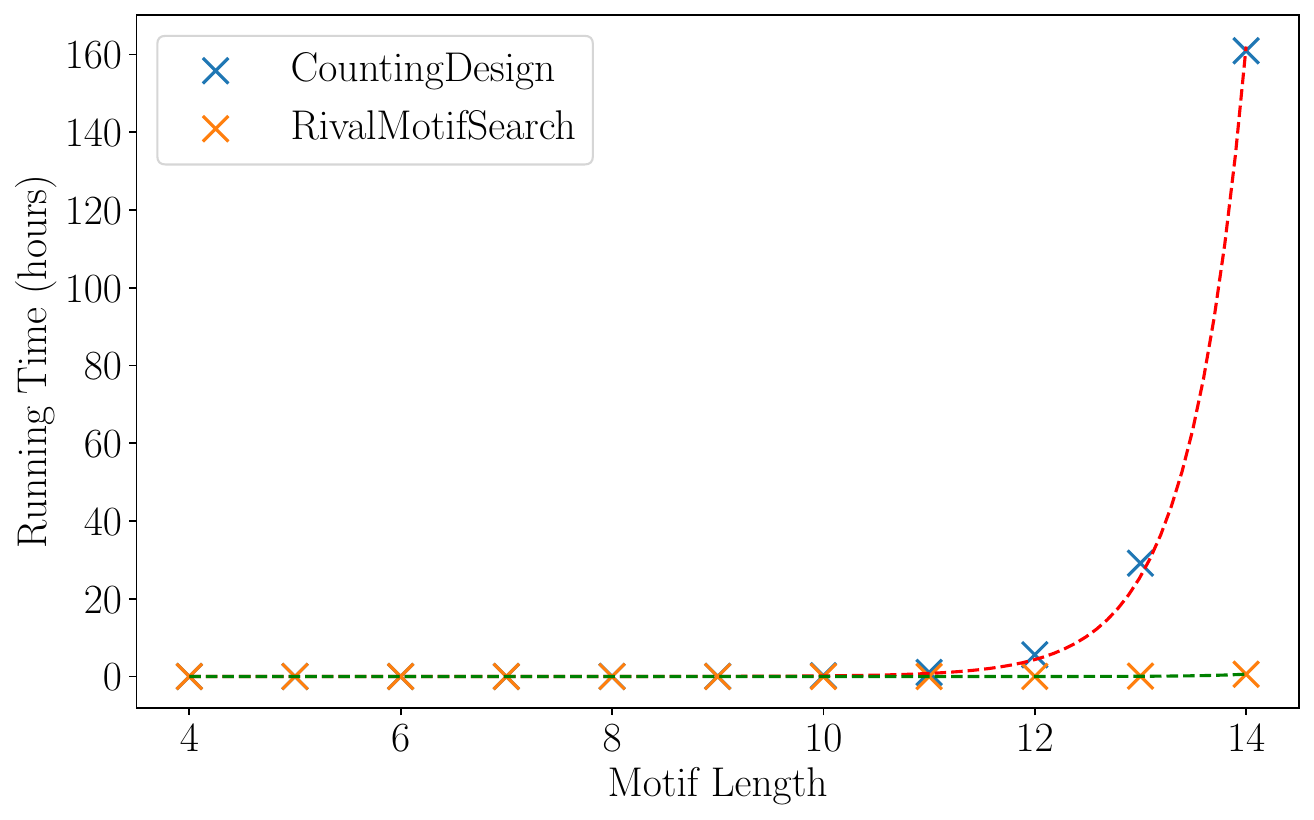}
    \hspace{-0.2cm}
    \raisebox{2cm} {
    \resizebox{0.37\textwidth}{!}
    {
    \begin{tabular}{|c|c|}
        \hline
         & Cumulative Runtime\\  \hline
         CountingDesign & 1.17 weeks \\ \hline
         {RivalMotifSearch}{} & 0.7 hours\\ \hline
    \end{tabular}
    }
    }
    \caption{Running time comparison between RivalMotifSearch (Algorithm~\ref{alg:rivalmotif}) and CountingDesign.} % for evaluating motifs of lengths up to 14.
    \label{fig:fmvscd_time}
\end{figure}

From Table~\ref{tab:stats}, we can see the algorithm costs only a few seconds or minutes to scan an entire structure. 
Such efficiency is much superior to the previous work CountingDesign. As a comparison, identifying undesignable motifs of length 39, the average length of minimal undesignable motifs identified in our work, would take thousands of years per motif using CountingDesign (Fig.~\ref{fig:fmvscd_time}). 
We also apply our rival motif search algorithm (Algorithm~\ref{alg:rivalmotif}) to all motifs up to length 14, which is how CountingDesign was benchmarked. 
%We successfully identified all the minimal undesignable motifs (total: 4561; unique: 1805) in 0.7 hours, whereas CountingDesign takes more than a week.
%See Supplementary Section~\ref{sec:countds} for detailed comparison with CountingDesign.
We ran \Call{RivalMotifSearch}{} (Algorithm~\ref{alg:rivalmotif}) and the original CountingDesign~\cite{yao+:2019} program\footnote{\url{https://gitlab.com/htyao/countingdesign}} under the same setting (3.40 GHz Intel Xeon E3-1231 CPU and 32 GB of memory, with parallelization of 8 CPU cores).
Fig.~\ref{fig:fmvscd_time} shows the running time of the two methods for identifying undesignable motifs of different lengths. 
Both methods found all the undesignable motifs (total: 4561; unique: 1805) and the designable motifs up to length of 14.
However, the time cost of CountingDesign increases exponentially with motif lengths, highlighting its impracticality for longer motifs. 
In contrast, FastMotif only need 0.7 hours. 
More importantly, FastMotif identified a set of rival motifs for each undesignable motif, which is explainable and helpful for further understanding RNA folding. Therefore, FastMotif demonstrates significant advantages in terms of not only scalability but also interpretability.
\subsection{Web Server}
%We built a web server ({\tt\footnotesize\color{blue}\url{http://linearfold.org/motifs}}) to browse the undesignable motifs and structures we have identified. The short motifs enumerated in Section~\ref{sec:efficiency} are named as Enum. Users can also upload new structures to identify undesignability on the fly.
We developed a web server ({\tt\footnotesize\color{blue}\url{https://linearfold.eecs.oregonstate.edu/motifs}}) that allows users to explore the undesignable motifs and structures we have identified. The short motifs enumerated in Section~\ref{sec:efficiency} are labeled as ``Enum.'' Additionally, users can upload new structures to analyze their undesignability in real time.
\begin{table}[h!]
  \centering
  \begin{threeparttable}
    \caption{Minimal Undesignable Motifs in 18 Undesignable Eterna100 Structures}
    \label{table:puzzles}
    \begin{tabular}{|p{0.21\linewidth}|c|p{0.2\linewidth}|c|} \hline
        \small ID: Puzzle & Minimal Undesignable Motifs & \small ID: Puzzle & Minimal Undesignable Motifs\\ \hline
        \small 50: 1, 2, 3 and 4 bulges\tnote{a} & \begin{picture}(0,0)
                \put(-10, 8){\small \textbf{1}}  % Numbering the top-left corner
            \end{picture} \raisebox{-.5\height}{\includegraphics[scale=0.08, angle=0]{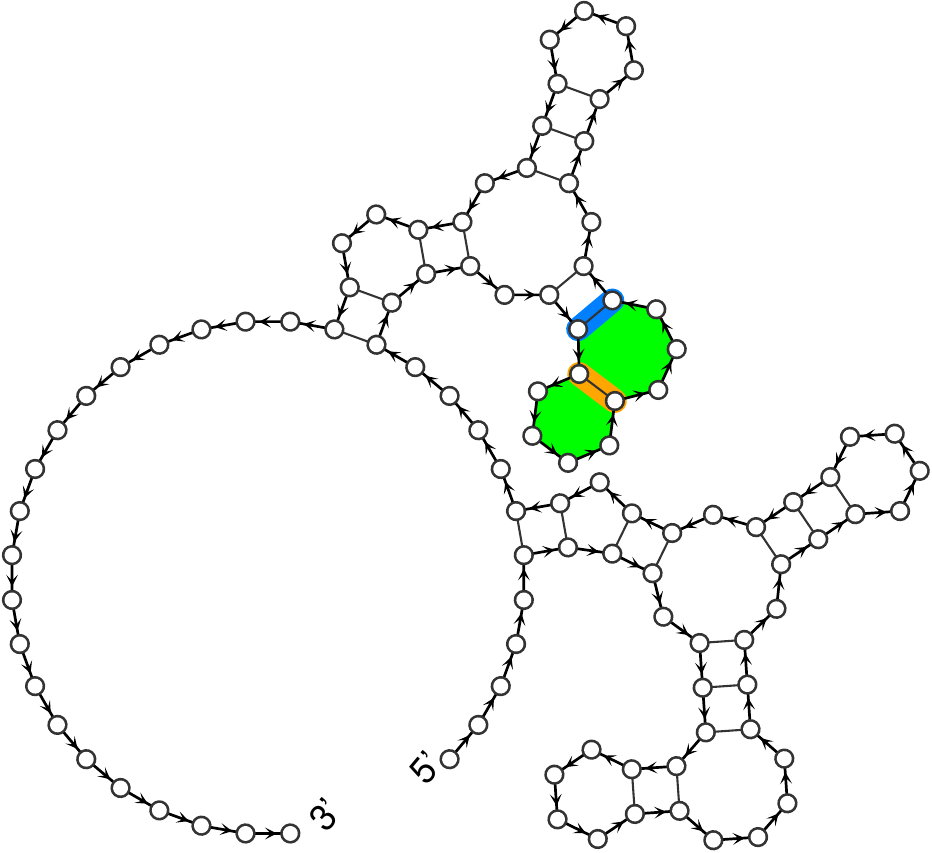}} &
        \small 52: [RNA] Repetitive~Seqs.~8/10 &  \begin{picture}(0,0)
                \put(-10, 8){\small \textbf{2}}  % Numbering the top-left corner
            \end{picture}  \raisebox{-.5\height}{\includegraphics[scale=0.07, angle=0]{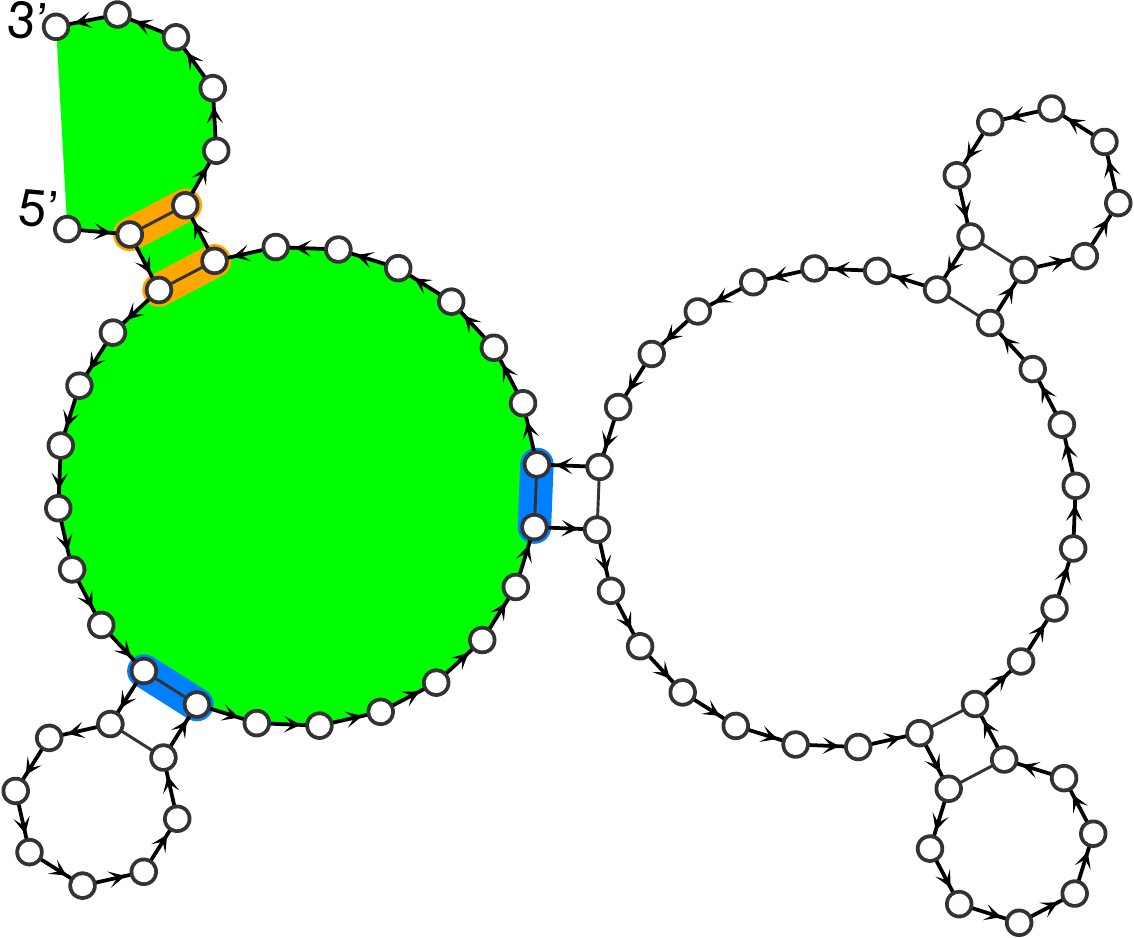}} 
        \raisebox{-.5\height}{\includegraphics[scale=0.07, angle=0]{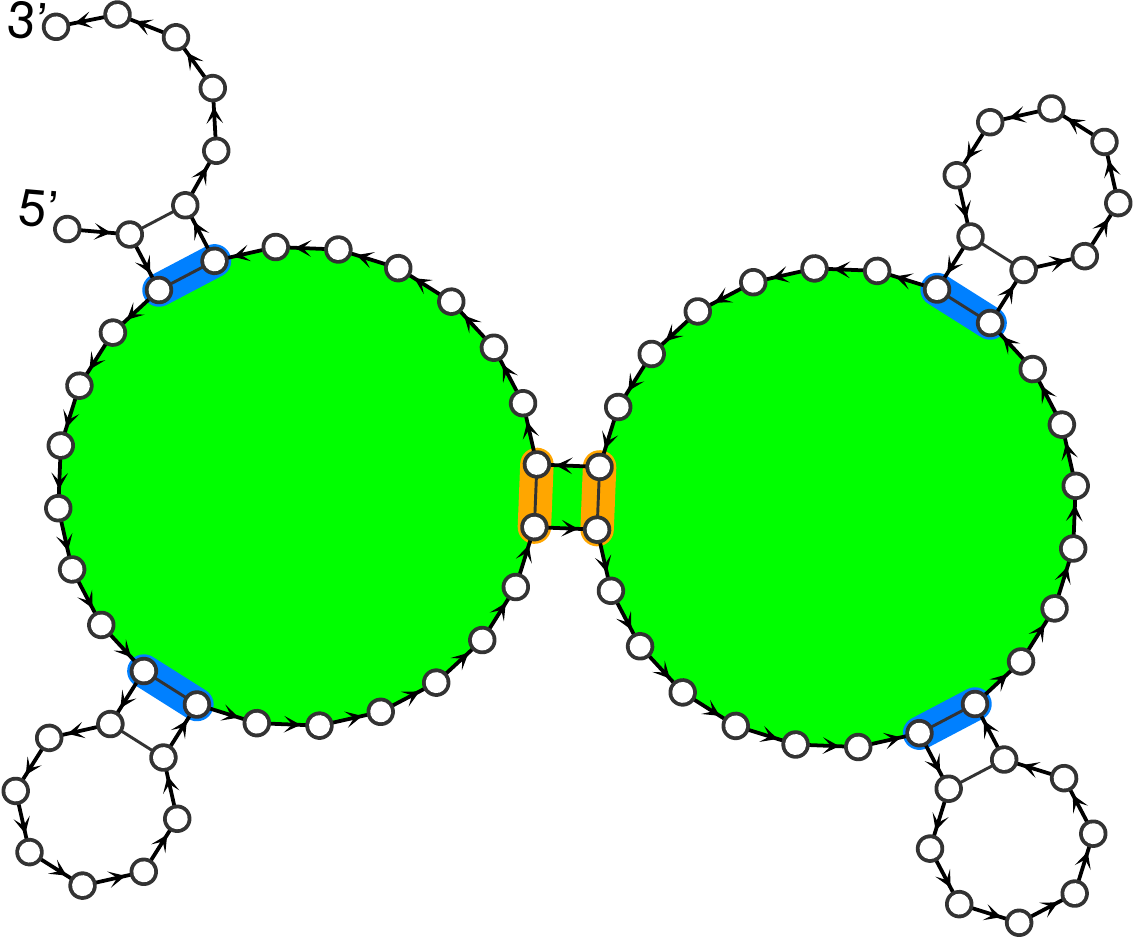}}\\ \hline
        \small 57: multilooping fun (see Fig.~\ref{fig:neighbors}) & \raisebox{-.5\height}{\begin{picture}(0,0)
                \put(-10, 12){\small \textbf{3}}  % Number for second image
            \end{picture}\includegraphics[scale=0.09]{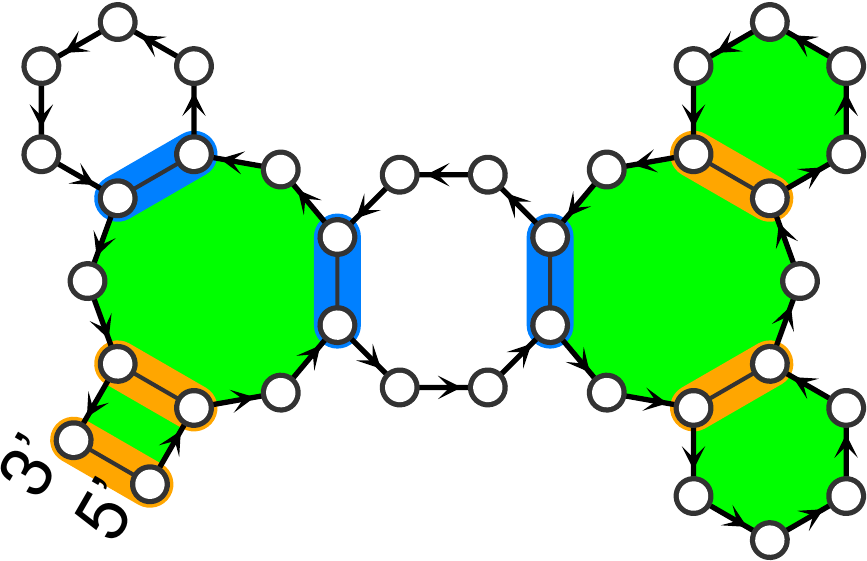}} \raisebox{-.5\height}{\includegraphics[scale=0.09]{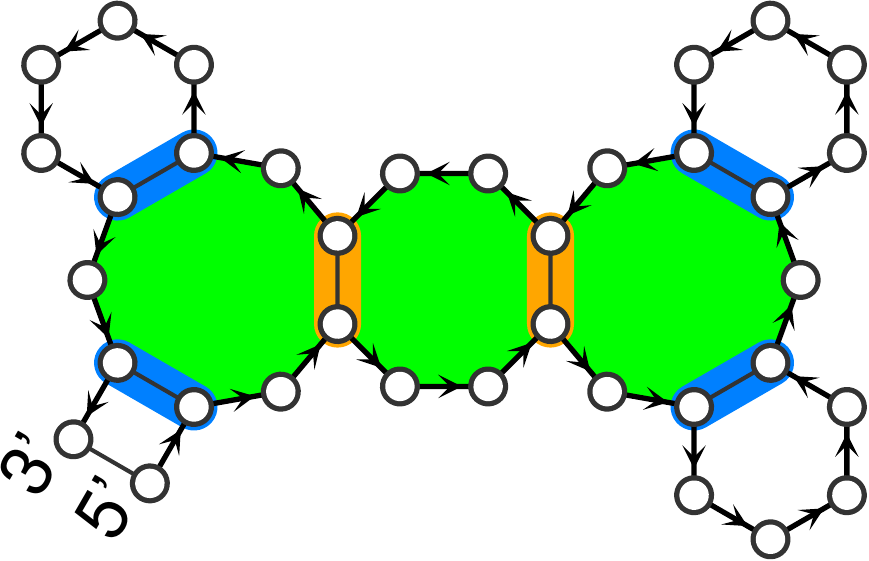}} &
        \small 60: Mat - Elements \& Sections & \raisebox{-.5\height}{\begin{picture}(0,0)
                \put(-10, 13){\small \textbf{1}}  % Number for second image
            \end{picture}\includegraphics[scale=0.16, angle=0]{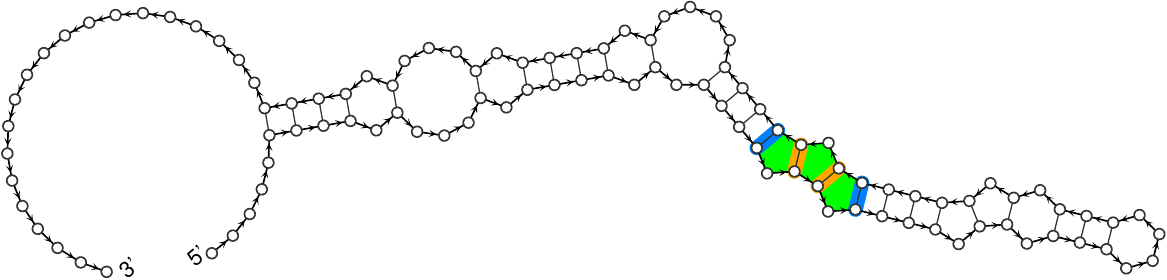}}\\ \hline
        \small 61: Chicken feet & \raisebox{-.5\height}{\begin{picture}(0,0)
                \put(-10, 20){\small \textbf{2}}  % Number for second image
            \end{picture}\includegraphics[scale=0.08, angle=0]{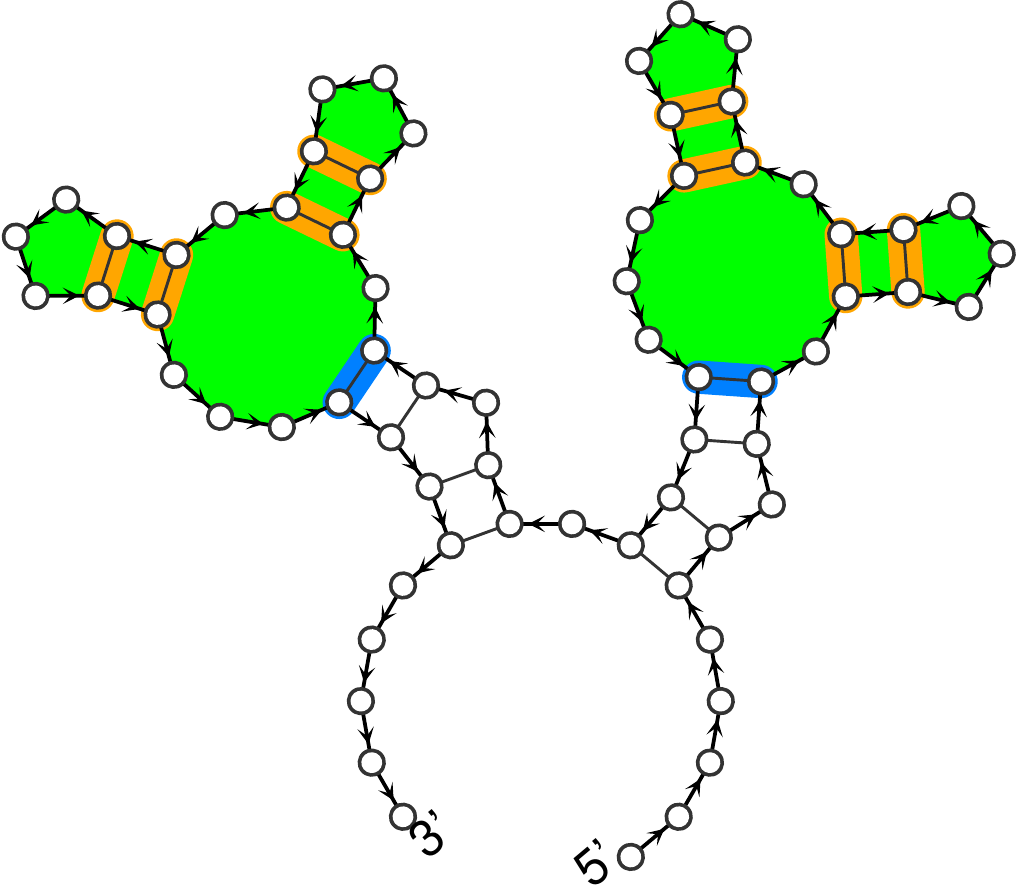}} &
        \small 67: Simple Single Bond & \raisebox{-.5\height}{\begin{picture}(0,0)
                \put(-10, 20){\small \textbf{1}}  % Number for second image
            \end{picture}\includegraphics[scale=0.11, angle=0]{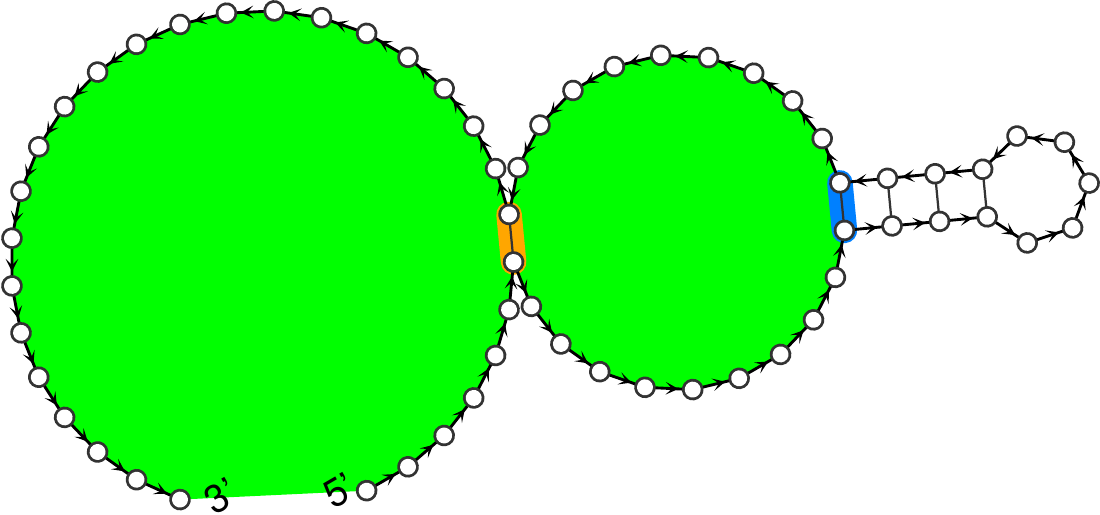}}\\ \hline
        \small 72: Loop next to a Multiloop & \raisebox{-.5\height}{\begin{picture}(0,0)
                \put(-10, 20){\small \textbf{1}}  % Number for second image
            \end{picture}\includegraphics[scale=0.12, angle=0]{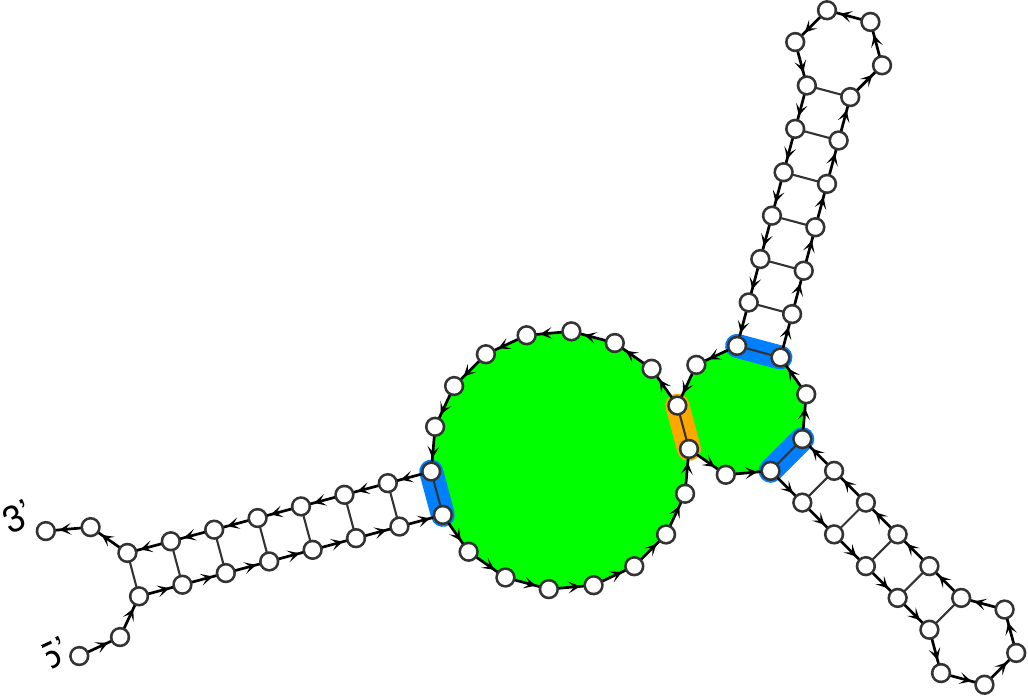}} &
        \small 78: Mat - Lot 2-2 B & \raisebox{-.5\height}{\begin{picture}(0,0)
                \put(-10, 20){\small \textbf{1}}  % Number for second image
            \end{picture}\includegraphics[scale=0.12, angle=0]{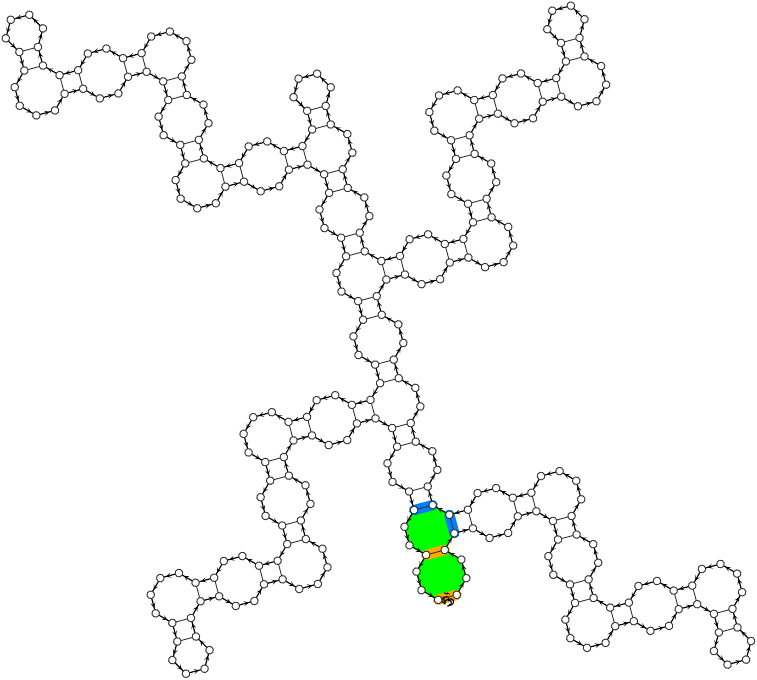}}\\ \hline
        \small 80: Spiral of 5's & \raisebox{-.5\height}{\begin{picture}(0,0)
                \put(-10, 20){\small \textbf{1}}  % Number for second image
            \end{picture}\includegraphics[scale=0.11, angle=0]{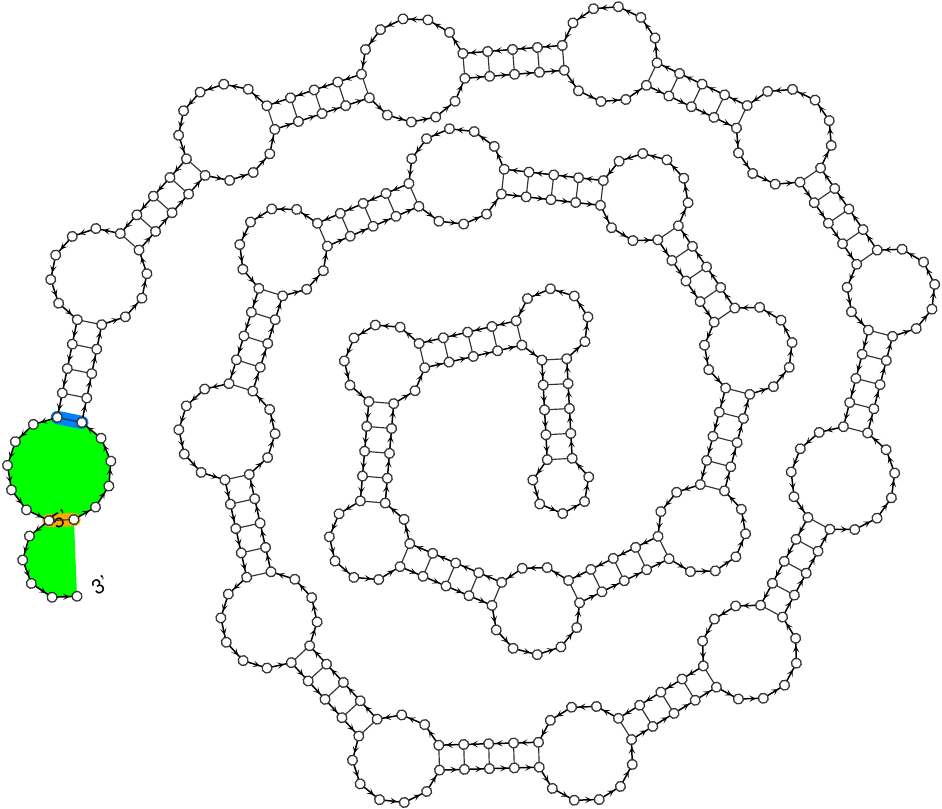}} &
        \small 81: Campfire & \raisebox{-.5\height}{\begin{picture}(0,0)
                \put(-10, 20){\small \textbf{1}}  % Number for second image
            \end{picture}\includegraphics[scale=0.12, angle=0]{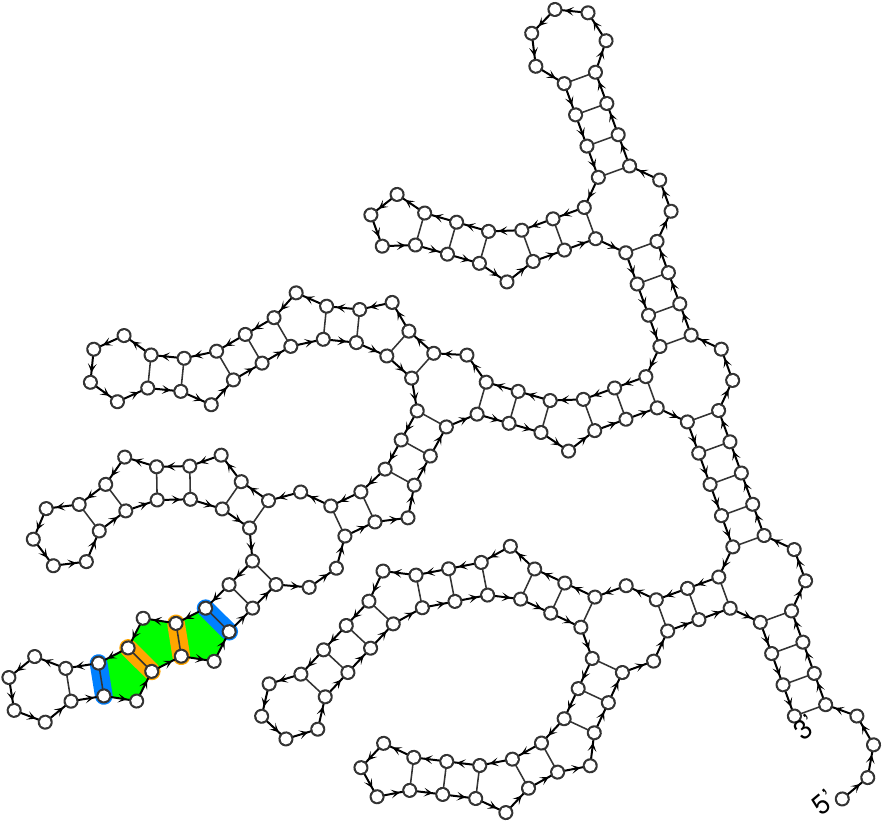}}\\ \hline
        \small 86: Methaqualone C$_{16}$H$_{14}$N$_2$O & \raisebox{-.5\height}{\begin{picture}(0,0)
                \put(-10, 20){\small \textbf{3}}  % Number for second image
            \end{picture}\includegraphics[scale=0.18, angle=0]{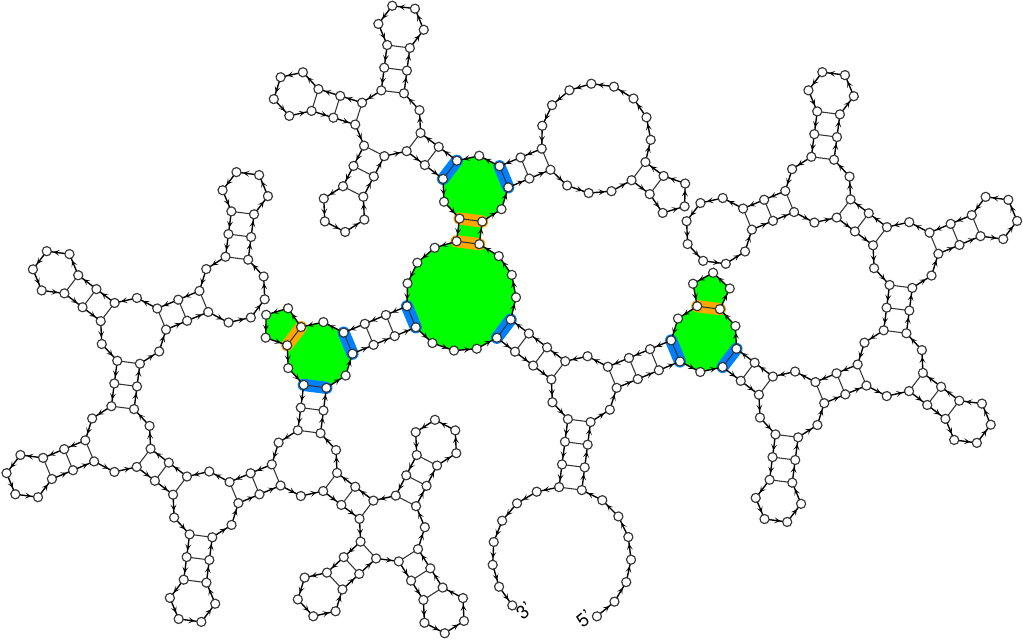}} &
        \small 87: Cat's Toy 2 & \raisebox{-.5\height}{\begin{picture}(0,0)
                \put(-10, 20){\small \textbf{1}}  % Number for second image
            \end{picture}\includegraphics[scale=0.15, angle=-25]{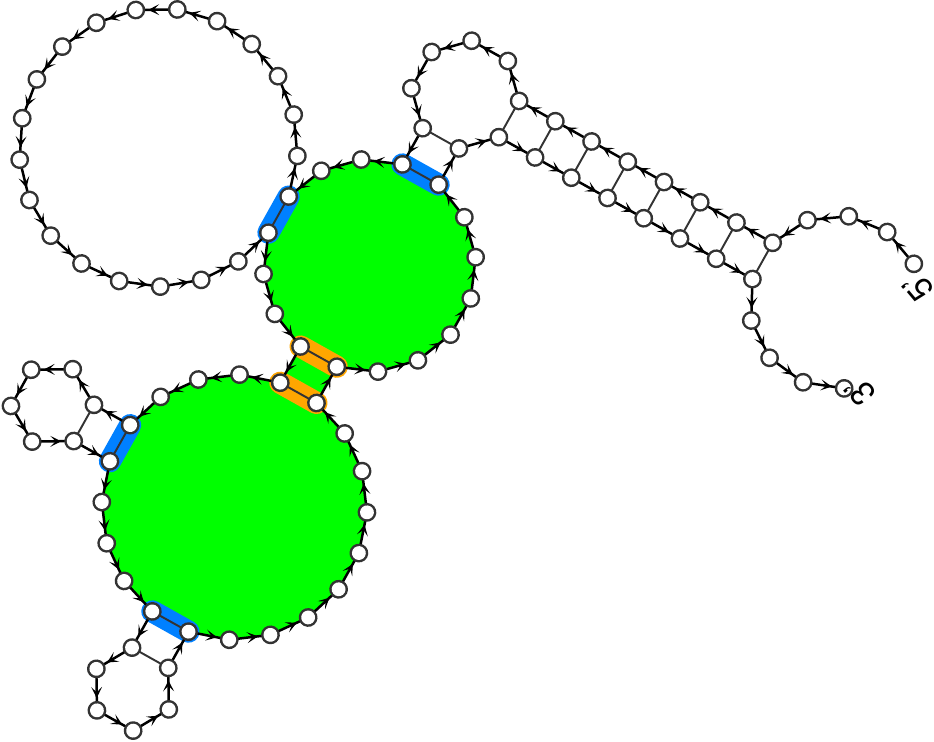}}\\[-0.7cm] \hline
        \small 88: Zigzag Semicircle & \raisebox{-.7\height}{\begin{picture}(0,0)
                \put(-10, 20){\small \textbf{1}}  % Number for second image
            \end{picture}\includegraphics[scale=0.10, angle=0]{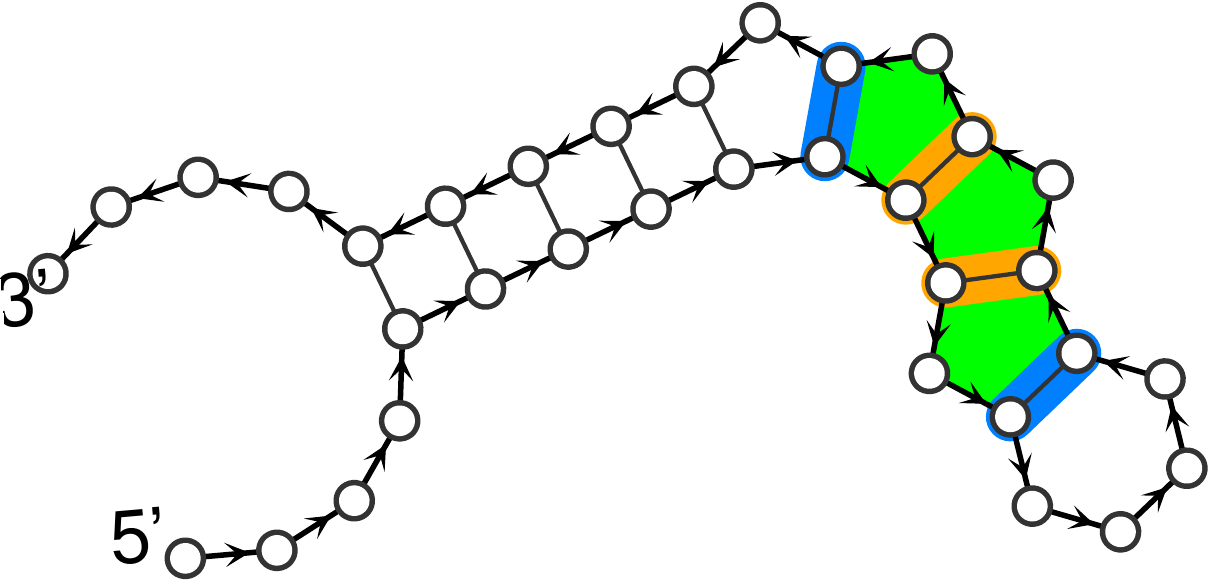}} &
        \multirow{2}{*}{\small 90: Gladius} & \hspace{-1cm}\raisebox{-.5\height}{\begin{picture}(0,0)
                \put(-5, 13){\small \textbf{2}}  % Number for second image
            \end{picture}\includegraphics[scale=0.14, angle=0]{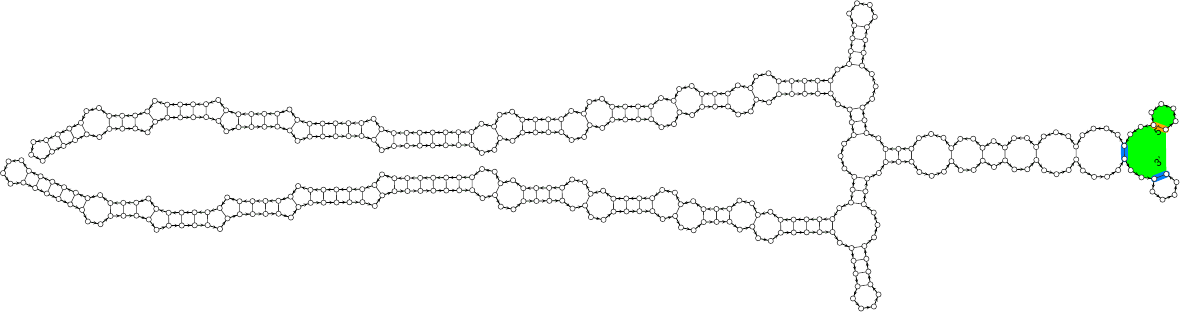}}\\[-0.5cm]
        ~  ~ & ~ & ~  & \hspace{1cm}\raisebox{-.0\height}{\includegraphics[scale=0.14, angle=0]{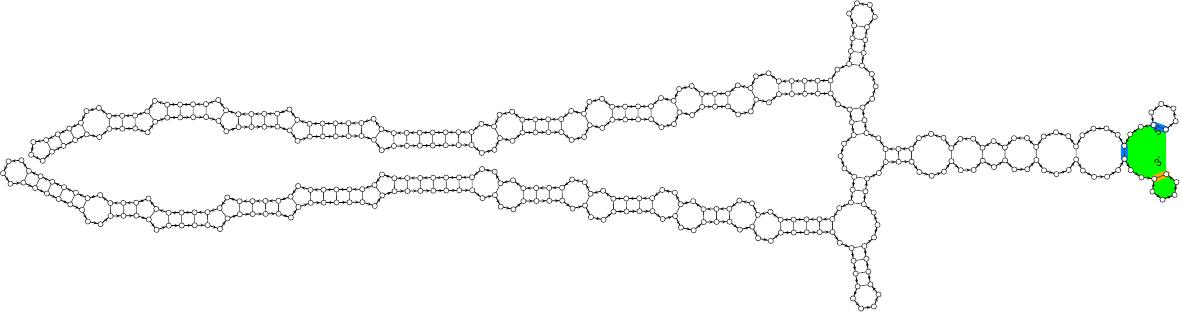}}\\[-0.1cm] \hline
        \multirow{2}{*}{\small 91: Thunderbolt} & \hspace{-1cm}\raisebox{-.4\height}{\begin{picture}(0,0)
                \put(0, 20){\small \textbf{3}}  % Number for second image
            \end{picture}\includegraphics[scale=0.17, angle=0]{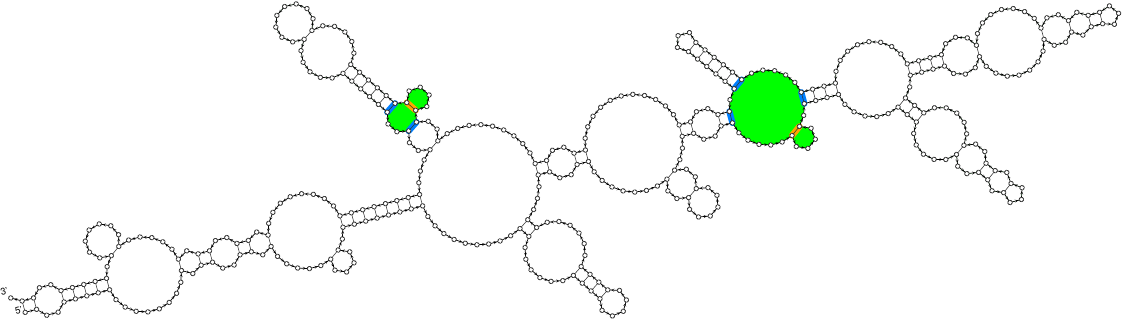}} &
        \small 92: Mutated chicken feet & \raisebox{-.5\height}{\begin{picture}(0,0)
                \put(-10, 20){\small \textbf{3}}  % Number for second image
            \end{picture}\includegraphics[scale=0.10, angle=0]{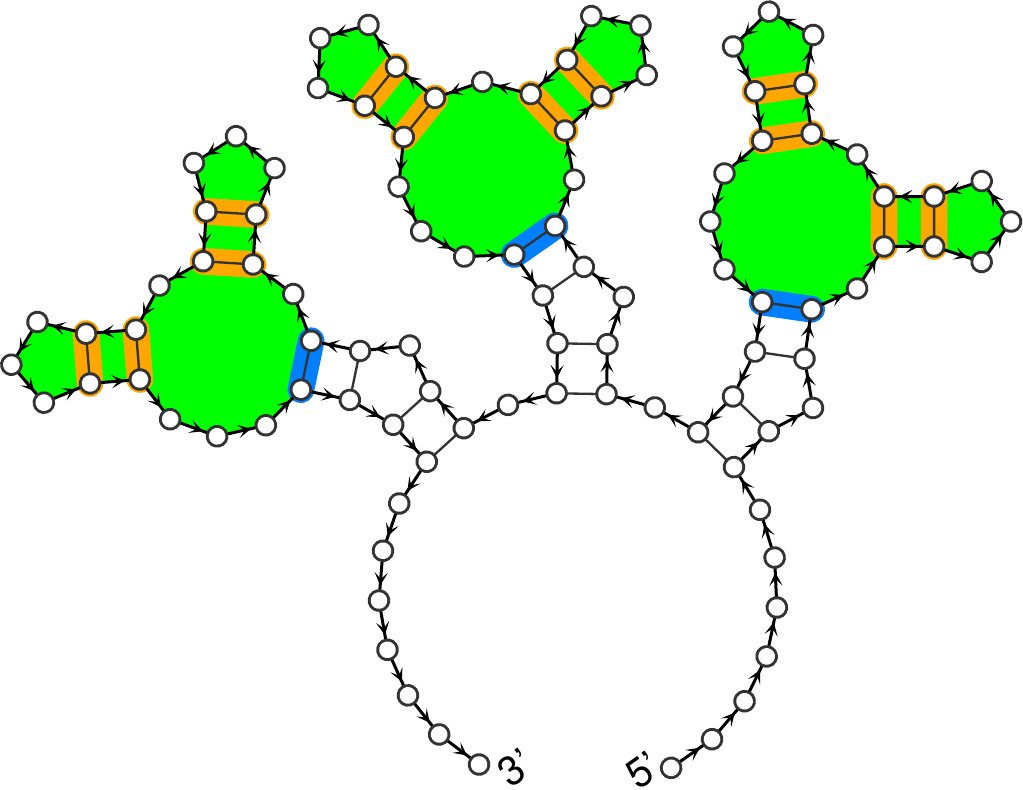}}\\[-0.5cm]
        ~  ~ & \hspace{1.0cm}\raisebox{-.0\height}{\includegraphics[scale=0.14, angle=0]{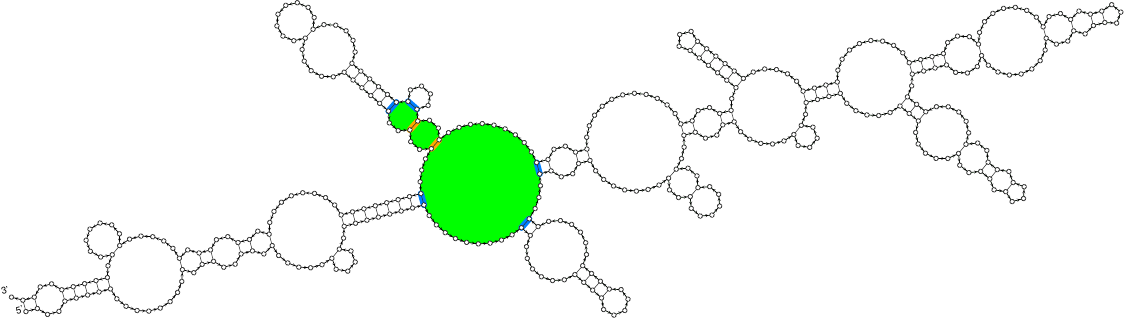}} & ~  & ~\\[-0.1cm] \hline
        \multirow{2}{*}{\small 96: Cesspool} & \raisebox{-.4\height}{\begin{picture}(0,0)
                \put(-5, 28){\small \textbf{8}}  % Number for second image
            \end{picture}\includegraphics[scale=0.12, angle=0]{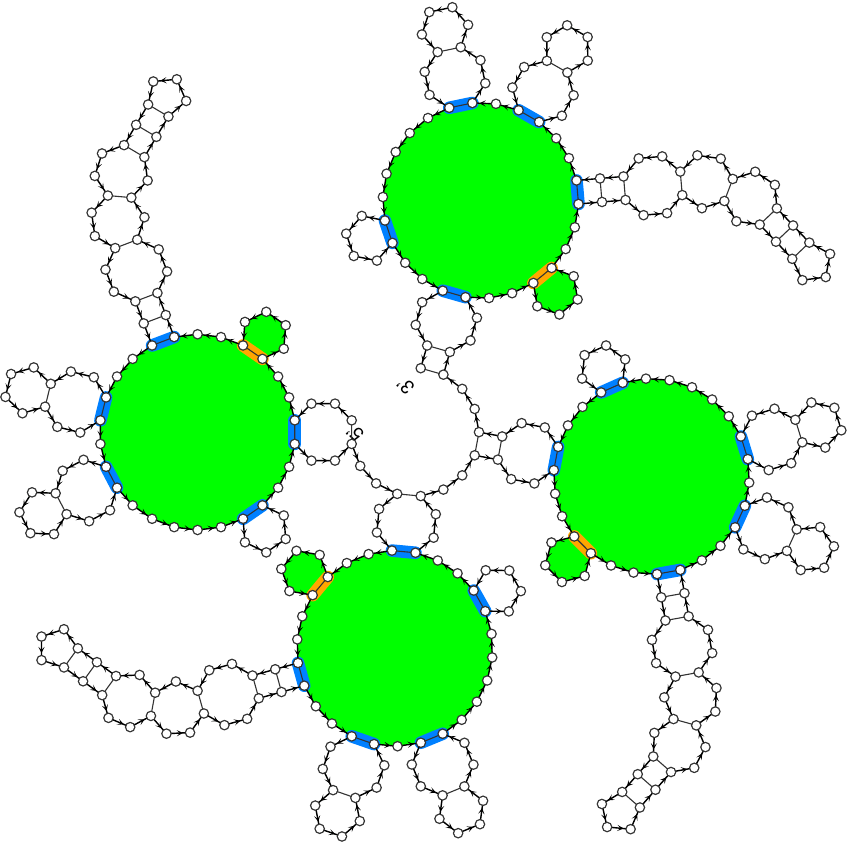}}
        \quad
        \raisebox{-.4\height}{\includegraphics[scale=0.12, angle=0]{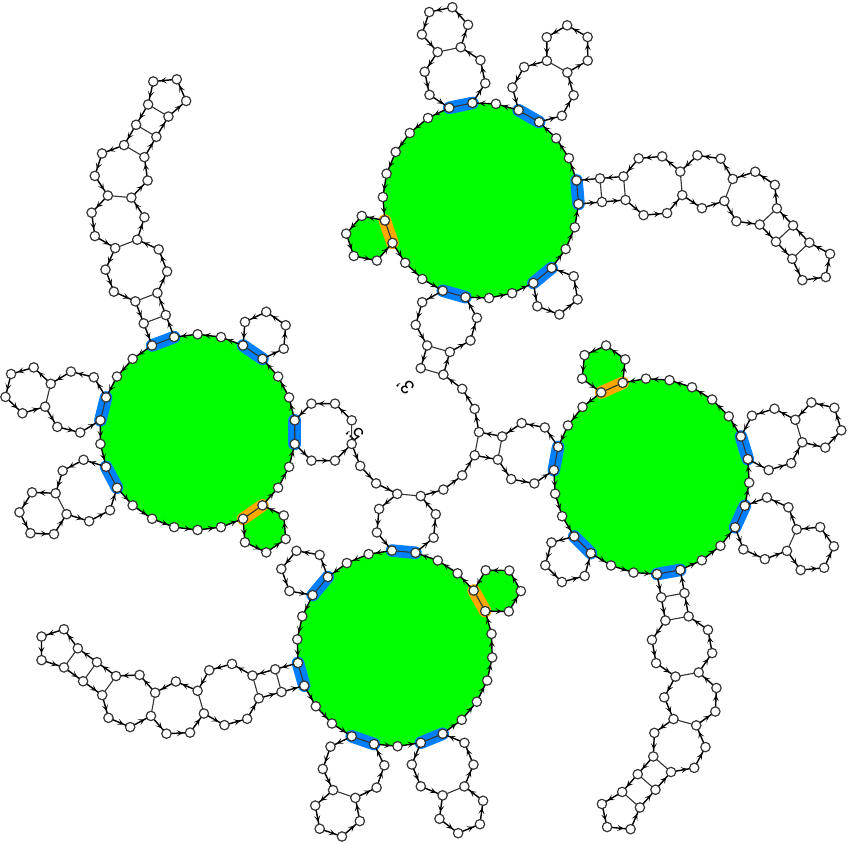}} &
        \small 99: Shooting Star & \raisebox{-.4\height}{\begin{picture}(0,0)
                \put(-10, 20){\small \textbf{1}}  % Number for second image
            \end{picture}\includegraphics[scale=0.18, angle=0]{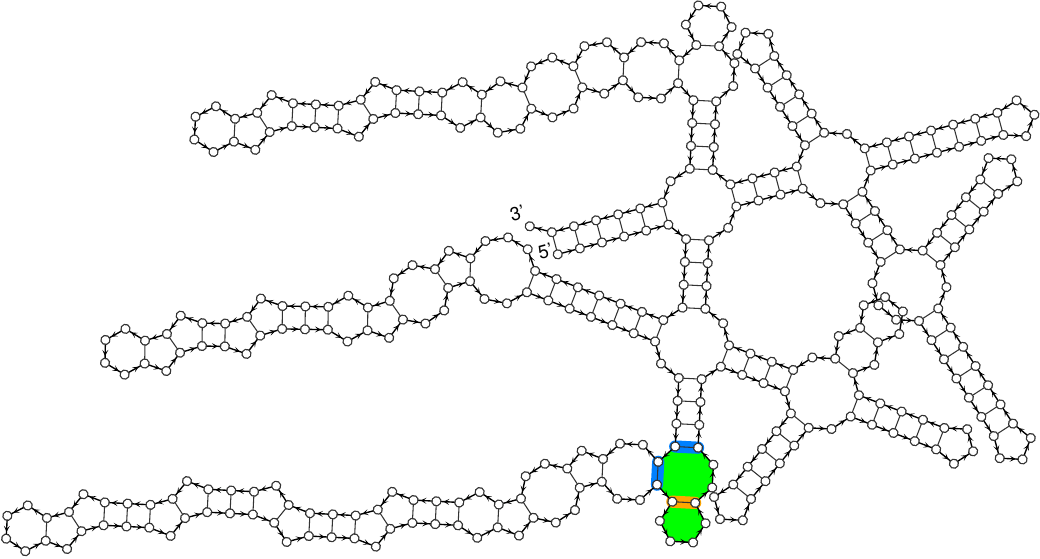}}\\ \hline
    \end{tabular}
    \begin{tablenotes}
      \footnotesize
      \item[a] This is a special puzzle with unique features.
    \end{tablenotes}
  \end{threeparttable}
\end{table}
% !TEX root = main.tex
%\vspace{-0.4cm}
\section{Limitations}
We acknowledge several limitations of this work. First of all, this work focuses on the widely used Turner RNA folding model \cite{Mathews+:2004,turner+:2010nndb} where a structure can be decomposed into loops. Theoretically, theorem~\ref{theorem:onerm} and theorem~\ref{theorem:multi-rival} (for fixed $k$) provide sufficient but not necessary conditions for motif undesignability. Algorithmically, while rival motif search (Algorithm~\ref{alg:rivalmotif})  has demonstrated strong performance, it includes stop conditions to prevent excessive running time. As a result, it does not guarantee the identification of satisfying rival motifs. Additionally, for the undesignable motifs found in the native structures from ArchiveII, some instances of undesignability may be influenced by tertiary interactions such as pseudoknots. However, theses interactions fall outside the scope of this work. 

\section{Conclusions and Future Work}\label{sec:conclusion}
%\vspace{-0.25cm}
We introduced a theoretical framework for loop-based motifs,  and fast algorithms with a loop-pair graph representation to identify unique minimal undesignable motifs in RNA structures. 
By searching for rival motifs, the undesignability of motifs can be efficiently confirmed and explicitly explained.  Future exploration could involve implementing DFS/BFS-based algorithms to search for a broader range of undesignable motifs.

%The results with ArchiveII  suggest that the current thermodynamic parameters are deficient.  We speculate there is room to improve the parameterization~\cite{zuber+:2019estimating,ward+:2019determining} from the perspective of designability and undesignability.  In future work, it would be interesting to compare the set of minimal undesignable motifs using parameter sets beyond that implemented in the ViennaRNA package, including complete parameterizations that include coaxial stacking~\cite{reuter+:2010rnastructure} or parameterizations that are informed by known structures~\cite{wayment+:2022rna,andronescu+:2010computational,rivas+:2012range}. 
The results with ArchiveII  suggest the current thermodynamic parameters are deficient. 
We hypothesize that improvements in parameterization~\cite{zuber+:2019estimating,ward+:2019determining} could be made, particularly from the perspectives of  undesignability. 
Future work could  involve comparing the sets of minimal undesignable motifs using alternative parameter sets beyond those implemented in the ViennaRNA package, including comprehensive parameterizations that account for coaxial stacking or that are informed by experimentally known structures~\cite{wayment+:2022rna,andronescu+:2010computational,rivas+:2012range}. In addition, the methodology of this work can also be extended to other loop-based RNA folding models such as Contrafold~\cite{do+:2006}.
%In addition, it is also possible to refine the current folding models by addressing issues related to illegitimate undesignable motifs.

\section*{Acknowledgements}
This work was supported in part by NSF grants 2009071 (L.H.) and 2330737 (L.H. and D.H.M.) and NIH grant R35GM145283 (D.H.M.). We thank the anonymous reviewers for feedback.
%\clearpage
\bibliographystyle{splncs04}
\bibliography{references}

\clearpage

\pagenumbering{roman}

\appendix
\title{Supplementary Information}
\author{}
\institute{}
\maketitle
% !TEX root = main.tex

\setcounter{figure}{0}
\renewcommand{\thefigure}{SI\,\arabic{figure}} % spacing
\setcounter{table}{0}
\renewcommand{\thetable}{SI\,\arabic{table}}
\setcounter{page}{1}

\setcounter{section}{0}
\renewcommand\thesection{\Alph{section}}
\setcounter{subsection}{0}
\renewcommand\thesubsection{\Alph{section}.\arabic{subsection}}

\iffalse %%%%%%%%%%%%%%% lhuang: merged into Fig. 1
\section{A List of Loops}\label{sec:loops}
\begin{figure}
 \centering
 \includegraphics[width=0.4\textwidth]{figs/loops.pdf}
 \caption{An example of secondary structure $y$ and loops $\LP(y)$.} \label{fig:loop_supp}
\end{figure}
\begin{enumerate}
    \item Stack: $S\blangle (3, 50), (4, 49) \brangle$
    \item Stack: $S\blangle (4, 49), (5, 48) \brangle$
    \item Multiloop: $M\blangle (5, 48), (9, 24), (28, 44)\brangle$
    \item Stack: $S\blangle (9, 24), (10, 23) \brangle$
    \item Bulge: $B\blangle (10, 23), (11, 19) \brangle$
    \item Stack: $S\blangle (11, 19), (12, 18) \brangle$
    \item Hairpin: $H\blangle (12, 18) \brangle$
    \item Stack: $S\blangle (28, 44), (29, 43) \brangle$
    \item Internal Loop: $I\blangle (29, 43), (32, 39)\brangle$
    \item Stack: $S\blangle (32, 39), (33, 38) \brangle$
    \item Hairpin: $H\blangle (33, 38) \brangle$
    \item External Loop: $E\blangle (3, 50) \brangle$
\end{enumerate}
\fi
\vspace{-1.0cm}
\section{Projection and Intersection}\label{sec:op}
\begin{algorithm}[H]
%\NoCaptionOfAlgo
\caption{Projection $\hat{\vecx} = \vecx \proj I$}\label{alg:proj}
\begin{algorithmic}[1]
    \Function{Projection}{$\vecx, I$} \Comment{$I = [i_1, i_2, \ldots, i_n]$ is a list of critical positions}
    \State $\hat{\vecx} \gets \text{map}()$
    \For{$i$ in $I$}
        \State $\hat{\vecx}[i] \gets \vecx_i$ \Comment{Project the $i$-th nucleotide to index $i$}
    \EndFor
    \State \Return $\hat{\vecx}$
    \EndFunction
\end{algorithmic}
\end{algorithm}
\vspace{-1.0cm}
\begin{algorithm}
%\NoCaptionOfAlgo
\caption{Contraint Intersection $C' = \text{Intersection}(C_1, C_2)$}\label{alg:contract}
\begin{algorithmic}[1]
    \Function{Intersection}{$C_1, C_2$} \Comment{$C_1, C_2$ are sets of constraints}
     \State $(I_1, \setxi_1) \gets C_1$ \Comment{ $I$: critical positions; $\setxi$: a set of nucleotides compositions}
     \State $(I_2, \setxi_2) \gets C_2$
     \State $I' \gets I_1 \cap I_2$

     \If{$I' = \emptyset$} \Comment{No overlapping positions; return original constraints}
        \State \Return $C_1, C_2$ 
     \EndIf

     \State $\setxi'_1 \gets \{ \hat{\vecx} \proj I' \mid \hat{\vecx} \in \setxi_1 \}$
     \State $\setxi'_2 \gets \{ \hat{\vecx} \proj I' \mid \hat{\vecx} \in \setxi_2 \}$

     \For{$\hat{\vecx} \in \setxi_1$} \Comment{Remove nucleotides compositions from $\setxi_1$ that is not in $\setxi_2$}
        \If{$\hat{\vecx} \proj I' \notin \setxi'_2$}
        $\setxi_1 \gets \setxi_1 \setminus \{\hat{\vecx}\}$
        \EndIf
     \EndFor

     \For{$\hat{\vecx} \in \setxi_2$} \Comment{Remove nucleotides compositions from $\setxi_2$ that is not in $\setxi_1$}
        \If{$\hat{\vecx} \proj I' \notin \setxi'_1$}
        $\setxi_2 \gets \setxi_2 \setminus \{\hat{\vecx}\}$
        \EndIf
     \EndFor

     \State $C'_1 \gets (I_1,~\setxi_1)$
     \State $C'_2 \gets (I_2,~\setxi_2)$
    %  \State $C' \gets (I_1 \cup I_2,~\setxi_1 \cup \setxi_2)$
     \State \Return $C'_1 \cup C'_2$     \Comment{Return updated constraints}
     \EndFunction
\end{algorithmic}
\end{algorithm}
\section{Brute-Force Enumeration and  Folding}\label{sec:bf}
Given a target motif $\mstar \subseteq \vecy$, the most straightforward method is to enumerate all possible nucleotides compositions , and check whether there exists at least one composition that can fold into $\mstar$ under the constraint $\vecy\setminus\mstar$. However, this is impractical in reality because of high time cost. The design for $\mstar$ should at least satisfy that nucleotides at the paired position should be matchable, the number of brute-force enumeration is $6^{|\pairs(\mstar)|}\times4^{|\unpaired(\mstar)|}$, as there are $6$ choices for a pair and $4$ types of nucleotides. Constrained folding algorithms typically have a cubic time complexity with respect to  length, the overall complexity~$$\mathcal{O}(6^{|\pairs(\mstar)|}\times4^{|\unpaired(\mstar)|}\cdot (2|\pairs(\mstar)|+|\unpaired(\mstar)|)^3)$$ makes  exhaustive search impractical even for small structures.

\end{document}